%% file: 00_main.tex
\documentclass[acmtog]{acmart}

\usepackage{xargs}  
\usepackage[colorinlistoftodos,prependcaption,textsize=tiny,backgroundcolor=white,linecolor=white,disable]{todonotes}
\usepackage{eso-pic}
\usepackage{subcaption}
\usepackage{dirtytalk}
\usepackage{makecell}
\usepackage{booktabs}
\usepackage{array}
\usepackage{microtype}
\usepackage{ragged2e}
\usepackage{multirow}
\usepackage{cleveref}
\crefname{figure}{Fig.}{Figs.}
\crefname{section}{Sec.}{Secs.}
\crefname{subsection}{Sec.}{Secs.}
\crefname{subsubsection}{Sec.}{Secs.}
\crefname{table}{Tab.}{Tabs.}

\AtBeginDocument{%
  \providecommand\BibTeX{{%
    \normalfont B\kern-0.5em{\scshape i\kern-0.25em b}\kern-0.8em\TeX}}}

\setcopyright{acmlicensed}
\copyrightyear{2022}
\acmYear{2022}
\acmDOI{10.1145/1122445.1122456}

\acmJournal{TOG}

\usepackage{xcolor}
\usepackage{colortbl}

\usepackage{hyperref}

\begin{document}

\definecolor{colorMC}{rgb}{0.0, 0.8, 0.0}
\newcommand{\MC}[1]{\textcolor{colorMC}{#1}}
\newcommandx{\MCx}[2][1=]{\todo[linecolor=colorMC,backgroundcolor=colorMC!15,bordercolor=colorMC,#1]{MC: #2}}

\definecolor{colorAS}{rgb}{0.6, 0.0, 0.8}
\newcommand{\AS}[1]{\textcolor{colorAS}{#1}}
\newcommandx{\ASx}[2][1=]{\todo[linecolor=colorAS,backgroundcolor=colorAS!15,bordercolor=colorAS,#1]{AS: #2}}

\definecolor{colorDB}{rgb}{0.0, 0.0, 0.9}
\newcommand{\DB}[1]{\textcolor{colorDB}{#1}}
\newcommandx{\DBx}[2][1=]{\todo[linecolor=colorDB,backgroundcolor=colorDB!15,bordercolor=colorDB,#1]{DB: #2}}

\definecolor{colorNP}{rgb}{0.9, 0.0, 0.0}
\newcommand{\NP}[1]{\textcolor{colorNP}{#1}}
\newcommandx{\NPx}[2][1=]{\todo[linecolor=colorNP,backgroundcolor=colorNP!15,bordercolor=colorNP,#1]{NP: #2}}

\newcommand{\cino}[1]{\textcolor{magenta}{cino: #1}}
\newcommandx{\cinox}[2][1=]{\todo[linecolor=magenta,backgroundcolor=magenta!15,bordercolor=magenta,#1]{cino: #2}}

\definecolor{colorFL}{rgb}{0.9, 0.1, 0.1}
\newcommand{\FL}[1]{{\textcolor{colorFL}{(franck)~#1}}}
\newcommandx{\FLx}[2][1=]{\todo[linecolor=colorFL,backgroundcolor=colorFL!15,bordercolor=colorFL,#1]{FL: #2}}

\definecolor{colorGMC}{rgb}{0.9, 0.5, 0.2}
\newcommand{\GMC}[1]{\textcolor{colorGMC}{#1}}
\newcommandx{\GMCx}[2][1=]{\todo[linecolor=colorGMC,backgroundcolor=colorGMC!15,bordercolor=colorGMC,#1]{Gian: #2}}

\definecolor{colorGMC}{rgb}{0.6, 0.9, 0.1}
\newcommand{\alla}[1]{\textcolor{colorGMC}{#1}}
\newcommandx{\allax}[2][1=]{\todo[linecolor=colorGMC,backgroundcolor=colorGMC!15,bordercolor=colorGMC,#1]{Alla: #2}}

\definecolor{colorRS}{rgb}{0.5, 0.0, 0.0}
\newcommand{\RS}[1]{\textcolor{colorRS}{#1}}
\newcommandx{\RSx}[2][1=]{\todo[linecolor=colorRS,backgroundcolor=colorRS!15,bordercolor=colorRS,#1]{Ric: #2}}

\newcommand{\TODO}{\textcolor{red}{\textbf{TODO}}}
\newcommand{\Q}{\textcolor{red}{\textbf{Q}}}

\newcommand{\minrev}[1]{\textcolor{black}{#1}}

\title{Hex-Mesh Generation and Processing: a Survey}


\author{Nico Pietroni}
\email{nico.pietroni@uts.edu.au}
\affiliation{University of Technology Sydney, Australia}

\author{Marcel Campen}
\email{campen@uos.de}
\affiliation{Osnabr\"uck University, Germany}

\author{Alla Sheffer}
\email{sheffa@cs.ubc.ca}
\affiliation{University of British Columbia, Canada}

\author{Gianmarco Cherchi}
\email{g.cherchi@unica.it}
\affiliation{University of Cagliari, Italy}

\author{David Bommes}
\email{david.bommes@inf.unibe.ch}
\affiliation{University of Bern, Switzerland}

\author{Xifeng Gao}
\email{gxf.xisha@gmail.com}
\affiliation{Tencent America, USA}

\author{Riccardo Scateni}
\email{riccardo@unica.it}
\affiliation{University of Cagliari, Italy}

\author{Franck Ledoux}
\email{franck.ledoux@live.fr}
\affiliation{CEA, France}

\author{Jean-Fran\c{c}ois Remacle}
\email{jean-francois.remacle@uclouvain.be}
\affiliation{Universit\'e catholique de Louvain, Belgique}

\author{Marco Livesu}
\email{marco.livesu@gmail.com}
\affiliation{CNR IMATI, Italy}

\renewcommand{\shortauthors}{, et al.}

\begin{teaserfigure}
	\includegraphics[width=.99\linewidth]{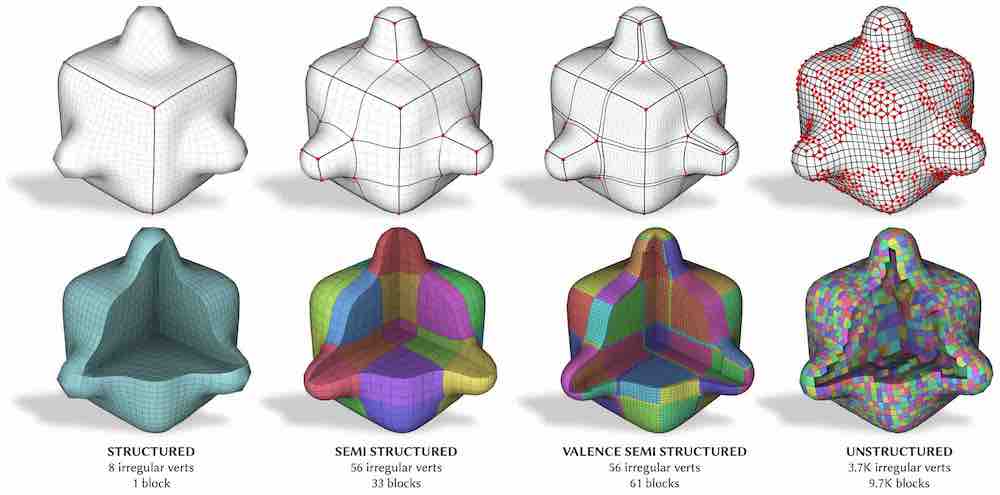}
	\caption{\minrev{Hex-meshes can be categorized according to their structural regularity, which depends on the amount of irregular edges and vertices present in the mesh (red dots) and on how they are connected to each other (thick lines).
	The first three meshes were computed with a polycube method~\cite{polycut_livesu}, the latter with an octree method~\cite{gao2019feature}.}}
	\label{fig:structure}
\end{teaserfigure}

\begin{abstract}
In this article, we provide a detailed survey of techniques for hexahedral mesh generation. We cover the whole spectrum of alternative approaches to mesh generation, as well as post processing algorithms for connectivity editing and mesh optimization. For each technique, we highlight capabilities and limitations, also pointing out the associated unsolved challenges. Recent relaxed approaches, aiming to generate not pure-hex but hex-dominant meshes, are also discussed. The required background, pertaining to geometrical as well as combinatorial aspects, is introduced along the way.
\end{abstract}



\maketitle

\input{01_intro}

\input{02_fundamentals}

\input{03_elem_quality}
\input{04_meshing}
\input{05_topological_operations}

\input{06_refinement_coarsening}
\input{07_optimization_untangling}

\input{08_visualization}

\input{09_outro}

\bibliographystyle{ACM-Reference-Format}
\bibliography{biblio}


\end{document}

%% file: 01_intro.tex
\section{Introduction}
\label{sec:intro}
\minrev{Volume meshes explicitly encode both the surface and the interior of an object, thus offering a richer representation than surface meshes. They are primarily used in industrial and biomedical applications, where volume elements are exploited to encode various information, such as structural and material properties, permitting to simulate and precisely estimate the physical behavior of an object subject to external or internal forces, or the dynamics involving multiple objects interacting in the same environment.
Alongside tetrahedra, hexahedral elements are the most prominent solid elements used to represent discrete volumes in computational environments. Meshes entirely or partially made of hexahedra have been used for many years as the computational domain to solve partial differential equations (PDEs) that are relevant for the automobile, naval, aerospace, medical and geological industries to name a few,
and are at the core of prominent software tools used by such industries, such as~\cite{meshgems,ansys,hypermesh,cubit,coreform,Tessael}.\\}

\minrev{In academic research, the generation and processing of hexahedral meshes have been studied for more than 30 years now.} Despite the huge effort that various scientific and industrial communities have spent so far, the computation of a high-quality hexahedral mesh conforming to (or approximating) a target geometry remains a challenge with various open aspects for which no fully satisfactory solutions have been provided yet. Some of the known methods are extremely robust and scale well on complex geometries; some others produce high-quality meshes; some others are fully automatic. But no known method successfully combines all these properties into a single product. \minrev{The hex-meshing problem had been so elusive that it was even once termed the \say{holy grail} of mesh generation~\cite{blacker2000meeting}. Ever since, many advancements in the field have been made, while major challenges still remain.\\}

In the last decade, the Computer Graphics community has contributed \minrev{significantly} to the hex-meshing problem, proposing both seminal ideas and practical algorithms. In this survey, we wish to summarize this work, also reporting on previous methods developed by other scientific communities. 

The engineering community has already produced a few surveys on this topic, but they are either no longer up to date~\cite{schneiders2000algorithms,owen1998survey,tautges2001generation,blacker2000meeting} or focus just on a particular narrow subset of the available techniques~\cite{Shepherd2008,survey1,sarrate2014unstructured}. We wish to create a comprehensive entry point for researchers and practitioners dealing with hexahedral meshing. We therefore embrace the whole field, \minrev{revisiting} and structuring a vast amount of literature, and covering basic topological (\cref{sec:fundamentals}) and geometrical (\cref{sec:quality}) concepts, all kinds of approaches to hexahedral mesh generation (\cref{sec:methods}), operators to edit mesh connectivity and to perform refinement or coarsening (\cref{sec:operators}), mesh optimization and untangling (\cref{sec:untangling}), visual exploration (\cref{sec:visualization}), and also addressing the recent trend of methods for hex-dominant meshing (\cref{sec:hexdominant}). Last but not least, in the final part of the survey, we highlight the current challenges the field is facing and indicate interesting directions for future work (\cref{sec:future}).

%% file: 02_fundamentals.tex
\section{Hex-Mesh Structure}
\label{sec:fundamentals}

\minrev{A hexahedral mesh has structural aspects (concerning the connectivity of mesh elements) and geometric aspects (concerning the elements' shape and their embedding or immersion in space).
In this section we focus on the diverse set of structural aspects, and consider geometry in \cref{sec:quality}.}

\subsection{Primal structure} \todo{\DB{(David)}}
\label{sec:primalstructure}

\minrev{In terms of connectivity, a hexahedral mesh} is a \minrev{3-dimensional} \emph{cell complex}, $\mathcal{H} = (V,E,F,C)$, consisting of vertices $V$ (0-cells), edges $E$ (1-cells), facets $F$ (2-cells), and cells $C$ (3-cells). The facets $F$ are also often referred to as faces, and the 3-cells $C$ are, given the context, often referred to as hexahedra or hexes. In a \emph{pure hexahedral mesh}, \minrev{each facet is a topological quadrilateral (i.e., incident to four edges) and each cell is a topological cube (i.e., incident to six such facets)}. If a relatively small number of facets and cells are of different type (e.g., tetrahedra, prisms, or pyramids) a mesh is called \emph{hexahedral dominant}. 

\minrev{On top of this connectivity structure, a mesh is equipped with a geometric structure, typically an embedding (or immersion) in $\mathbb{R}^3$ (\cref{sec:quality}).}

\minrev{Often, instead of assuming fully generic CW or $\Delta$ cell complexes \cite{Hatcher:AT}, more restricted connectivity definitions are used for practical purposes \cite{erickson2013theoretical}. A very common one is to assume that each cell has eight \emph{distinct} vertices, i.e., no hexahedron is self-adjacent at a vertex, edge, or facet. Similarly, pairs of edges, facets, or hexes being adjacent via more than one vertex, edge, or facet, respectively, may be ruled out. This simplifies data structures and algorithms; furthermore, many applications assume each hex to be embedded in a geometrically simple way (e.g. straight edges, ruled facets, cf.~\cref{sec:quality}) which rules out such self-adjacency and multi-adjacency anyway. \cref{sec:outputreq} discusses further application-dependent structural assumptions and requirements.}

\subsubsection{Singularities}
The most \emph{regular} hexahedral mesh is an (infinite) Cartesian grid, where each vertex, edge, and facet is incident to 8, 4, and 2 hexahedra, respectively.
General hexahedral meshes contain elements of different local connectivity, which are accordingly called \textit{irregular} or  \textit{singular}. Irregular facets simply correspond to the \textit{boundary} of the mesh, i.e., all facets that are incident to a single hexahedron.

Since interior facets cannot be irregular and vertex singularities are never isolated~\cite{liu2018singularity}, structurally most interesting is the set of irregular edges, which forms the so-called \textit{singularity graph}.

\begin{figure}[tb]
	\includegraphics[width=.8\linewidth]{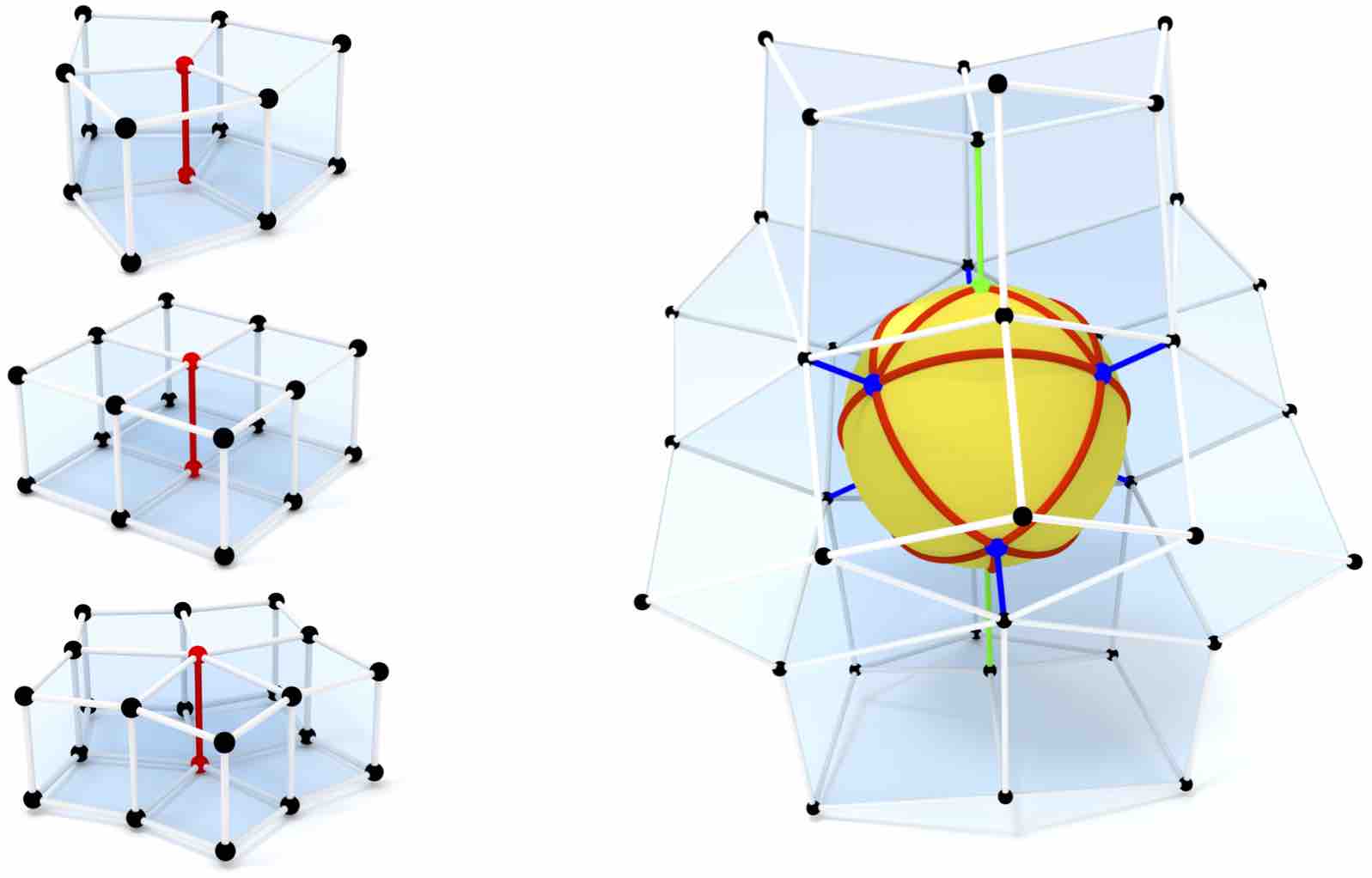}
	\caption{Left: Singular edges of valence $k=\{3,4,5\}$. Right: There is a 1:1-correspondence between configurations of vertices in a hexahedral mesh and triangulations of the sphere. Image from~\cite{liu2018singularity}.}
	\label{fig:singularities}
\end{figure}

\begin{figure}[tb]
	\includegraphics[width=.7\linewidth]{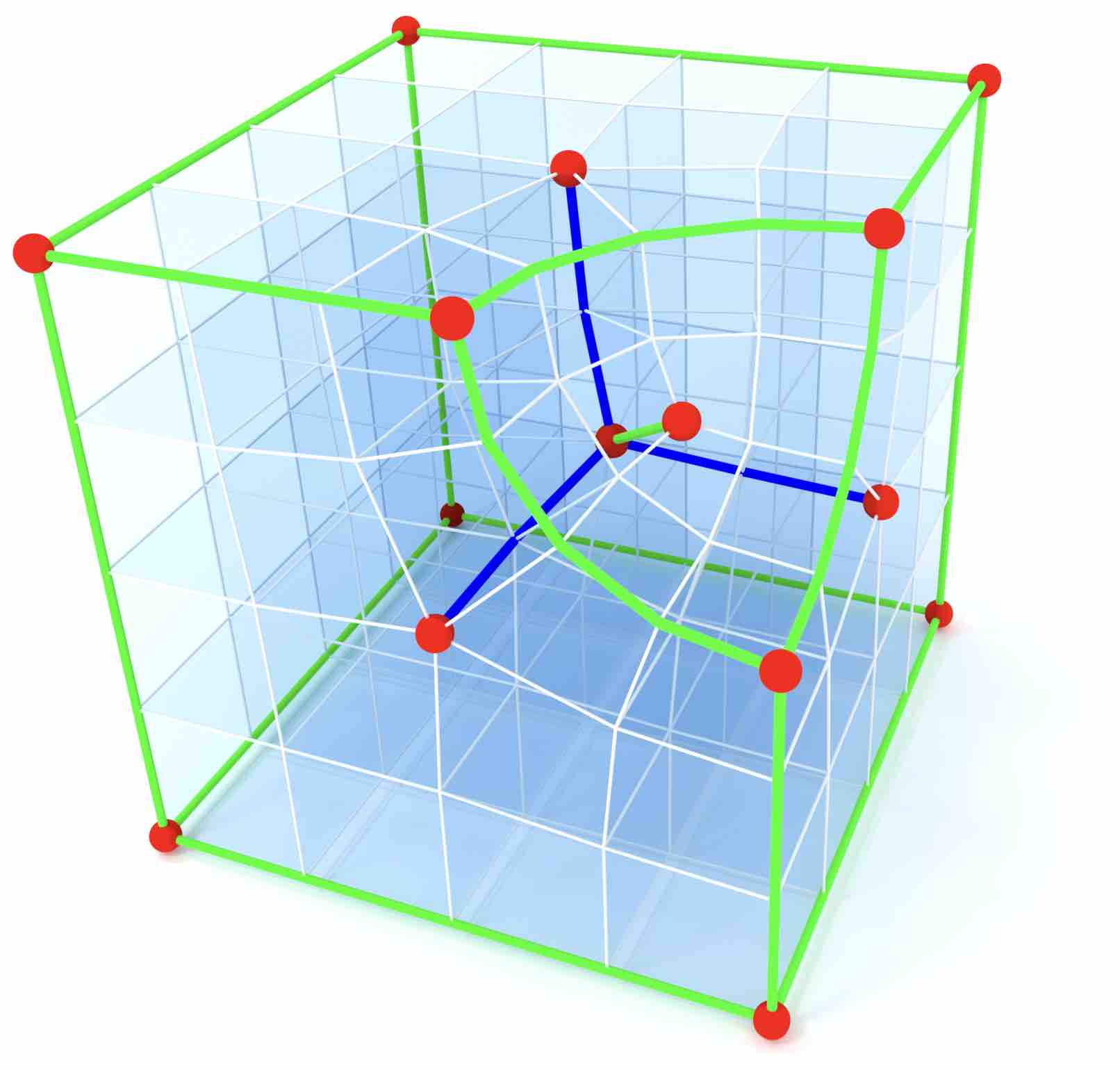}
	\caption{The singularity graph is formed by interior singular edges of valence $k=3$ (green) and $k=5$ (\minrev{blue}) and boundary edges of valence $k=1$ (green). Chains of edges with homogeneous type connect or terminate at singular vertices (red). Image from~\cite{liu2018singularity}.}
	\label{fig:singularity_graph}
\end{figure}

\paragraph{Singularity Graph}
The fundamental building block of the singularity graph is a \textit{singular edge} of valence $k$, i.e., an interior edge incident to $k\neq 4$ hexahedra, or a boundary edge incident to $k\neq 2$ hexahedra (cf.~\cref{fig:singularities}). While a single integer $k$ is sufficient to characterize the structural type of an edge, the specification of vertex types is more complex. As observed by Nieser et al.~\shortcite{Nieser2014} there is a 1:1-correspondence between vertex configurations in a hexahedral mesh, and triangulations of the 2-sphere. This can be understood by observing that the intersection of a cube with a sphere centered at one of it's corners results in a triangular patch (\cref{fig:singularities}). Hence, different vertex types can be specified by enumerating all triangulations of the sphere. Restricting to (the practically most relevant) edge valences $\{3,4,5\}$, it turns out that only $11$ different configurations of interior vertex types exist~\cite{sabin1997spline,liu2018singularity}. Specifically, for an interior vertex it is impossible to be incident to a single singular edge, and in case of two incident singular edges they can only be of identical type. Consequently, the singularity graph is formed by \textit{singular arcs}, which are chains of singular edges with identical type. These singular arcs either terminate at the boundary, or connect to other singular arcs at singular vertices, cf.~\cref{fig:singularity_graph}.

\subsection{Dual structure} 
\todo{\MC{(Marcel)}}

\label{sec:dual}

In a hexahedral mesh, regardless of its level of structural regularity (\cref{sec:regularity}), each cell has a constant number of 6 facets and each facet has a constant number of 4 edges. Conversely, however, each vertex may have a varying number of incident edges, and each edge a varying number of incident facets.

One may consider the (polyhedral) cell complex that is \emph{dual} to a hexahedral mesh: For each $k$-cell of the primal mesh there is a $(3-k)$-cell in 1:1-correspondence in the dual mesh and incidence relationships are adopted. The above regularity of cells and facets in the primal mesh translates into regularity of vertices and edges in the dual. Concretely, except at the mesh's boundary, each dual vertex has a constant number of 6 incident dual edges, and each dual edge has a constant number of 4 incident dual facets. Conversely, dual facets and dual cells are polygons and polyhedra of varying structure. Further details and facts about the dual complex can be found in~\cite{tautgesa2003topology}.

Depending on the algorithmic context, it may be advantageous to consider the primal or this dual view of a hexahedral mesh. A key reason is the following: 
\minrev{While vertices and edges of the primal mesh may be regular or singular, the vertices and edges of the dual mesh are all regular; this is due to the fact that all primal facets are quadrilaterals and all primal cells are hexahedra. The following useful definition of \emph{opposite} edges at a regular vertex and \emph{opposite} facets at a regular edge therefore applies everywhere in the dual mesh.}

\paragraph*{Opposite Elements} At each regular interior vertex $v$, there are~6 incident edges. For each edge $e_1$ of these, there is exactly one edge $e_2$ among these~6 that does not share a facet with $e_1$; the edges $e_1$ and $e_2$ are called \emph{opposite at} $v$. At each regular interior edge $e$, there are 4 incident facets. For each facet $f_1$ of these, there is exactly one facet $f_2$ among these 4 that does not share a cell with $f_1$; the facets $f_1$ and $f_2$ are called \emph{opposite at} $e$. \minrev{In the primal setting, this} concept of opposite edge is relevant for algorithms that trace internal arcs in the mesh, e.g., connecting pairs of singular vertices. Similarly, opposite facets are useful to flood internal facet sheets bounded by singular arcs, e.g., to perform a coarse block decomposition of a given mesh (\cref{sec:blockstructure}).

\begin{figure}[tb]
	\includegraphics[width=.70\linewidth]{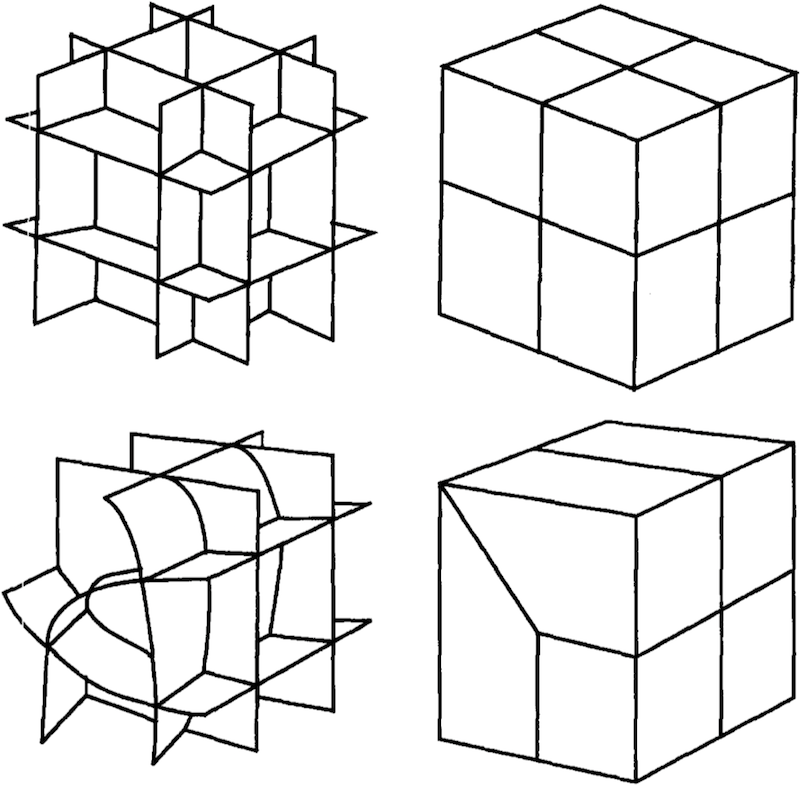}
	\caption{Two examples (top and bottom) of the relation between dual and primal mesh representation. Left: dual cell complex, formed by an arrangement of 2-manifold sheets. Right: corresponding hexahedral mesh. Image from~\cite{Murdoch97}.}
	\label{fig:STC}
\end{figure}

\minrev{In the dual setting, this opposite relation can be used to define the following:}

\paragraph*{Sheets} Consider the transitive closure of the dual facets' opposite-relation. Its equivalence classes are called \emph{sheets} (also referred to as twist planes~\cite{Murdoch97} or (pseudo-)hyperplanes~\cite{tautgesa2003topology}). These sheets are 2-manifold surfaces (with boundary, possibly self-intersecting) formed by dual facets.
\paragraph*{Chords} Analogously, the equivalence classes of the dual edges' opposite-relation's transitive closure are referred to as \emph{chords}~\cite{Murdoch97,borden2002hexahedral} (or \emph{polychords}~\cite{daniels2008quadrilateral}).  

The combinatorial  \minrev{\say{continuity}} of opposite dual facets across dual edges has inspired the early name \emph{spatial twist continuum} for this dual sheet based perspective.

It follows that the entire dual complex can be viewed as an arrangement of intersecting manifold surfaces (sheets): dual vertices are formed by three intersecting sheets, chords are formed by two intersecting sheets and split into dual edges by transversely crossing sheets, sheets are split into dual facets by crossing sheets, and dual cells are the spatial compartments enclosed by sheets.
Conceptually, a sheet corresponds to one \emph{layer} of hexahedra in the primal mesh; how this layer is composed of individual hexahedra, however, is not defined by this sheet itself but by sheets that cross this sheet transversely. \cref{fig:STC} illustrates this primal-dual relationship.

All this is in close analogy to dual complexes in the case of quadrilateral meshes. These can be viewed as arrangements of intersecting 1-manifolds~\cite{campen2012dual,campen2014dual}.
Quite differently, though, sheets can be topologically quite complex, they may have arbitrary genus and an arbitrary number of boundary loops, whereas in the quadrilateral mesh case each 1-manifold may only be either a closed loop curve, or an open-ended curve starting and ending at the mesh boundary.

\subsection{Block structure }

\todo{\MC{(Marcel)}}
\label{sec:blockstructure}
Each hexahedral mesh can be decomposed into disjoint blocks, where each block is a regular grid of hexahedra. Conversely, the mesh can be viewed as disjoint union of such blocks. As an extreme example, each hexahedron could be considered an individual block (of size $1\times 1\times 1$). We can distinguish conforming and non-conforming block decompositions: a block decomposition is conforming iff each side of each block coincides with one other block side (except at the mesh boundary).

Of particular practical relevance are decompositions that are conforming, and among these those that are coarse, i.e., that consist of a small number of blocks. Meshes rarely have a unique conforming block decomposition. The \emph{coarsest} conforming block decomposition is sometimes referred to as the mesh's \emph{base complex}~\cite{bommes2011global,gao2015hexahedral,razafindrazaka2017optimal}.

It is worth pointing out that the term base complex is sometimes used with alternative meanings~\cite{polycut_livesu,eck1996automatic,Dong2006,sheffer2007mesh}, for instance to refer to a coarse cell complex that is used as a domain for (cross)-parametrization. Note that this is not entirely unrelated though: a common use case of these parametrizations is structured remeshing; the resulting meshes typically exhibit a block structure induced by the underlying domain complex.

The base complex has the following defining property: a facet is part of a block side if and only if it is transitively incident to a singular edge via opposite facets (as defined in~\cref{sec:dual}). This suggests a simple way to extract the base complex of a given hexahedral mesh: starting from all facets incident to any singular edge, iteratively \minrev{expanding through} opposite facets across regular edges until termination.
Due to the practical relevance of semi-regular hexahedral meshes (\cref{sec:regularity}) mesh generation algorithms that take the coarseness of the implied base complex into account are of particular interest.

\subsection{Structure regularity}
\label{sec:structure_regularity}
\label{sec:regularity}
Similarly to quad-meshes~\cite{bommes2013quad}, hex-meshes can be roughly organized into four classes depending on the degree of regularity of their topological structure. The concept of mesh regularity is closely related with the relative amount of irregular vertices present in the mesh and with how these vertices are connected to each other.

\begin{itemize}
	\item \textbf{regular} (or structured) meshes have the topology of a gridded cube (\cref{fig:structure} left). These meshes are extremely convenient for storing and processing because of their simple connectivity: each internal vertex has exactly the same number of neighbors, with a consistent ordering. This allows for efficient storage and optimal query time, and also makes the computation of local quantities (e.g., finite differences) straightforward. There are, however, severe limits in the class of shapes they can represent: mapping an object containing long protrusions or deep cavities to a cube requires to dramatically distort the grid, likely resulting in a mesh with no practical usefulness due to the poor shape of its elements. Moreover, the rigid global structure does not allow for localized refinement: if more vertices are necessary around a specific area, the entire grid must be refined in order to maintain the \emph{pure hex} property;\\

	\item \textbf{semi-regular} (or semi-structured, also block-structured) hex-meshes are obtained by gluing in a conforming way several regular grids (also called \emph{patches} or \emph{blocks}). In a semi-regular hex-mesh all vertices that are internal to a patch are regular. Only vertices that lie at the edges or corners of a block may possibly be irregular. Semi-regular meshes represent the most important class in terms of applications, and are often the result of a manual or semi-manual meshing process. Differently from regular meshes they allow for higher flexibility and can be used to represent shapes of arbitrary complexity. At the same time, they contain a limited amount of irregular vertices, connected to each other so as to define a coarse block layout (\cref{fig:structure}, middle left) which can be exploited by dedicated data structures for cheaper storage and fast querying~\cite{tautges2004moab}, and is also useful in a variety of applications that exploit the tensor product structure of its elements (e.g., IGA~\cite{hughes2005isogeometric});\\ 
	
	\item \textbf{valence semi-regular} meshes also contain a limited amount of irregular vertices, but they are not connected in a way that induces a coarse block decomposition into few regular grids (\cref{fig:structure}, middle right). Meshes of this kind are often produced by modern hex-meshing algorithms such as frame field based methods, which introduce few singularities, but do not specifically address their connectivity pattern;\\ 
	
	\item \textbf{irregular} (or unstructured) hex-meshes contain a large fraction of irregular vertices (\cref{fig:structure}, right). Meshes of this kind are often produced via voxelization or other grid-based methods: portions of the object that do not align with the ambient Cartesian grid exhibit a typical staircase effect, triggering a proliferation of irregular vertices on the surface. Irregular meshes are not suited for applications that exploit the coarse block structure induced by the mesh connectivity, because the number of blocks is close to the number of hexahedra in the mesh. 

\end{itemize}

As for the quad-mesh case~\cite{bommes2013quad} the boundaries between semi-regular, valence semi-regular, and irregular meshes are blurred. Nevertheless, from an applicative perspective there is a substantial difference between these three classes and there exists a variety of structure enhancement algorithms that are specifically designed to improve mesh regularity (\cref{sec:struct_enahncement}). The whole taxonomy can be understood in terms of the ratio between the number of irregular vertices and the total amount mesh vertices ($r_V$), and the ratio between the number of blocks and the total number of mesh elements ($r_B$). If both $r_V$ and $r_B$ are high, the mesh is irregular; if $r_V$ is low but $r_B$ is high, the mesh is valence semi regular; if both $r_V$ and $r_B$ are low the mesh is semi-regular. Finally, if the number of blocks is exactly 1, the mesh is regular. Providing actual thresholds to precisely define what \textit{high} and \textit{low} mean, is an application dependent matter.

\begin{figure}[t]
	\includegraphics[width=\linewidth]{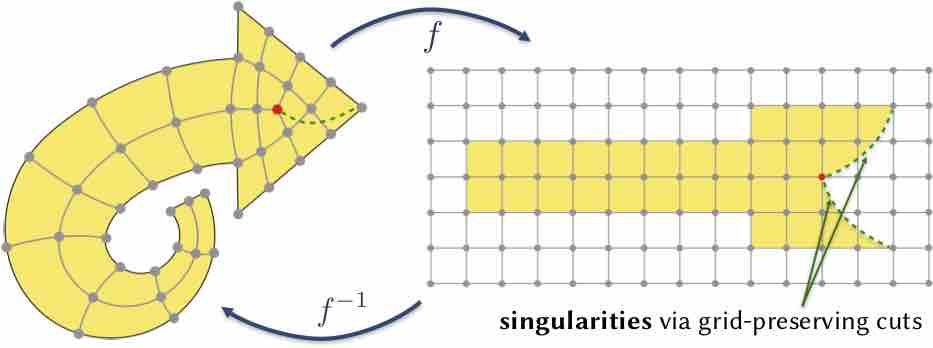}
	\caption{An integer-grid map $f$ deforms the shape (left) such that its boundary aligns with a Cartesian grid of integer isolines (right). Consequently, its inverse $f^{-1}$ deforms the covered grid cells into a structure-aligned mesh. Grid-preserving cuts (dashed green) enable irregular vertices (red) in the mesh.}
	\label{fig:IGM1}
\end{figure}

\subsection{Integer-Grid Maps }
 \todo{\DB{(David)}}
\label{sec:igm}
Integer-grid maps (IGM) are a class of maps that exist in arbitrary dimensions and which by construction induce structured meshes. The central idea\minrev{, as illustrated in \cref{fig:IGM1},} is to embed an $n$-dimensional shape into an 
$n$-dimensional voxel grid such that the inverse map deforms the set of covered voxels into a shape-aligned structured mesh. 
So far, integer-grid maps have been studied for 2-manifolds to generate quadrilateral meshes~\cite{kalberer2007quadcover,Bommes2013} and for 3-manifolds to generate hexahedral meshes~\cite{Nieser2014,liu2018singularity}. Similar to the parametrization of a general manifold, an integer-grid map can be decomposed into multiple charts. However, in order to guarantee that the inversely mapped voxels stitch conformingly, it is necessary to require specific transition functions that preserve the voxel grid. Assuming that the vertices of the voxel grid are given by integer coordinates $\mathbb{Z}^n$, the grid-preserving transition functions are exactly (i) integer translations and (ii) symmetry transformations of an $n$-cube. Such transition functions are essential to generate meshes with interior singularities, as for instance, the singular vertex (red) in \cref{fig:IGM1}.

\minrev{Mathematically, a map requires three properties to be an \emph{integer-grid map}: (i) grid-preserving transition functions, (ii) local injectivity, and (iii) singularities and boundaries mapping to integer-grid entities. A thorough definition can be found in \cite{liu2018singularity}.}

Integer-grid maps are sufficiently expressive to describe all potential hexahedral meshes. We can trivially generate a chart for each hexahedron that maps it to the voxel $[0,1]^3$. In this sense, integer-grid maps can be seen as an alternative representation of hexahedral meshes that has proven highly valuable for designing powerful generation algorithms.

\minrev{Reformulating the hexahedral mesh generation task as a map optimization problem offers many advantages. First of all, the optimization of low-distortion maps is a well-studied topic with a rich body of theory and algorithms that serve as a strong foundation. Moreover, the map optimization perspective enables multiple geometrically motivated continuous relaxations that are crucial for efficiently finding good approximate solutions of the hard underlying mixed-integer problem, e.g.~frame-fields to find suitable singularities (cf.~\cref{sec:frame_fields}), or seamless maps to estimate the required integer translations (cf. \cite{Nieser2014}, \cite{Brueckler:2021}).}

While a naive direct optimization formulation for a hexahedral mesh needs to explicitly encode and deal with the full set of (inherently discrete) elements and their connectivity, most of that becomes implicit in the map formulation, \minrev{enabling not only straightforward continuous relaxations but moreover a reduced set of discrete variables}. 
\minrev{A simple but instructive example consists of a regular block covering $n\times m \times o$ voxels in the IGM image. Stretching the image along the first coordinate axes corresponds to a continuous relaxation of the discrete action of changing the integer dimension $n$. Note that from the map perspective, the block is indeed fully characterized by only three integers $n$, $m$, and $o$, while a direct mesh optimization would need to deal with $(n+1)\times (m+1) \times (o+1)$ (discrete) vertices and their nontrivially-constrained connectivity. The number of integer degrees of freedom of a general integer-grid map is proportional to the number of singularities and topological handles. 
Consequently, in case of pre-determined singularities the resulting discrete search space is comparatively small since typically highly regular meshes with only few singularities are desired.}

\begin{figure}[tb]
	\includegraphics[width=\linewidth]{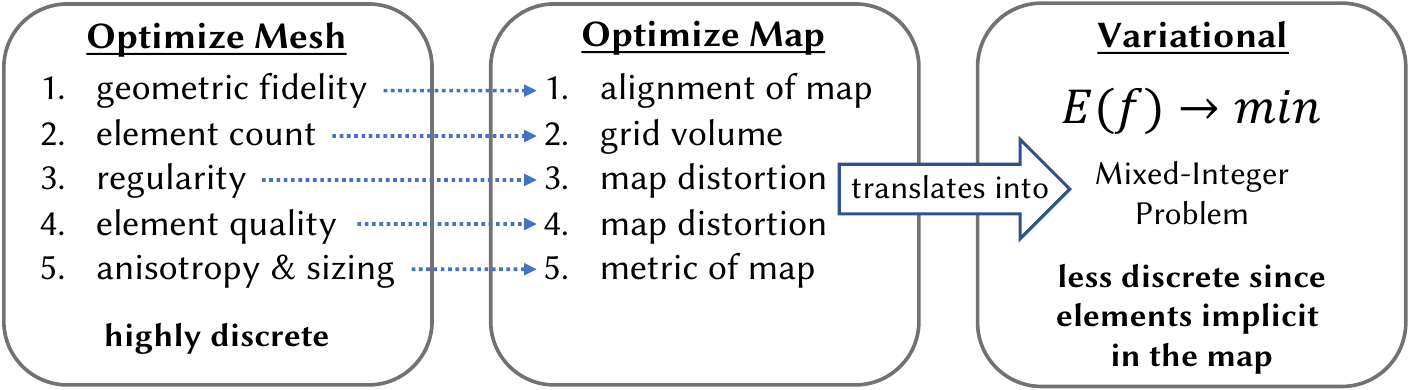}
	\caption{Integer-grid map optimization algorithms turn the highly discrete mesh optimization task into \minrev{a more tractable mixed-integer} map optimization. All mesh quality criteria \minrev{are directly related to} properties of the map\minrev{, as indicated by blue dashed arrows}. }
	\label{fig:IGM2}
\end{figure}

\minrev{Optimizing for a low-distortion map has two positive effects, (i) it directly promotes
well-shaped elements of high quality in the output hex-mesh, and (ii) it demotes the occurrence of spurious singularities.}

\minrev{A conceptual overview of interpreting mesh optimization as map optimization is shown in \cref{fig:IGM2}. The advantages
 related to superior continuous relaxations and compact discrete search spaces explain the popularity and success of integer-grid map based approaches in the automatic generation of structured meshes.}

A special case of integer-grid maps with a single chart and no interior singularities are \textit{polycube maps}, which are discussed in more detail in \cref{sec:polycubes}. The optimization of polycubes, therefore, targets a deformation of the input shape such that its surface aligns with the surface of the voxel grid. Despite the continuous nature of the deformation, the resulting optimization problem is nonetheless of mixed-integer type. The discrete degrees of freedom are the choices of how to map surface normals since valid solutions are only the coordinate axes $\{\pm x, \pm y, \pm z \}$ and a low-distortion assignment is not known a priori.

Frame-field based methods, which are discussed in more detail in \cref{sec:frame_fields}, target the generation of hexahedral meshes in two stages. From a high-level perspective, the first stage estimates the rotational part of the Jacobian of an integer-grid map, i.e., a frame-field, while the second stage constructs the map by inheriting the frame-field singularities. The decomposition is beneficial because the first stage can be formulated in a representation that automatically deals with the symmetry of the hexahedron. As a consequence, frame field singularities can be optimized solely with continuous degrees of freedom, converting the extremely difficult direct optimization of integer-grid maps into a tractable form.

%% file: 03_elem_quality.tex
\section{Hex-Mesh Geometry}
\label{sec:quality}

\todo{\MC{(Marcel -- partly)}}

Besides its combinatorial and topological structure (\cref{sec:fundamentals}), a hexahedral mesh's geometry, i.e., its embedding or immersion, typically in $\mathbb{R}^3$, plays an essential role in most applications.

This concerns the question of \minrev{geometric fidelity (to what extent the mesh conforms to the target shape)} 
and the question of element quality. This latter question is concerned with the \emph{shape} of a mesh's individual hexahedra or the distortion of \emph{maps} defining these hexahedra as deformations of an ideal (\emph{reference} or \emph{master}) element. 

\minrev{Depending on the application context, various geometric requirements may be in place: the mesh may be required to conform to a given boundary mesh or to interpolate it within some prescribed tolerance; facets may be required to be planar or to be convex; the above maps may be required to be locally injective or even to have bounded distortion in some particular sense. In the context of mesh generation (\cref{sec:methods}) the concrete requirements can have a significant influence on the hardness of the meshing problem. Many methods so far are unable to provide strict guarantees regarding such requirements, especially when  they are asked for in combination.}

Also the relevant notion of element quality, and the effect of low or high-quality elements, \minrev{are application dependent}. 
In the context of simulations by means of finite element methods (FEM), element quality can have a crucial impact on error estimates and convergence rates, thus simulation speed and accuracy~\cite{ciarlet2002finite,zlamal1968finite}, \minrev{and relevant quality measures depend on the type of simulation. In \cref{sec:outputreq} these varying requirements are discussed further.}

\paragraph{The Trilinear Element}
The geometry or embedding of hexahedral meshes is often represented by means of coordinates assigned to their vertices. This alone is sufficient only for simple applications. More commonly, the geometry of edges, faces, and cells has to be defined as well. A particularly simple (and common) scenario is the assumption of \emph{trilinear} elements (linear edges, bilinear faces, trilinear cells), as this does not require the specification of any further information---all other mesh elements' geometric embedding in $\mathbb{R}^3$ are derived from the vertex positions via multilinear interpolation. Precisely, a hexahedron's embedding (with vertex positions $p_{ijk}$) is defined via a \emph{geometric map} $\tau : [0,1]^3 \to \mathbb{R}^3$ (also called \emph{isoparametric map}) as follows:
$$\tau(u,v,w) = \sum_{i=0}^1\sum_{j=0}^1\sum_{k=0}^1 p_{ijk} B_{k}(w)B_{j}(v)B_{i}(u),$$
where
$$B_0(t) = 1-t \text{ and } B_1(t) = t.$$
The hexahedral element effectively is the image of an ideal cube $[0,1]^3$ under this map.
Note that the edges are straight line segments under this map; the faces are ruled surfaces (planar iff the four corner vertices are coplanar).
More generally, this definition can be extended to higher-order elements using higher-order basis functions $B_i^n$ (e.g., Bernstein polynomials of degree $n$, giving rise to tensor-product B\'ezier elements~\cite{prautzsch2002bezier}). In these higher-order cases, additional control points (besides the vertex points) come into play as coefficients for a higher number of basis functions.

The assessment of these elements' quality (or even just validity) is an application-dependent matter. In some cases it may be just the \emph{shape} of the region $\tau([0,1]^3) \subset \mathbb{R}^3$ that is of relevance, in others its concrete \emph{parametrization}, given by the map $\tau$, is crucial.

\subsection{\minrev{Geometric map}}

A particularly common measure of quality is the determinant of the geometric map's Jacobian $J_\tau$. It quantifies to what extent the hexahedron, defined through $\tau$, deviates (in terms of volume distortion) from the cube $[0,1]^3$. Note that $\det J_\tau$ depends on parameters $(u,v,w)$. Due to this dependence, the quality of an element (in contrast to the quality at a particular point) rather needs to be assessed by the extremal \minrev{value $\min_{[0,1]^3} \det J_\tau$.}

Note that $\det J$, while measuring volume distortion, is blind to angle distortion; it cannot distinguish sheared cubes from cubes. Additional angle-aware measures are thus often taken into account (\cref{sec:shapequality}).

\subsubsection{Element Validity}
If $\min_{[0,1]^3} \det J_\tau \leq 0$, the geometric map $\tau$ is non-injective and the implied element is said to be irregular. Sometimes a distinction is made between degeneration ($\det J_\tau = 0$) and inversion or fold-over ($\det J_\tau < 0$).

In the context of the finite element method, irregular elements must be considered invalid~\cite{mitchell1971forbidden,Knupp_1999}; with such elements, depending on the concrete setting, one may yield \say{inaccurate solutions or no solutions at all}~\cite{barrett1996jacobians}, solutions are \say{invalidated}~\cite{roca2011defining}, or \say{calculations cannot be continued}~\cite{salagame1994distortion}. Due to this crucial importance, specialized \emph{untangling} methods for the purpose of irregular element removal in hexahedral meshes have been proposed (\cref{sec:untangling}), that attempt to achieve $\min_{[0,1]^3} \det J_\tau > 0$.

\subsubsection{Computation}
\label{sec:detbounds}

The evaluation of $\det J_\tau$ at a concrete parameter point $(u,v,w)$ is quite easy. For the computation of the extrema $\min/\max_{[0,1]^3} \det J_\tau$, however, there is no closed-form expression. As this is particularly relevant to \emph{certify} regularity, simply probing at a number of well-distributed parameter points is a risky approach.

\paragraph{Determinant bounds:} Like $\tau$, the Jacobian determinant is a polynomial in $(u,v,w)$. It can thus be expressed in the  Bernstein basis as well:
$$\det J_\tau(u,v,w) = \sum_{ijk} d_{ijk} B_{k}(w)B_{j}(v)B_{i}(u).$$
Due to this basis' implied convex hull property (due to $0\leq B_i(t) \leq 1$ for $t\in [0,1]$) the function value is bounded from below by the smallest coefficient $\min_{ijk} d_{ijk}$ and from above by the largest coefficient $\max_{ijk} d_{ijk}$. The coefficients $d_{ijk}$ are easily computed from the vertex points $p_{ijk}$.
For the particular case of trilinear hexahedral elements, this is discussed in~\cite{johnen2017robust}. The same principle applies to higher-order elements as well as to simplicial (rather than tensor-product) elements~\cite{johnen2013geometrical,dey1999curvilinear,luo2002p,gravesen2012planar,mandad2020efficient}.

These bounds can be quite loose. They can, however, be tightened arbitrarily by re-expressing $\det J_\tau$ piecewise over subdomains of $[0,1]^3$~\cite{hernandez2006local}. This is accomplished (via affine reparametrization) using B\'ezier subdivision~\cite{prautzsch2002bezier}. Under repeated subdivision, the coefficients (and thus the derived bounds) converge to the actual function value~\cite{prautzsch1994convergence,leroy2008certificates}.

For use cases where precise knowledge of the Jacobian determinant's value range is not relevant but only injectivity is to be certified, simpler (possibly loose) conservative tests can be employed \cite{zhang2005subtetrahedral}. Various even simpler hypothetical tests (trying to derive bounds from determinant values at vertices or along edges) were shown to be false \cite{knupp1990invertibility,zhang2005subtetrahedral}.

\paragraph{Relaxation:} Through sum-of-squares (SOS) relaxation, the non-convex problem of finding the Jacobian determinant polynomial's global minimum (i.e., $\min_{[0,1]^3} \det J_\tau$) can be replaced by a convex problem \cite{Marschner2020SOS}. If a sufficiently high degree is chosen for the formulation of this replacement problem, the global minima coincide. A sufficient degree was determined empirically; a formal guarantee is outstanding.

\begin{table}
\centering
\caption{List of alternative metrics for hexahedral elements, from the Verdict library~\protect\cite{stimpson2007verdict}. Normal ranges are intended for elements not having degeneracies.}
\label{tab:quality_metrics}

\begin{tabular}{@{}lccc@{}}

\toprule
\multirow{2}{*}{\textbf{Metric}}   & \textbf{Overall}    & \textbf{Acceptable}  & \textbf{Value for}\\
& \textbf{range} & \textbf{range}         & \textbf{unit cube}\\
\midrule
Diagonal & $[0,1]$ & $[0.65,1]$ & 1\\
Dimension & $[0,\infty) $& app. dep. & 1\\
Distortion & $[0,1]$ &  $[0.5,1]$ & 1\\
Edge Ratio & $[1,\infty)$&  --- & 1\\
Jacobian & \minrev{$(-\infty,\infty)$}  & $[0,\infty)$ & 1\\
Max. Edge Ratio & $[1,\infty)$  &  $[1,1.3]$ & 1\\
Max. Asp. Frobenius  & $[1,\infty)$  &  $[1,3]$  & 1\\
Mean Asp. Frobenius    & $[1,\infty)$  &  $[1,3]$  & 1\\
Oddy                      & $[0,\infty)$  &  $[0, 0.5]$  & 0\\
Relative Size Squared    & $[0,1]$  &  $[0.5,1]$  & ---\\
Scaled Jacobian            & $[-1,1]$ &  $[0.5,1]$  & 1\\
Shape                     & $[0,1]$  &  $[0.3,1]$  & 1\\
Shape and Size        & $[0,1]$  &  $[0.2,1]$  & ---\\
Shear                    & $[0,1]$  &  $[0.3,1]$  & 1\\
Shear and Size           & $[0,1]$  &  $[0.2,1]$  & ---\\
Skew                      & $[0,1]$  &  $[0,0.5]$  & 0\\
Stretch                   & $[0,1]$  &  $[0.25,1]$  & 1\\
Taper                    & $[0,\infty)$  &  $[0,0.5]$  & 0\\
Volume (signed)   & $(-\infty,\infty)$  &  $[0,\infty)$  & 1 \\
\bottomrule
\end{tabular}
\end{table}

\subsection{Shape quality}
\label{sec:shapequality}
Besides metrics based on the pointwise assessment of the geometric map, there exist a variety of metrics based simply on the vertex positions that have been proposed in the literature to assess the quality of hexahedral elements or have been exploited in specific applications. The documentation of the Verdict library~\cite{stimpson2007verdict} -- a de facto standard for finite element mesh quality assessment --  exhaustively reports per-hex metrics, as well as associated bounds and commonly acceptable ranges. We succinctly report these metrics in~\cref{tab:quality_metrics}. For more details on how each metric is formulated, we point the reader directly to the original source. \minrev{It must be noted, though, that the question whether an element is good or at least acceptable can be highly application dependent; in FEM, for instance, elements far from being cube-shaped (in particular anisotropically stretched elements) can be ideal -- if they are aligned suitably, in a PDE-guided or even solution-adaptive manner \cite{knupp2007remarks}.}

%% file: 04_meshing.tex
\section{Hex-Mesh generation}
\label{sec:methods}

In this section, we survey all mesh generation techniques present in the literature to date. We firstly provide a general introduction about input and output requirements. Then, algorithms will be organized according to the meshing paradigm they implement.
The generation of hybrid, in particular hex-dominant, meshes containing spurious non-hexahedral elements is also discussed (\cref{sec:hexdominant}). \minrev{Finally, \cref{tab:summary} summarizes the main properties of each class of hex-meshing algorithms reported in this survey.}

\subsection{Input}
Input data can be either a surface or a volume mesh describing the target geometry. Methods that take a surface mesh or other surface description and produce a conforming hexahedralization are often called \emph{direct}~\cite{Shepherd2008}, as opposed to \emph{indirect} methods, which typically operate on a supporting tetrahedral mesh and produce hexahedra by modifying this mesh (through splitting, clustering, etc.) or by computing some volumetric mapping encoded on the vertices of this supporting mesh. 

The most trivial form of indirect hex-meshing consists of splitting each tetrahedron into four hexahedra via midpoint refinement~\cite{li1995hexahedral}. This technique is trivial to implement and always guarantees a correct result. However, it produces an unstructured mesh with an overly dense singular structure, also containing four times more \minrev{elements} than the input mesh. Therefore, this approach is unsuitable for real applications.
As will come clear in the remainder of this section, indirect hex-meshing has evolved significantly since these early days and now comprises highly advanced tools to convert a tet-mesh into a much coarser hex-mesh with cleaner singularity structure. Notably, indirect approaches that cluster tetrahedra to form hexahedra are quite predominant in hex-dominant meshing (\cref{sec:hexdominant}).

Most of the techniques discussed in this section make assumptions on the topology and geometry of the input mesh and are not able to operate on meshes containing topological (e.g., open boundaries, holes, or non-manifold elements) or geometric (e.g., intersecting or degenerate elements) defects. Methods that operate on a supporting tetrahedral mesh may leverage robust tetrahedralization techniques such as~\cite{hu2018tetrahedral,hu2020fast,diazzi2021convex}. Methods that operate on surface meshes can sanitize their inputs with known robust surface processing algorithms, such as~\cite{attene2013polygon,cherchi2020fast,zhou2016mesh,attene2010lightweight}.

In addition to the target geometry, algorithms may optionally take as input a variety of other desiderata, such as target edge lengths or density fields to control local element size, or a list of features that the output mesh should conform to. Typical features are geometric curves on the outer surface (i.e., sharp creases), but there may also be additional ones -- both internal and external -- such as separation membranes between different materials, or other forms of semantic attributes. Finally, methods based on guiding fields (see \cref{sec:frame_fields}) may also take as input some additional parameters that control the field generation, or may even assume the whole guiding field as an input by itself.

\begin{figure*}
\centering
\includegraphics[width=\linewidth]{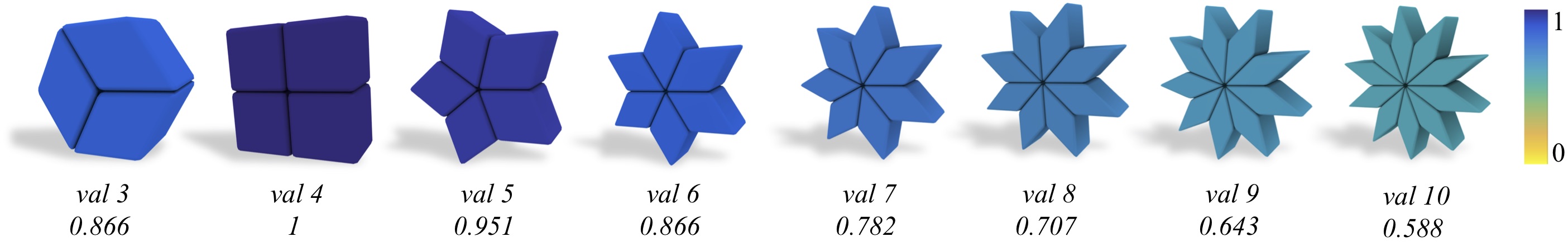}
\caption{\minrev{Topology and geometry are tightly coupled: the number of hexahedra incident to a singular edge directly bounds the inner angles, thus affecting the geometric quality of the elements (numbers below each configuration refer to the Scaled Jacobian of the geometric map). Image from~\cite{livesu2021optimal}.}}
\label{fig:val}
\end{figure*}

\subsection{Output}
\label{sec:outputreq}
\begin{figure}
	\centering
	\includegraphics[width=\columnwidth]{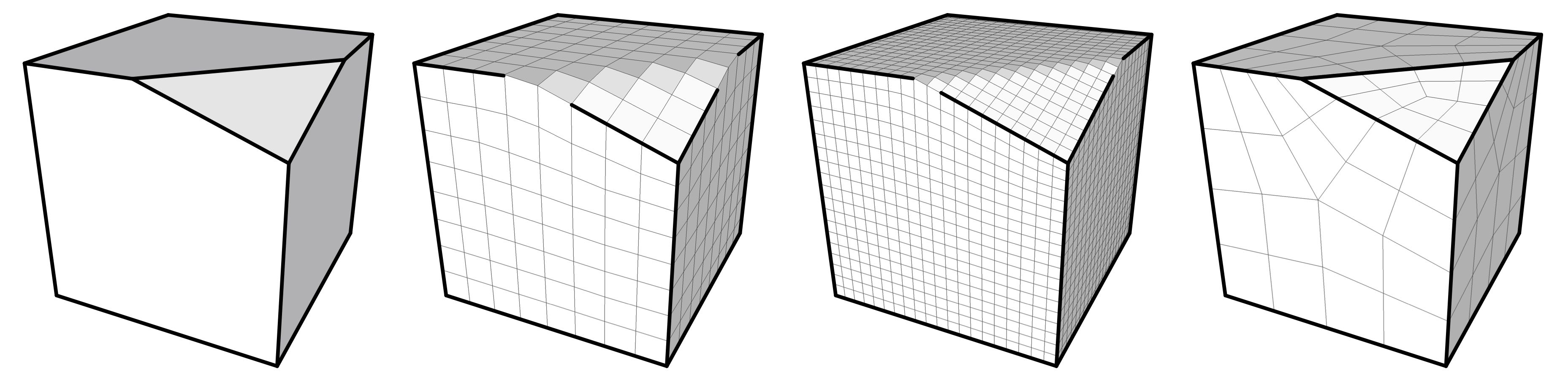}
	\caption{Incorporating an input feature network (left, bold lines) into the output hex-mesh is not possible if the connectivity does not align to it (middle left), even refining the mesh (middle right). Key to feature preservation is the ability to align surface edges to the input network, carefully positioning mesh singularities (right). Image from~\cite{LoopyCuts2020}.}
	\label{fig:feat_alignment}
\end{figure}

Output meshes must satisfy a variety of requirements, some of them strictly, some others loosely.
\minrev{In the following we list the most important topological and geometric requirements, also connecting them with specific applications that demand their fulfillment. The main \textbf{topological} desiderata are:
\begin{itemize}
\item\textbf{element type: } methods that strive for pure hexahedral meshing must ensure that all their cells are topological cuboids made of 8 vertices, 12 edges, and 6 quadrilateral faces. This requirement is loosened for hex-dominant methods, where spurious non-hex elements may be present in the output mesh. This topological freedom is not unlimited, and may be bounded by the specific application. In fact, methods for the numerical solution of PDEs often require non-hex elements to belong to a restricted class of polyhedra. For example, the Poly-Spline Finite Element Method~\cite{SchneiderDGBPZ19} demands that all mesh elements (non-hexahedra included) have quadrilateral faces, and enforces this property through mesh subdivision if the input mesh does not fulfill this requirement. Similar restrictions are also imposed by alternative methods;\\
\item\textbf{local structure: } topological limitations may apply not only at a local (per element) level, but also involve clusters of adjacent cells. For instance, the Poly-Spline Finite Element Method \cite{SchneiderDGBPZ19} requires that two non-hex cells are not face-, edge-, or vertex-adjacent, and also that non-hex cells are not exposed on the boundary. More generally, many methods that employ higher order basis functions can handle just a few local configurations, and put constraints on the local mesh patterns. This holds for both hex and hex-dominant meshes. For example, the blended spline method for unstructured hexahedral meshes  proposed in~\cite{wei2018blended} embraces only a small fraction of the possible singularities that are created by the meshing methods surveyed in this section. To this end, the intricate mesh connectivity generated by grid-based methods can be extremely challenging~\cite{livesu2021optimal};\\
\item\textbf{global structure: } depending on how the singular elements align, the mesh may or may not have a coarse block structure (\cref{sec:structure_regularity}). While basic numerical schemes like the Finite Element Method operate at a local (per element) level and may not exploit this property, block-structured meshes may be highly important for methods that employ tensor product constructions per block, for multi-grid solvers that rely on a hierarchy of nested meshes, and also for mesh compression~\cite{tautges2004moab};\\
\item\textbf{conformity: } some hex-dominant methods restrict their output to a narrow class of alternative polyhedra (e.g., permitting only tetrahedra and hexahedra). On the positive side, this restricts the alternative types of cells that applications must handle. On the negative side, the resulting meshes may be \emph{non-conforming}, meaning that structural discontinuities arise between elements that are geometrically but not topologically adjacent (due to T-junctions). Topological continuity can be restored using special \emph{connectors}. For example, a mesh containing two tetrahedra that are jointly adjacent to a hexahedron can be made conforming by adding a zero volume element containing one quadrilateral (at the hex side) and two triangular (at the tet side) facets. Nevertheless, the resulting meshes (with or without connectors) are not supported by all numerical solvers, and dedicated numerical schemes (e.g., Discontinuous Galerkin~\cite{chan2016gpu}) must be used.
\end{itemize}
From the geometric point of view, the output meshes should faithfully represent the target shape, preserve its prescribed features (if any), and be composed of well-shaped elements. More precisely, the main \textbf{geometric} desiderata are:
\begin{itemize}
\item\textbf{fidelity: } geometric fidelity is achieved by construction by methods that conform to an input quadrilateral mesh. Conversely, many other methods typically deviate from the target geometry and may only produce a geometric approximation of it. Just a handful of methods provide strict guarantees on the maximum (Hausdorff) distance from the reference geometry (e.g., ~\cite{gao2019feature}), whereas in the majority of the cases an unbounded approximation of the input geometry is produced. Depending on the complexity of the input shape, significant deviation from the target geometry may be present;\\
\item\textbf{features: } special care must be paid for input features such as sharp creases. While the general requirement is the same as for geometric fidelity, imprecision in the geometric approximation of features is both aesthetically much more evident and may also have a significant impact in the solution of the PDE (e.g., when studying the aerodynamic flow around creased objects). Feature alignment requires that sequences of edges of the hex-mesh conform to feature curves, otherwise some deviation is inevitable, regardless of resolution (\cref{fig:feat_alignment});\\
\item\textbf{quality: } the assessment of the quality of a mesh is a major topic by itself~\cite{knupp2007remarks} that is only touched upon in this survey (\cref{sec:quality}). It is important to note that the relation between mesh quality and, e.g., the quality of a numerical solution of a PDE may heavily depend on the concrete PDE as well as on the solver at hand. While a common requirement is that all mesh elements are \emph{valid} (everywhere positive Jacobian determinant of the geometric map), different numerical schemes may demand the fulfillment of additional requirements. Shape regularity criteria for the Finite Element Method (FEM) are mostly concerned with star-shapedeness and avoidance of large angles~\cite{ciarlet2002finite,zlamal1968finite,shewchuk2002good}. As recently shown, these methods can be modified in order to even cope with badly shaped elements, locally selecting higher order basis that compensate for the lack of geometric quality~\cite{schneider2018decoupling}. In Computational Fluid Dynamics (CFD) it can be beneficial to use meshes that are \emph{orthogonal}, meaning that the interface between two shared elements and the line connecting their centroids form a right angle~\cite{moraes2013analysis,aqilah2018study}. The Virtual Element Method~\cite{beirao2014hitchhiker} assumes that all mesh faces are planar. Considering this jungle of metrics that are relevant for one numerical method or the other, general purpose algorithms are often not suited to address these specific criteria at the mesh generation stage, but mainly strive to create meshes with valid elements, possibly addressing further quality concerns in post processing (\cref{sec:untangling}). 
\end{itemize}
The methods surveyed in the following typically aim to create \say{good} meshes according to a subset of the criteria above. Fully and equally embracing both topological and geometric requirements at once can be a huge challenge, and many methods put a stronger focus on one aspect over the other. Some methods focus more on the topological aspects and may produce well structured meshes containing (near-)degenerate or even invalid elements. Some others may guarantee valid elements or even lower bounds on certain geometric quality measures but produce meshes with a highly irregular topological structure. Both flaws can potentially be alleviated to some extent in post-processing, using dedicated algorithms for structure simplification (\cref{sec:struct_enahncement}) or geometric enhancement (\cref{sec:untangling}). Certainly, topology and geometry are coupled to some extent. For instance, a mesh with poor topological structure often inevitably contains poorly shaped elements as well (\cref{fig:val}).
}

\subsection{Advancing/Receding front}
\label{sec:advancingfront}
\todo{\FL{}}
First attempts to algorithmically generate hexahedral meshes were made by extending 2D advancing-front algorithms that generated full quadrilateral meshes. Starting from a quad-meshed boundary, algorithms like~\cite{Blacker93,blacker97} incrementally insert hexahedra starting from the boundary. The volume is progressively filled until final small voids are solved with simple patterns made of a few hexahedral cells. Such an approach is challenging on two main points. First, fronts can collide during their generation, and geometrical intersection must be performed. Owen and Sunil~\shortcite{owen00} solve this issue by preserving a hybrid mesh during the whole process. Every created hex is inserted into this mesh, and front collisions are easily detected. The second point is much more problematic: 
there is no guarantee that the process will eventually generate a usable full hexahedral mesh. Starting from an even number of quads on the boundary of a remaining void, a structural decomposition into a set of hexahedral elements is guaranteed to exist~\cite{MitchellCharacterization1996}, but the geometrical quality of hexes can be very low. And if one ends up with an odd number of quads surrounding a remaining void, one cannot fill it up with hexahedral elements at all, necessitating a backtracking of the front propagation (with no general guarantee to perform better the next time).
The main reason for this inflexibility lies rooted in the fact that one cannot easily perform structural modifications on a 3D hexahedral mesh in a local manner (cf.~\cref{sec:operators}).

Considering the problem as being over-constrained, the next generation of advancing-front algorithms do not start from a quadrilateral boundary mesh, but rather from the geometric surfaces~\cite{Plastering1,Plastering, staten10}. Complete layers of hexahedral cells are inserted in the domain until they collide. Final cavities are easier to fill, but this process can fail, too. In~\cite{RuizGironesRecedingFront}, the authors adopt the advancing-front technique. Considering that the final cavities that remain may be difficult to mesh, they use an inside-outside mesh generation approach that requires as an extra input an inner seed, which is a hexahedral mesh of a possible final cavity. Two solutions of the Eikonal equation are then computed: one going inward from the boundary of the geometric domain; another one going outward starting from the surface mesh of the inner seed. Both solutions are then combined to define a smooth distance function, and an advancing-front algorithm is performed to expand the quadrilateral surface mesh of the inner seed towards the unmeshed external boundary using the distance function to locate points of each layer of cells. This process is used in practice to mesh the outside of objects like aircraft (for aerodynamics problems, for instance). But it remains limited to geometric domains that are homeomorphic with the sphere, and the domain must not have sharp features.

\begin{figure}
	\centering
	\includegraphics[width=\columnwidth]{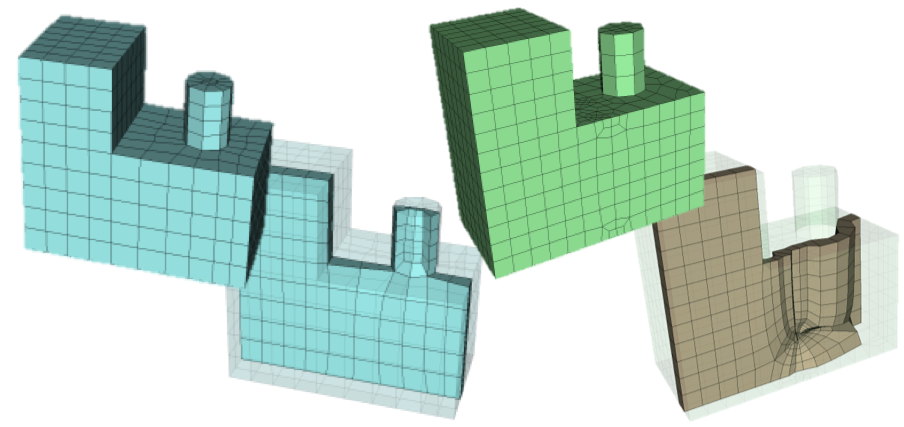}
	\caption{\minrev{
Example of advancing-front progression to fill in a geometric 3D shape starting from different pre-meshed boundary. On the left, it succeeds in getting a valid hex-mesh, while it fails on the right. Image from \cite{Ledoux08}.}}
	\label{fig:adv_front_failure}
\end{figure}

In general, advancing-front approaches are not reliable enough to generate a good quality hexahedral mesh for general domains. They strongly depend on the boundary mesh structure and the compatibility of this structure with the restrictive structure of hexahedral meshes. Often, this compatibility is not given, since the boundary mesh generation process is unaware of the structural and geometric constraints imposed by the to-be-created hexahedral mesh.
As a consequence, they, e.g., fail to connect fronts when they collide \minrev{(see Figure~\ref{fig:adv_front_failure})}. Moreover, most of the proposed works deal with the extra constraint of starting from a pre-meshed boundary. This constraint is strongly related to the meshing process, which consists of meshing a complex assembly of parts where meshes must be conforming along part interfaces.

\subsection{Dual approaches }
\todo{\MC{((Marcel))}}

Taking a dual perspective in the context of mesh generation, i.e., focusing on the dual representation of a hexahedral mesh (\cref{sec:dual}), has proven to offer certain benefits.

\paragraph{Dual advancing front} For one, the interpretation of the advancing front approach (discussed in \cref{sec:advancingfront}) in the dual domain can reveal interesting structures, simplify the formulation of constraints and rules, and provide additional intuition.
This dual view is taken in the so-called Whisker Weaving~\cite{tautges1996whisker} method and its variants~\cite{folwell99,Ledoux08,kss08}. These start from a prescribed surface quad-mesh that is to be matched by the hexahedral mesh to be constructed. Accordingly, the quad-mesh's dual loops form the prescribed boundaries of the hex-mesh's dual sheets. The algorithms' objective thus is to determine the dual sheets -- in particular their mutual intersection combinatorics -- inside these prescribed boundary curves.

The addition of a next hexahedron in the course of an advancing front approach can be interpreted in the dual as (combinatorially) fixing the intersection of three dual sheets~\cite{tautges1996whisker} or as locally (combinatorially) contracting one of the sheet boundary loops, conceptually fixing part of the dual sheet and leaving the loop as the boundary of that part of the sheet that is yet to be determined~\cite{folwell99}. The dual view enables the formulation of local and semi-local rules and heuristics to more favorably steer the incremental mesh construction process~\cite{Ledoux08,folwell99}. Nevertheless, issues such as poorly shaped elements, inverted elements, or high valence vertices in the result are not easy to avoid in general, even with this dual perspective.

A particular challenge for this approach is posed by the (very common) existence of self-intersecting dual loops in the prescribed boundary quad-mesh. While there is no general theoretical obstacle to the successful meshing of these, such loops need to be brought into pairwise or manifold correspondence and be filled by common sheets of non-trivial topology. It is unclear how the process can be steered to naturally establish this required structure in general; therefore, degenerate elements (so-called knives, \cref{fig:verdict}) and inverted elements are common in the result in these cases. Various strategies (with more or less severe negative side effects on quality) have been proposed to modify the quad-mesh to get rid of such self-intersections in advance~\cite{folwell99,kss08,muller2001shelling,muller2002}.

\paragraph{Dual sheet-by-sheet} Besides these alternative interpretations of advancing front methods, the dual perspective gives rise to a further class of methods, less local and incremental.
A general challenge faced by algorithms that attempt to construct hex-meshes in an incremental fashion (like those discussed in \cref{sec:advancingfront}) is to ensure that \minrev{\say{things work out in the end}}. Without careful look-ahead, one may easily end up in intermediate configurations that cannot be completed in either a valid or a qualitatively reasonable manner.
Algorithms that, by contrast, approach the problem of mesh generation in a global manner, e.g., via global optimization formulations (cf.~\cref{sec:frame_fields}), on the other hand, can be computationally much more intensive.

The dual perspective permits an interesting incremental approach on a \emph{semi-local} level.
Instead of individual cells, entire dual sheets can be considered as the atomic entities for incremental mesh generation in the dual domain.
For the case of quadrilateral mesh generation, which is in close analogy to the hexahedral mesh generation scenario, the advantages of this semi-local dual view for the purpose of incremental construction have been discussed in depth \cite{campen2012dual,campen2014dual}. 
Similar properties hold in the hexahedral case, as is exploited by a number of algorithmic approaches.
However, while in the quad case the dual is formed by chords, which are 1-manifolds (i.e., either a loop or a curve with two endpoints, possibly self-intersecting in points), in the hex case the dual consists of sheets, which are 2-manifolds of arbitrary genus and with an arbitrary number of holes, possibly self-intersecting in curves. Therefore, the problem is of significantly higher complexity and algorithms often restrict to sub-classes of problem instances for simplicity, such as objects of genus 0, sheets with a single boundary loop, or dual loops without self-intersections.
An idea of inserting dual sheets in a divide-and-conquer manner was outlined by~\cite{calvo00}. A concrete algorithm for incremental hex-mesh construction based on sequential dual sheet generation is described by~\cite{muller2001shelling}. The boundary geometry along an entire candidate sheet is assessed in the decision-making process. In contrast to related methods that can be interpreted as operating in a sheet-by-sheet manner~\cite{folwell99,Ledoux08}, this algorithm preserves an invariant through all intermediate stages that strictly avoids combinatorially invalid configurations. This obviates the need for intermediate repair operations and guarantees the absence of degenerate elements such as knives or wedges (\cref{fig:verdict}). On the downside, the more restrictive sheet selection rules that are in place to ensure the invariant can bring the algorithm to an early halt. Rather expensive back-tracking strategies can be used as a remedy to some extent.
By the introduction of additional rules for the selection of sheet operations~\cite{Kremer2014} in particular non-convex shapes can be handled in a more geometry-aware manner, commonly leading to less distorted (or less inverted) mesh elements.

\begin{figure}
    \centering
    \includegraphics[width=\linewidth]{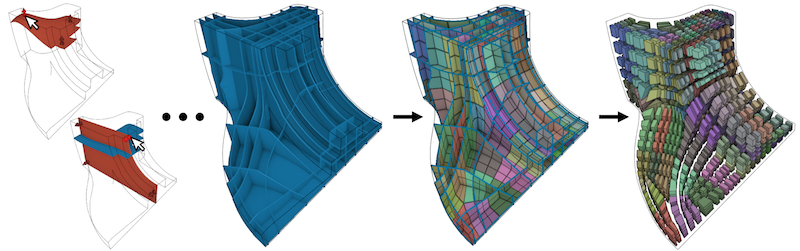}
    \caption{Interactive sheet-based hex-mesh modeling. Image from~\cite{Kenshi:2019}.}
    \label{fig:kenshi}
\end{figure}

\paragraph{Free boundary}
The above methods assume that a quadrilateral mesh of the domain boundary is given, effectively as a starting point for the incremental construction.
The ability to prescribe a boundary mesh can be seen as an advantageous feature in some scenarios (e.g., when adjacent domains are to be meshed separately but compatibly). In others, it rather is a limitation: it restricts the meshing algorithm from the set of all hex-meshes suitable for the domain to a (small) subset. Recently there have been first attempts to construct hex-meshes on a sheet-by-sheet basis \emph{without} predetermined boundary structure. Instead, they exploit interactive user guidance along the domain boundary~\cite{Kenshi:2019} (\cref{fig:kenshi}), or loosely follow principal curvature directions~\cite{LoopyCuts2020} to construct loops which then serve as sheet boundaries.
It is worth remarking that the latter method essentially outputs a subdivided version of the primal mesh that is implied by the sheets; this has the effect that the sheets appear as primal facet sheets in the output mesh. Nevertheless, conceptually both methods are to be viewed as dual approaches.

Due to the larger search space compared to methods with prescribed boundary mesh, they conceptually have the potential to achieve results of better quality -- but at the same time are computationally more expensive and require a user in the loop~\cite{Kenshi:2019} or make simplifications sometimes leading to meshes that contain some non-hex elements~\cite{LoopyCuts2020}.

In this context, the interesting question is that of efficient geometric sheet representation -- while in the above methods assuming a prescribed boundary-mesh, a non-geometric combinatorial representation was employed for simplicity. An implicit representation by means of a level set formulation has proven efficient~\cite{Kenshi:2019}. It, however, does not support self-intersecting sheets, which would grant higher flexibility and enable better mesh quality in various cases. Another, discrete sheet representation space is described by~\cite{rs08}, embedding sheets in the facets of a particular tessellation of the domain; a concrete algorithm that operates in this space has not been addressed yet.

\paragraph{Dual validity}
Generally, when constructing hex-meshes out of dual sheets, it needs to be considered that not any arrangement of intersecting sheets implies a primal hex-mesh. A number of conditions need to be satisfied so as to avoid non-manifold configurations and self-adjacent elements, as detailed by~\cite{MitchellCharacterization1996}. Violating sheet arrangements can be modified, often through the insertion of additional sheets, to ensure these conditions are met~\cite{folwell99}. As these modifications not rarely have a negative impact on (geometrical and structural) mesh quality, a relevant challenge is to avoid the need for them right from the start.

\subsection{Domain decomposition}
\label{sec:dom_decomp}

Early proposals for automatic domain decomposition
relied on simple topological operations like submapping and sweeping~\cite{white1995automated}, that were mainly trying to incorporate the knowledge of the users upon the two-dimensional domain to expand the decomposition to the third dimension with a sweeping step. 
\minrev{
\paragraph{Sweeping.}
\todo{\textcolor{cyan}{(Xifeng)}} Given a volume represented by a closed surface, by identifying two patches where one serves as the \minrev{\emph{source}} and the other one as \minrev{\emph{target}}, a hexahedral mesh can be generated through  \minrev{\say{sweeping}} the quad-meshed \minrev{\emph{source}} over the volume to the \minrev{\emph{target}}~\cite{Shih1996AutomatedHM}. This simple idea is very suitable for CAD models since many shapes are formed by extrusion. The first batch methods using such an idea focus on shapes that can be easily meshed by identifying one source and one target, which are called one-to-one methods~\cite{blacker97,Liu97automatichexahedral,LIU1999413}. However, for slightly complex CAD models, more source or target patches have to be involved in decomposing the extrusion geometry into simpler one-to-one sub-volumes for easy processing.}

A step ahead towards automatic decomposition is presented by Lu and colleagues~\shortcite{lu2001feature} that suggest recognizing in a CAD model the characteristics of portions that can be treated as submappable. The pipeline uses first a feature recognition, then a cutting plane identification, and, finally, a decomposition to mesh each portion with predetermined schemes. \minrev{Along this direction, a set of many-to-one and many-to-many approaches are developed~\cite{many2many2000,White2004,scott06,WU2014136,wu2018fuzzy}. These methods often rely on specific rules to detect line and planar features, such as various angle thresholds, so that the 3D model can be decomposed into sub-volumes having the same sweeping direction. If the decomposition is successful, various node insertion tricks for the sweeping can be employed to ensure the high quality of the generated hex-mesh~\cite{knupp98,BMSweep1999,RuizGironesMultisweep2011}. There are also approaches that allow multiple sweeping directions by computing a hierarchical sub-geometry structure~\cite{BlackerMulti}.} 

Kowalski and colleagues~\shortcite{kowalski2012fun} introduce the notion of fundamental sheets (fun-sheets), noticing that a hexahedral mesh is layered, in opposition to the lack of reference surfaces typical of tetrahedral meshes. Starting from a tet-mesh, converted in a hex mesh and identifying these fun-sheets, using topology and geometry of the shape, they obtain a better decomposition that catches the intrinsic characteristics of the shape. This approach is further enhanced in~\cite{WANG2017}.

An interesting approach to the problem is the one presented by Lu and colleagues~\shortcite{lu2017evaluation}. They design and implement a sketch-based decomposition tool and evaluate its performance on a group of beginners and experienced users. They conclude that visual assistance and a geometric reasoning engine can help to obtain excellent results from a semi-automatic decomposition.

\begin{figure}
\centering
\includegraphics[width=\columnwidth]{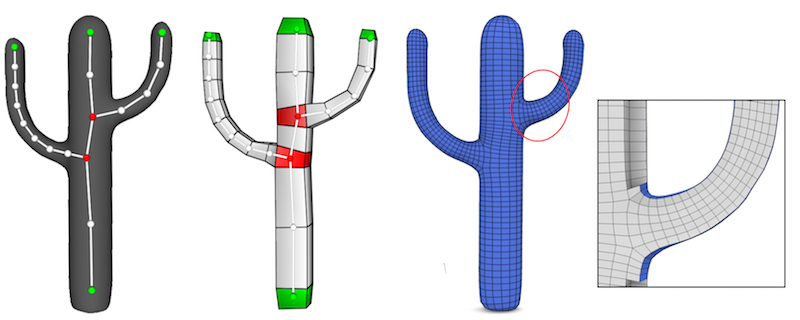}
\caption{\minrev{Skeleton driven hex-meshing starts from an input surface mesh and line skeleton (left), around which a tubular structure composed of hexahedral boxes is initialized (middle left). Refining this structure and projecting it on the target surface yields a hexahedral mesh (middle right) where the distribution of the mesh elements aligns with the skeleton guiding curves (right closeup). Image from~\cite{LMPS16}.}}
\label{fig:skelhex}
\end{figure}

\paragraph{Medial descriptors}
Medial descriptors are a valid proxy in helping to realize a domain decomposition. Both mechanical objects and free-forms are possible to identify characteristics, mainly the skeleton catching the crucial elements of the shape's mutual relations. A three-dimensional shape's \textsl{skeleton} is, in fact, a topological representation of the shape capable of providing information regarding the various boundary entities' relative positions. The skeleton has been used in multiple methods to help in generating a hexahedral mesh inside the shape \minrev{(see e.g., \cref{fig:skelhex})}. When dealing with mechanical objects, usually containing boxes, it is vital to use the general skeleton (or medial object), including surfaces. Shapes more related to biology approximable with a collection of generalized cones can be easily represented by their curve-skeleton \minrev{(or medial axis)}. Both these proxies have been used to guide the hex-meshing.

Price and colleagues introduced the possibility to use the topological skeleton of the shape to produce a hex-mesh. They apply it first on convex shapes~\cite{Price1995HMG}, and then on solids with flat and concave edges~\cite{Price1997HMG}. The idea is to decompose the domain so that each sub-domain can be hex-meshed using a midpoint subdivision scheme~\cite{li1995hexahedral}.
Each sub-domain is meshed using basic primitives that can be placed using the skeleton and used as elementary blocks to mesh the original domain.
The topological information guides the choice of the correct primitive. There are limitations in the approach since high-valence boundary vertices do not have elementary schemes placing them.

Instead of using the skeleton, Sheffer and colleagues~\shortcite{sheffer1999hexahedral} start from the embedded Voronoi graph of the domain, which is simpler to create. Using a set of configurations that include the Voronoi graph's local topology, it can decompose the domain in sweepable subdomains that can be combined and smoothed to yield the final decomposition of the whole domain. 
\minrev{Through the computation of a harmonic field, a general 3D model can be decomposed into 2D curved slices where quad-mesh templates can be used to form a large structure decomposition of the 3D model~\cite{Gao16}.}

Zhang and colleagues~\shortcite{zhang2007patient} exploit the particular shape of the vascular structure to devise a method that uses the curve skeleton as a basis for the meshing. It is the first proposal in which there is decomposition in tubular subdomains that are quite simple to mesh via sweeping. The uniform diameter of the typical vases treated in the application does not pose the problem of resolution in the elements. Usai and colleagues~\shortcite{ULPTS15} use the curve-skeleton to derive a quadrilateral base complex given the triangular mesh of shape. The surface decomposition can be expanded to the domain's interior and lead to a method for hex-meshing~\shortcite{LMPS16}. 
In this work, a scheme for keeping the mesh elements uniform while the diameter of the subdomains changes is introduced and applied. Another similar approach~\cite{LAPS17} employs solid cylindrical parameterizations to map from the curve-skeleton to the cylindrical subdomains. This choice allows a simple but effective way to use the topological information to generate the hex-mesh.

All the methods described in the previous paragraph work fine only for models resembling collections of generalized cones.

Quadros~\shortcite{Quadros} also uses the skeleton as a starting point for meshing and, combining it with an advancing front approach, can create hex-dominant meshes. The surface and the skeleton jointly contribute to form what the author calls corridors that are the basis for meshing the domain with an advancing front method.

Cai and Tautges~\shortcite{cai2015optimizing} propose an approach that heavily relies on integer programming due to the classification of the edges for their parameterization. It is in line with the topological methods since it introduces a new set of templates that, once applied to the class of objects they use in their experiments: mechanical parts.

Another interesting approach \cite{liu2015feature} mixes skeletal representation of the shape and polycubes to guide the creation of the hex-mesh. The resulting meshes are non-conforming, including T-junctions.
\minrev{Another type of non-conforming decomposition, the so-called motorcycle complex \cite{Brueckler:2021}, can be constructed guided by a seamless parametrization (cf.~\cref{sec:igm}). This decomposition has hexahedral subdomains only, and can be refined into a conforming hexahedral mesh.}\smallskip

\minrev{Once a suitable decomposition is computed, submapping or sweeping approaches can likely generate hex-meshes with satisfactory quality. For example, \cite{wu2017global} can be employed to first generate a quadrilateral mesh for interfacing surfaces while ensuring conformity among adjacent sub-volumes, and then apply the straightforward sweeping to generate the final hex-mesh. However, up to now, the critical issue still lies in how to robustly decompose a 3D model into sweepable sub-volumes while ensuring the necessary conformity and feature preservation at the interfaces of different parts. Industry resolved this issue by putting the user in the loop, relying on manual block decomposition and automatic sweeping as a workhorse for hex mesh generation in commercial software~\cite{hypermesh,ansys}. Automatizing the process and freeing the user from tedious critical work is an open challenge for future methods in this family.}

\begin{figure*}
	\centering
	\includegraphics[width=\linewidth]{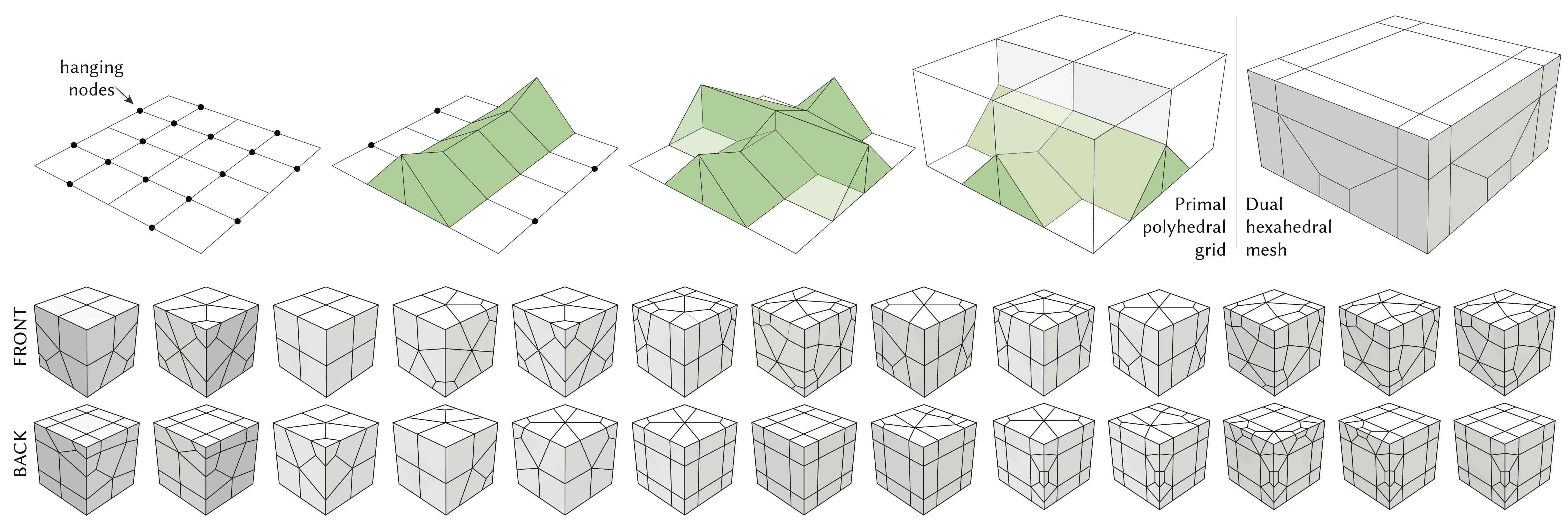}
	\caption{\minrev{Dual methods regularize the valence of hanging nodes (black dots) by connecting them pairwise along triangular bridges, so that the dual is a hex-mesh. Top: the flat transition firstly introduced in~\cite{Marechal2009}. Bottom: the set of atomic schemes introduced in~\cite{livesu2021optimal} to handle all possible transitions in strongly and weakly balanced grids. Images partly from~\cite{livesu2021optimal}.}}
	\label{fig:duak_schemes_stringbal}
\end{figure*}

\subsection{Grid based} \label{sec:grid_based_gen}
\todo{\textcolor{cyan}{(Xifeng)}}
A hex-mesh can be trivially created by voxelizing the interior of a closed surface \minrev{and then projecting its boundary onto the target geometry~\cite{Schneiders1996}. Geometric fidelity} can be controlled by tuning the resolution of the voxelization. Since the size of regular grids grows cubically, to reduce element count a set of adaptive spatial partitioning approaches that rely on \minrev{hierarchical} structures \minrev{have been} proposed.
However, \minrev{adaptive grids do not define a conforming hex-mesh}
because adjacent \minrev{grid elements may have different size}, generating spurious (\emph{hanging}) nodes. 
\minrev{Grid-based methods differ to each other for the refinement policy they use, for the technique used to suppress hanging nodes, or for the method used to project the mesh on the target geometry.}\\
\minrev{Methods in this class are among the firsts that were introduced in the field. From a mesh quality standpoint, they are typically considered inferior to other methods because: (i) the grid is fixed in space and the result depends on the orientation of the model; (ii) the connectivity they generate is intricate and rich of singular edges with high valence~\cite{livesu2021optimal}; (iii) the meshes they generate are highly unstructured and do not endow a coarse block decomposition (see \cref{fig:structure} and Fig. 21 in~\cite{LoopyCuts2020}). Nevertheless, when compared with alternative options grid-based methods really stand out in terms of robustness. To date, they are the only fully automatic methods capable of successfully hex-meshing any input shape, regardless of its geometric or topological complexity. For this reason, they are the only automatic methods currently implemented in professional software~\cite{meshgems,coreform,cubit}. Despite the most prominent methods were developed more than 10 years ago and the field remained quiet for some years, major improvements have been proposed in recent years, also opening avenues for further research.}

\minrev{
\paragraph{Refinement. } Grids should satisfy both local and global criteria. At a local level, cell size must be compatible with the local size of the input object, ensuring geometric fidelity. At a global level, it must be possible to select a subset of grid elements (e.g., the ones completely internal to the input shape) such that the topology of this arrangement matches the one of the original object. In case the grid and the input mesh are not homotopic, a bijective mapping between them is not possible. Local criteria are easier to enforce. The most typical split rules used in the literature are normal similarity~\cite{ito2009octree}, local thickness~\cite{livesu2021optimal,PLCGS21,Marechal2009}, surface approximation~\cite{gao2019feature} or a combination of these and other indicators~\cite{bawin2021automatic}. The fulfillment of global criteria is more complex and demands to preprocess the input shape~\cite{mitchell1992quality}. For this reason, the vast majority of methods do not guarantee  that the output hex-mesh will have the same genus and number of connected components of the input model~\cite{Marechal2009,livesu2021optimal,PLCGS21}, or ensure this property at the cost of severe over refinement (e.g., iteratively splitting all grid elements until topological equivalence is obtained~\cite{gao2019feature}). Refined cells can be split in two alternative ways: \emph{2-refinement} splits each edge in two, thus obtaining 8 sub-cells for each adjacent hexahedron; \emph{3-refinement} splits each edge in three, thus obtaining 27 sub-cells. In both cases, the sequence of splits is encoded in a hierarchical tree structure, which corresponds to an octree for the 2-refinement, and to a 27-tree for the 3-refinement. Approaching this body of literature for the first time may be confusing, because all methods generally refer to these data structures as \say{octrees}, even though this is not always correct. The use of 27-trees for 3-refinement is explicitly mentioned in~\cite{Schneiders96octree-basedgeneration} and a few other articles, and is only implicitly assumed in other articles that refer to these ones.}

\begin{figure}
	\includegraphics[width=.95\columnwidth]{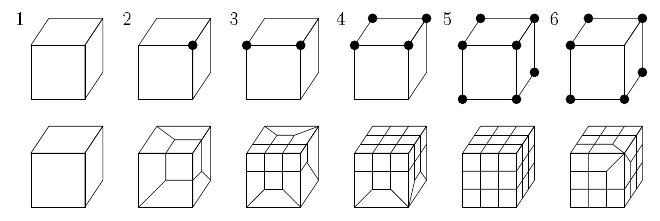}
	\caption{\minrev{The set of 6 out of 22 schemes of 3-refinement introduced in~\cite{schneiders1996refining}. The 1 and the 5 are respectively the empty and the fully refined hexahedra, while the other templates are used to manage the transition between them.
	Image taken from~\cite{schneiders2000algorithms}.}}
	\label{fig:ref_schemes}
\end{figure}

\minrev{
\paragraph{Hanging nodes. } The removal of hanging nodes is obtained by substituting elements of the grid with templated topological transitions that locally restore mesh conformity. 
If adjacent grid elements differ by at most one level of refinement there exist $2^8$ alternative configurations which, discarding symmetries, reduce to 20 unique cases~\cite{weiler1996automatic}. Existing methods can be broadly categorized into two families: \emph{primal} methods aim to directly incorporate the hanging nodes in the output hex-mesh; \emph{dual} methods aim to modify the input grid such that its dual mesh contains only hexahedral cells.\\
Primal methods often operate on 3-refined grids and 27-trees, because it is easier to suppress their hanging nodes~\cite{Schneiders96octree-basedgeneration}. However, handling all the possible 20 configurations is provably impossible, because many concave transitions are bounded by an odd number of quadrilateral elements, a condition for which it is known that a hexahedralization of the interior does not exist~\cite{MitchellCharacterization1996}. Transition schemes for 4 flat and convex transitions (see \cref{fig:ref_schemes}) appeared in multiple articles~\cite{schneiders2000algorithms,Schneiders2002QuadrilateralAH,schneiders97,tack1994two} and were successfully used to compute hexahedral meshes, prescribing additional refinement to convert unsupported transitions into the supported ones. Over the years additional schemes were introduced to handle concave edges~\cite{ito2009octree,ZHANG2006942,ELSHEIKH201486}, but a correct handling of concave corners remains elusive.
Several works, like \cite{ebeida2011isotropic, ZHANG201388, owen2017template}, exploit the 2-refinement schemes introduced in \cite{Schneiders96octree-basedgeneration} to remove hanging nodes. Unlike from the 3-refinement approaches, the grid needs to satisfy more strict constraint as those described below for dual methods. Note that, as for the 3-refinement case, the schemes in \cite{Schneiders96octree-basedgeneration} do not allow to address all the possible configurations, often leading to an excessive over-refinement of the grid.\\
Dual methods operate on 2-refined grids and octrees, and are superior to primal methods because they can handle all possible transitions. All known schemes operate on \emph{balanced} grids, that is, grids where the refinement mismatch between adjacent elements is at most one. However, not all methods agree on the definition of \say{adjacent}. For the majority of methods two cells are adjacent if they share one face, edge or vertex (\emph{strong balancing}). In~\cite{livesu2021optimal} the authors relaxed this formulation, enlarging the class of balanced grids and limiting restrictions to size mismatch only for cells sharing a face (\emph{weak balancing}). Weakly balanced grids permit to greatly reduce refinement (up to $64\%$ less elements in their experiments), but require a slightly more complex scheme set. Mar{\'e}chal was the first to observe that if all grid vertices have valence 6 and all grid edges have valence 4, the dual of the grid is a pure hexahedral mesh~\cite{Marechal2009}. Based on this \minrev{observation} he proposed a set of cutting schemes that, regularizing the valence of grid elements, allow to obtain a pure hexahedral mesh via dualization (\cref{fig:duak_schemes_stringbal}).}

\begin{figure*}
	\includegraphics[width=\linewidth]{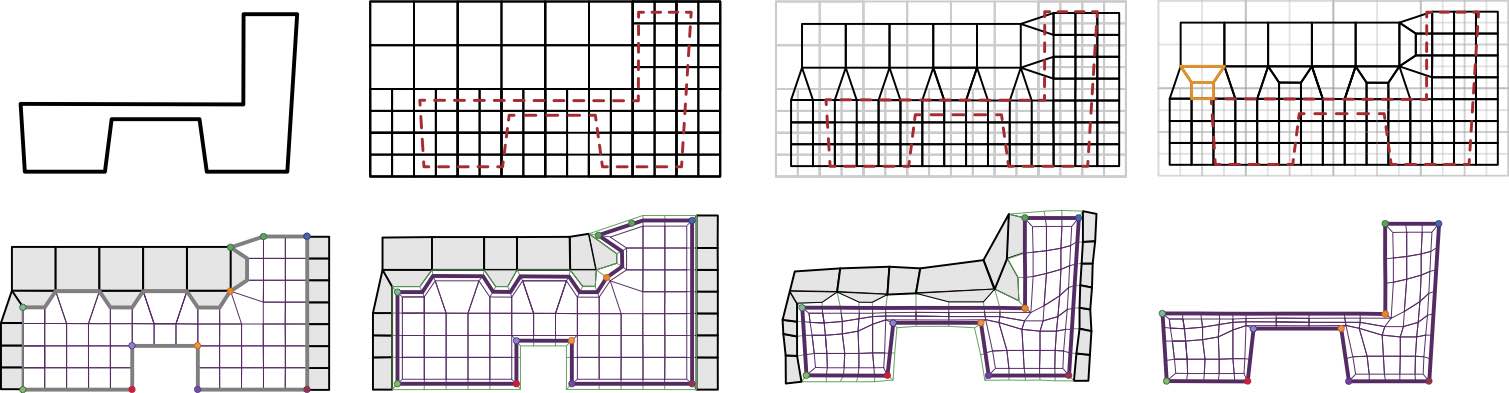}
	\vspace{-0.5cm}
	\caption{2D pipeline of the feature preservation octree-based hex-meshing. Top row: adaptive quadtree constructed from the input, dual of the quadtree, and quadrilateral (quad-) mesh including a scaffold mesh. Bottom row: topological matching of feature graphs, variational padding of both the target mesh and the scaffold, mesh deformation to fit the input, and the final pure quad-mesh. \minrev{Image from~\cite{gao2019feature}.}}
	\label{fig:octree}
\end{figure*}
\begin{figure}
	\centering
	\vspace{-5mm}
	\includegraphics[width=.99\linewidth]{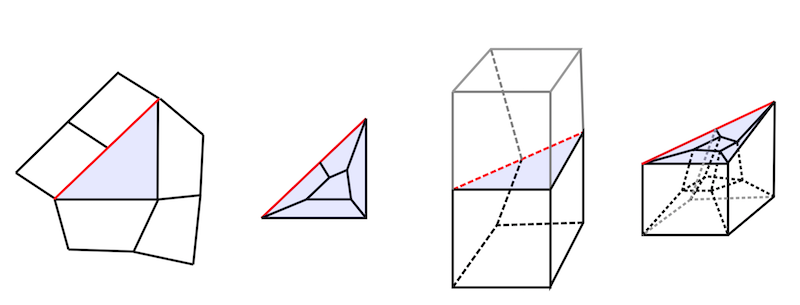}
	\vspace{-3mm}
	\caption{If a quad maps two of its four edges onto a linear feature line it becomes locally degenerate (left). Splitting it into 5 sub-quads ensures enough degrees of freedom to produce all valid mesh elements. Similar configurations may also occur on 3D meshes, and can be resolved with a special padding scheme that splits a hexahedron into 6 sub elements (right). \minrev{Image from~\cite{gao2019feature}.}}
	\label{fig:feat_padding}
\end{figure}

\minrev{Since the valence of hanging nodes is fixed pairwise, dual methods also require that the grid is \emph{pair}, that is, for each cluster of grid elements with same amount of refinement the number of hanging nodes must be even across all grid directions. Differently from balancing, the pairing condition is non local, hence difficult to enforce. Pairing is typically enforced directly in the octree, fully splitting parent nodes if their siblings have been split~\cite{Marechal2009,gao2019feature,Hu2013AdaptiveAM,livesu2021optimal}. As shown in~\cite{PLCGS21} all these methods operate in a restricted space of solutions and tend to severely over refine the input grid, even if it is already pair. The authors showed that pairing can be enforced directly in the grid by solving a sequence of linear problems, obtaining coarser grids that approximately halve the number of elements. Despite superior to tree-based methods, also this method does not cover the whole space of solutions, and may occasionally refine an already pair input grid (see Sec. 7 in ~\cite{PLCGS21}).
Even though dual approaches exist since 2009, the transition schemes they use were only vaguely described in the literature, making these methods hardly reproducible. Mar{\'e}chal~\shortcite{Marechal2009} pioneered this technique, but his paper describes in detail only one specific transition (\cref{fig:duak_schemes_stringbal}, top). Gao and colleagues proposed three alternative schemes based on similar ideas~\cite{gao2019feature}, also releasing their code, but these schemes were recently shown to be not fully exhaustive and may fail to produce a conforming hex-mesh starting from a balanced and paired grid~\cite{livesu2021optimal}. In~\cite{livesu2021optimal} the authors propose a comprehensive study of dual schemes, clarifying ambiguities and implementative choices, and ultimately deriving an exhaustive optimal set of transitions for both strongly and weakly balanced grids (\cref{fig:duak_schemes_stringbal}, bottom). CinoLib~\cite{cinolib} hosts an open source implementation of all such schemes, as well as the code necessary to install them in a given adaptive grid.}

\paragraph{Projection. } Considering the axis-aligned nature of grid-based methods, to approximate the input object well the boundary vertices have to be projected onto the target geometry. To this end, maintaining the inversion-free property of a hex-mesh poses a great challenge. While~\cite{Marechal2009,Lin2015QGA} rely on iterative vertex smoothing to slowly move the vertices onto the boundary so that a local smoothing can be backtracked if it causes flipped hexahedra,~\cite{gao2019feature} presents a global deformation method that can robustly align the generated hex-mesh with the input surface (including sharp features) within a distance bound. \cref{fig:octree} shows the 2D pipeline of the method presented in~\cite{gao2019feature}. 
\minrev{After grid refinement and removal of hanging nodes, the grid is partitioned into} two sub-meshes: an inside \say{target} mesh that will be optimized to be the final output, and an outside \say{scaffold} mesh that ensures the bijectivity of the map throughout the optimization process. Geometric fidelity is achieved by first building a topological bijectivity mapping between the input mesh and the boundary of the target mesh, and then geometrically deforming the target mesh towards the input surface shape using a locally injective mapping technique~\cite{rabinovich2017scalable}. Note that a variational padding technique \minrev{(see~\cref{sec:pillowing})} is also introduced for both the target mesh and the scaffold, so as to increase the number of degrees of freedom for optimization. The approach can robustly produce an all-hexahedral mesh with several guarantees: 1) the output is manifold and its boundary surface has the same genus with the input, (2) all hexahedral elements have positive scaled Jacobian (3) the boundary of the hex-mesh is error-bounded, i.e., within $\epsilon$ distance from the input mesh, and (4) the boundary of the mesh has no self-intersections thanks to the scaffold mesh. \minrev{All of this is obtained by trading robustness for efficiency, thus computational cost and memory resources can be prohibitive for commodity hardware. On the other hand, iterative methods such as~\cite{Marechal2009,Lin2015QGA} are quite efficient, although may occasionally fail to preserve the shape well. Further research is needed to devise an algorithm that optimally combines robustness, efficiency and geometric fidelity.}

\begin{figure*}[h]
	\includegraphics[width=\linewidth]{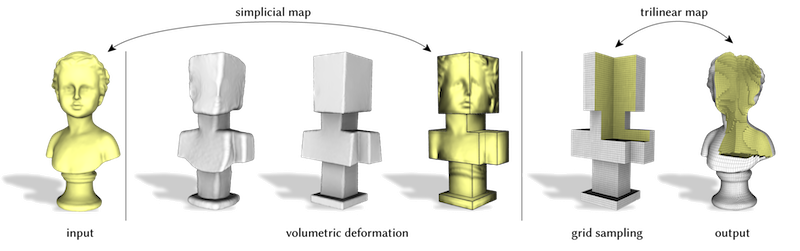}
	\vspace{-0.5cm}
	\caption{The pipeline for polycube based hexmeshing generates a locally injective simplicial map through volumetric deformation, and then uses it as a medium to transfer a regular grid sampling of the polycube to the target shape.}
	\label{fig:pc_map}
\end{figure*}

\paragraph{Features. } The preservation of sharp surface features is both geometrically and topologically challenging for grid-based approaches. First of all, since the mesh connectivity is derived by the underlying grid, surface vertices may not have enough incident edges to reproduce high valence feature points in the target mesh. Therefore only a subset of all possible feature networks can be faithfully reproduced. Moreover, hexahedra that have more than one facet exposed on the surface may easily be traversed by feature lines across more than one edge, becoming ill-shaped or even degenerate once projected onto the target geometry. To make sure that each element has at most one feature edge, specific padding schemes are used (\cref{fig:feat_padding} \minrev{and \cref{sec:pillowing}}). Finally, despite the fact that it works well in most cases, current algorithms for feature mapping are heuristic and do not offer guarantees. The most recent methods are based on ideas expressed in~\cite{gao2019feature}, and operate by iteratively processing each feature separately, projecting its endpoints to the closest vertices in the hex-mesh, and then finding the discrete path that connects them with a Dijkstra search that operates on a scalar field that encodes the euclidean distance from the input feature. Depending on the ordering of the features and the combinatorics of the hex-mesh, there can be conflicting configurations where a path that connects the two endpoints of a feature and does not conflict with any previously inserted feature does not exist. Furthermore, even if such a path exists, there may be cases in which the previously inserted features force a path to deviate from its geometric target significantly. 

\minrev{
\paragraph{Assemblies and multiple materials. } While all methods described so far assume as input a single model composed of a single material, grid-based techniques have been successfully extended to the multi material case~\cite{suleekumar04,zhang_jessica10}, and can also handle complex non manifold CAD assemblies~\cite{qian2012automatic}. From a grid processing perspective, these method rely on the processing techniques described in the previous paragraphs.}

\subsection{Polycube maps} \label{sec:polycubes}
A successful line of algorithms works by \minrev{volumetrically mapping a shape to} an orthogonal polyhedron (or \emph{polycube}~\cite{tarini2004polycube}) embedded in $\mathbb{Z}^3$.
Sampling the polycube at a dense integer lattice gives a regular all hex connectivity, whose nodes can be positioned inside the initial object following the inverse map (\cref{fig:pc_map}). 
Polycube methods are based on two fundamental building blocks: the definition of the polycube structure, and the generation of the volumetric map. These two objectives can be pursued separately (i.e., defining a valid polycube structure first, and then computing the map) or together, using mesh deformation to explore the space of shapes and find the orthogonal polyhedron closest to the input object.\\

\paragraph{Structure.} The structure of the polycube is typically defined by assigning to each surface element of the mesh a label that represents one of the six global axes ($\pm X, \pm Y, \pm Z$). Clusters of adjacent elements with same the label identify the facets of the polycube. 
A purely local naive approach consists of assigning to each surface element the axis closest to its normal~\cite{gregson_polycube_2011}. However, local surface orientation often does not produce a globally consistent structure, and heuristic post-processing is necessary to fix the labeling. Hu and Zhang~\shortcite{cvt_pc_zhang2016} use a modified CVT labeling in the space of normals, but 
still require manual post-processing to sanitize topological inconsistencies that may arise~\cite{yu2020hexgen,yu2021hexdom}. 
The authors of PolyCut~\cite{polycut_livesu} listed sufficient topological conditions for the existence of a valid polycube (following~\cite{eppstein2010steinitz}) and -- starting from naive labeling -- systematically explore the space of labelings in order to find the closest label assignment that satisfies all such criteria. Also this method is not guaranteed to converge to a valid polycube structure. \minrev{Interactive tools for user assisted polycube construction also exist~\cite{interactive_polycubes,yu2020hexgen,yu2021hexdom}.}

\begin{figure}
	\includegraphics[width=\columnwidth]{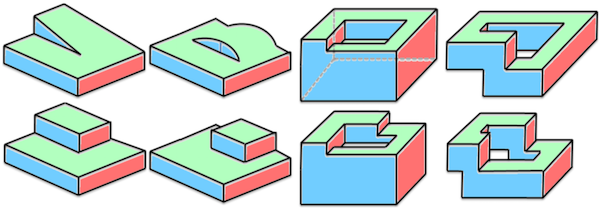}
	\caption{Top: a set of pathological shapes having sharp creases and surface normals that push polycube deformation energies towards the generation of globally inconsistent orthogonal polyhedra. Bottom: the associated polycubes. Additional corners were strategically introduced to define topologically sound global structures. Image from~\cite{sokolovHDR}.}
	\label{fig:pc_normal_issues}
\end{figure}

\paragraph{Mapping.} The volumetric map is obtained using dedicated deformation energies that iteratively rotate surface normals until they \emph{snap} to the global axes. These methods may input a pre-computed polycube segmentation~\cite{gregson_polycube_2011,polycut_livesu}, freely deform the shape until the polycube structure reveals itself \cite{l1pc2014,closedform_pc2016}, or interleave the two operations, updating the reference labeling after each iteration~\cite{fu2016efficient}. Techniques that are not driven by a guiding (valid) polycube labeling seldom converge to a topologically consistent polyhedron. As reported in multiple articles, these pipelines include heuristic post-processing steps that aim to remove topological artifacts
(e.g., removing facets with less than 4 sides). In~\shortcite{sokolov2015fixing} Sokolov and Ray observe that degenerate conditions are both local and global, hence difficult to sanitize. In particular, since axis-aligned features are naturally preserved during deformation, the presence of long and slightly diagonal creases may push the deformation in the wrong direction, resulting in globally inconsistent configurations which are extremely hard to recover from (\cref{fig:pc_normal_issues}). Tiny features such as protuberances, tunnels and handles are also critical, and may become intrinsically impossible to resolve if the mesh lacks the sufficient resolution to define a valid labeling.

Polycube deformation operates on a supporting tetrahedralization of the object. Early deformation energies defined in~\cite{gregson_polycube_2011,l1pc2014} did not sufficiently penalize distorted and flipped elements, producing non locally injective maps containing various foldovers, especially in the vicinity of concavities. Fu and colleagues~\cite{fu2016efficient} introduced a deformation energy that incorporates the AMIPS term~\cite{fu2015computing}, which grows to infinity in the presence of degenerate or inverted elements. Various similarly flip-preventing energies have been introduced in recent years~\cite{Fu-2016-SA,rabinovich2017scalable}, and could be adopted in this setting. \minrev{It is important to note, however, that the prevention of flips may reduce the deformation space to an extent that no map that respects the boundary-alignment constraints can be found.}
Alternatively to volumetric deformation, one could in principle use a surface method to define a polycube-surface map (e.g., with~\cite{yang2019computing}) and then solve for a compatible volumetric mapping between the two shapes. \minrev{Again, however, despite the high level of practical robustness showcased by recent approaches~\cite{garanzha2021foldover,du2020lifting}, the fully reliable automatic generation of constrained volumetric maps without flips remains an open problem~\cite{fu2021inversion}
Motivated by this difficulty, an interactive polycube construction pipeline that puts the user in the loop has been recently proposed~\cite{interactive_polycubes}. Users are allowed extensive control over each stage, such as editing the polycube structure, positioning vertices, and exploring the trade-off among competing quality metrics, while also providing automatic alternatives. The robust mapping energy proposed in~\cite{garanzha2021foldover} is internally used to discourage the generation of flipped elements.\\
The use of alternative mesh representations has also proved useful to robustly construct volumetric mappings. In~\cite{PRPSL15} the authors represent a tetrahedral mesh as a collection of dihedral angles, and propose a robust spectral reconstruction method to generate an explicit mesh up to a global similarity transformation. The use of reduced coordinates to represent and manipulate meshes (e.g. via curvature or edge lengths) is a broad topic and has been widely studied, especially for the surface case~\cite{Crane:2011:STD,Campen:2021}. Specifically, the aforementioned paper shows that any input polycube segmentation can be translated into a set of prescribed dihedral angles that encode the change of normal orientation along the surface. Using the proposed reconstruction method allows to convert such angles into an explicit mesh, obtaining a locally injective polycube map.}

\paragraph{Sampling.} The generation of the hex-mesh connectivity is based on a grid sampling in polycube space. This is typically done by snapping polycube corners to integer coordinates, and then sampling the space at the corresponding dense integer lattice. Current methods assume an \emph{ideal} mapping that is virtually free from distortion. In reality, hexahedra may undergo severe deformation, and occasionally flip their orientation even if the underlying simplicial map is injective. A valid trilinear mapping is guaranteed to exist, but it may require a very dense polycube sampling. In practice, concrete implementations often rely on a possibly imperfect coarse hex-mesh, and then apply untangling techniques in case such mesh contains inverted elements (\cref{sec:untangling}). No guarantees of correctness can be provided in this case.

Regarding sampling frequency, the density of the integer grid must be fine enough to ensure that no element of the polycube collapses. Cherchi and colleagues~\shortcite{cherchi2016polycube} showed that different samplings produce different meshes and that a coarse sampling yields a mesh with a simpler block-structure, because corners are more likely to snap to the same integer isolines. Finding a good balance between these ingredients is subject to optimization~\cite{cherchi2016polycube,protais2020robust,chen2019constructing}. Since snapping to coarser grids involves bigger displacements, it becomes more complicated to preserve the injectivity of the simplicial map. All existing methods are heuristic and can break the map in complex cases.

\begin{figure}
	\includegraphics[width=\columnwidth]{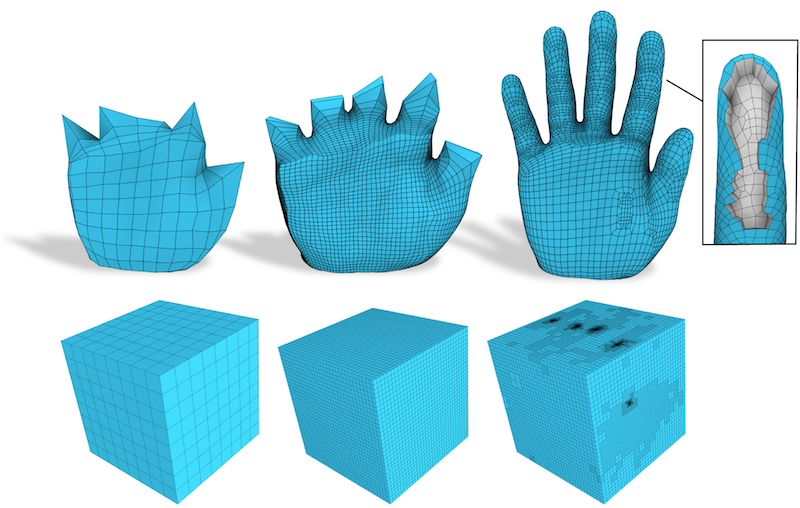}
	\caption{\minrev{Adaptively sampling a coarse polycube allows to restore major features that are not explicitly encoded in parametric space (right), and that could not be obtained with a regular sampling (left), not even with a dense one (middle). Image from~\cite{PLCGS21}.}}
	\label{fig:polycube_hand}
\end{figure}

While regular grids are widespread, smarter sampling schemes can be used to control element size and anisotropy, or even to counterbalance map distortion. In recent literature there have been a few attempts to address size control, obtained either thickening a region of interest in polycube space prior sampling~\cite{xu2017hexahedral} or adapting octree-based meshing to polycube space~\cite{cvt_pc_zhang2016}. \minrev{More recently, Pitzalis and colleagues~\shortcite{PLCGS21} showed how adaptive sampling can be unlinked from rigid octrees and extended to generalized grids of any shape or topology. Adaptively sampled polycubes can be used to improve geometric fidelity while keeping the mesh resolution low, or it can also be used to restore major features that were missing in polycube space and that could not be reproduced otherwise, not even with extremely dense regular samplings (\cref{fig:polycube_hand}). These preliminary results suggest that a tighter integration of adaptive sampling in polycube space could benefit the whole pipeline. Current methods heavily rely on the ability of the polycube finding module to catch \emph{all} the features of the object (at all scales) so as to secure the proper mesh connectivity. Finding a better balance between features that are explicitly encoded in the polycube and features that will be reproduced with adaptive sampling, algorithms may be able to better distribute the complexity throughout the whole pipeline, possibly increasing their robustness. On the negative side, any non regular sampling will introduce additional irregularities in the mesh, reducing its level of structural regularity (\cref{sec:structure_regularity}) and possibly making it unusable for applications that exploit the coarse block decomposition endowed in the mesh connectivity, such as IGA methods~\cite{hughes2005isogeometric}.}

\paragraph{Features.} Desirable properties such as curvature and feature alignment depend on how the polycube map orients these entities in $\mathbb{Z}^3$. In particular, since sharp creases are preserved only if they map to integer isolines in polycube space, there are intrinsic topological limitations to the class of feature networks that can be correctly reproduced (e.g., a convex vertex with more than three incoming feature lines cannot be correctly meshed). \minrev{Since geometric features are often characterized by surface normal discontinuities, labeling methods such as~\cite{polycut_livesu,cvt_pc_zhang2016} intrinsically promote their positioning along polycube edges. Nevertheless, these methods do not explicitly handle surface features, and may often fail to preserve them~\cite{guo2020cut}. To our knowledge, the only method that explicitly promotes feature alignment is CE-PolyCubeMaps~\cite{guo2020cut}. Given an input network, the authors attempt to transform each feature into a piece-wise linear curve that aligns with the global axes. Features that do not align (or conflict with other features) are discarded; the others are included in the polycube structure generation, with a feature-aware variant of PolyCut~\cite{polycut_livesu}. Despite practically superior to previous approaches, also this method does not provide strict guarantees. Furthermore, features are only mapped to polycube edges, and the possibility to align to integer isolines that are internal to polycube faces is not exploited.}

\paragraph{Maturity.} Polycube methods have received increasing attention from the meshing community and have now reached a discrete maturity level. The most recent algorithms allow to blindly process datasets composed of more than a hundred shapes, producing hex-meshes of good quality~\cite{fu2016efficient}. 
In terms of mesh structure, these methods typically produce valence semi-regular meshes, and may occasionally produce semi-regular meshes if singularities (i.e., polycube corners) align~\cite{cherchi2016polycube}. The singular structure of a polycube-based hex-mesh is fully exposed on the surface, and consists of all polycube edges and corners. This inability to position singularities in the interior inherently limits the map, and may occasionally be the source of unnecessary distortion. A recent work of Guo and colleagues~\shortcite{guo2020cut} proposes to enhance the singular structure with diagonal cut surfaces that penetrate the interior of the polycube, permitting further distortion reduction. 
Intuitively, these cuts can be thought of as analogous to cone singularities in surface mesh parameterization~\cite{soliman2018optimal}, although they are more constrained because the two copies of each cut surface must still obey to the constrained polycube structure. Another typical improvement consists in pushing the external singular structure one layer inside the volume, adding a global padding layer that avoids over-constrained hexahedra with more than one facet exposed on the surface (see \cref{sec:pillowing}). More sophisticated padding schemes that directly exploit the polycube map to optimally balance distortion with mesh growth are also available~\cite{cherchi2019selective}.

\paragraph*{Abstract polycubes}
Various techniques use skeletons or segmentation to partition a shape, generating atlas of maps to face-adjacent cuboidal domains~\cite{LMPS16,li2010generalized,li2013surface,liu2015feature,ULPTS15}. Some of these authors explicitly refer to the structures they produce as \emph{abstract} or \emph{generalized} polycubes, where these words mean that such structures may not have an embedding in $\mathbb{Z}^3$. The major difference between embedded and abstract polycubes is that in the former the whole polycube is considered as a global entity that lives in a unique space, whereas in the latter each cuboidal domain lives in its own space and has a dedicated map. For this reason, despite the name similarity, these techniques are more similar to domain decomposition approaches, and are discussed in \cref{sec:dom_decomp}.

\begin{figure*}[tb]
	\includegraphics[width=\linewidth]{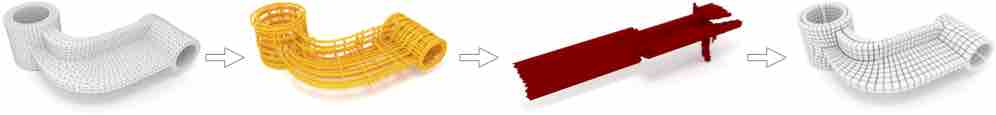}
	\caption{Frame field based hexahedral mesh generation: Given an input tetrahedral mesh (left), first a boundary aligned and smooth frame field is generated (middle, yellow). The frame field serves as a proxy for the local orientation of hexahedra and thus enables the efficient generation of an integer-grid map (middle, red), which induces a hexahedral mesh (right). Image from~\cite{liu2018singularity}.}
	\label{fig:frame_field_meshing}
\end{figure*}

\subsection{Frame fields}
\label{sec:frame_fields}
\todo{\DB{(David)}}

Frame fields offer a promising research direction for general hexahedral mesh generation. They can be seen as a continuous relaxation of an integer-grid map (cf.~\cref{sec:igm}) and thus a key component of the corresponding hex-mesh generation algorithms. A prototypical algorithm (see \cref{fig:frame_field_meshing}) consists of three major steps: (i) optimization of a boundary-aligned frame field, (ii) generation of an integer-grid map that resembles the frame field \cite{Nieser2014}, and (iii) extraction of the integer level-sets which form an explicit hexahedral mesh~\cite{lyon2016hexex}.

\paragraph{Frame Field.} A \textit{frame field} can be seen as a generalization of a vector field to a quantity that locally describes the shape of a (linearly deformed) cube. Locally, a frame consists of three linearly independent vectors, which represent a parallelepiped, i.e., the orientation and shape of a linearly deformed cube. It is important to understand that globally a frame field is significantly more complex than three superimposed vector fields since it can contain singularities where topologically the vector fields are nontrivially interconnected on a branched covering~\cite{Nieser2014}. As a consequence, the connection induced by a frame field might exhibit nonzero \minrev{monodromy}, i.e., a vector does not return to itself when transported along a cycle \minrev{around a singularity}. 

\subsubsection{Local frame representation.} 
There are various different representations to locally encode a frame. A straightforward choice that is capable of fully describing the shape of a linearly deformed hexahedron are three \textit{explicit} vectors $u,v,w \in \mathbb{R}^3$ bundled into a matrix $F = \left( u, v, w \right) \in \mathbb{R}^{3\times 3 }$. The local shape of the hexahedron then simply corresponds to the parallelepiped formed by $u$, $v$ and $w$.
In practice, often reduced representations are preferable, where $F$ is orthonormal or orthogonal, meaning that only rotations or rotations and scaling along the principal axes are encoded. For example, unit quaternions have been used as a compact representation for orthonormal frames~\cite{hybridHexa,liu2018singularity}. \minrev{Another common choice for orthonormal frames are Euler angles w.r.t.~either a global coordinate system \cite{huang2011boundary}, or alternatively a local coordinate system \cite{palmer2019algebraic,ray2016practical} to avoid gimbal locks.}

\paragraph{Handling Cube Symmetries.}
The explicit tangent vector representation of frames has one major disadvantage; it is not unique. For example, the matrix $(u,v,w)$ describes an identical parallelepiped as the matrix $(v,w,-u)$. Since there are $6$ potential permutations of the three vectors and $2^3$ potential choices of sign, in total, there are $48$ different matrices, which encode a single frame. Formally, equivalence is established by the binary octahedral group $BO$ with 48 symmetry transformations. Since the octahedron is dual to the cube, their symmetry transformations are identical.  By fixing the orientation (the sign of the determinant of $F$), it is possible to reduce the elements in one equivalence class to $24$ elements with octahedral symmetry. This explains why frame fields are often called \emph{octahedral fields} \cite{solomon2017boundary}, while \emph{3D cross fields} is another common name. The non-uniqueness of the $(u,v,w)$-representation significantly complicates the optimization of frame fields by inducing discrete variables with values from the octahedral group $O$. A remedy is offered by alternative representations that are equipped with a build-in symmetry invariance and thus a unique representation of frames.

The seminal work of Huang and colleagues~\shortcite{huang2011boundary}, expressed orthonormal frames as rotations of the polynomial $x^4+y^4+z^4$, which are by construction invariant under transformations by elements of $O$. By restricting the polynomial to a sphere, it can be expressed in the spherical harmonics basis, lifting a single orthonormal frame to a $9$-dimensional representation vector. Since rotations only possess three degrees of freedom, it is clear that the spherical harmonics representation is a relaxation, i.e.~not all $9$-dimensional vectors correspond to a rotation of the polynomial $x^4+y^4+z^4$. An identical representation can be derived from the perspective of 4th-order symmetric tensors~\cite{chemin127representing,golovaty2019variational}, which have been further generalized to a 15-dimensional representation of orthogonal frames offering independent scaling of axes~\cite{palmer2019algebraic}. 

\paragraph{\minrev{Differential Frame Representation.}} 
\minrev{Instead of representing a frame field by pointwise specification of frames, one valuable alternative consists in encoding its derivative, i.e.~the change of frames. After specifying one frame in the domain, the entire frame-field can then be re-constructed by integration. Such integration is path-independent given that the specified derivative is integrable, i.e.~all fundamental monodromies are elements of the octahedral group when expressed in the coordinate system of the frame itself. 
Corman and Crane \shortcite{Corman:2019:SMF} employ such a differential representation in a frame-field optimization setting. They extend the theory of moving frames to frame-fields with 
cube symmetry. Conceptually, the setting is analog to the 2D setting of trivial connections \cite{Crane:2010:TCD}. In both cases, all singularities and thus all fundamental monodromies are specified as input. However, while trivial connections in 2D are generated with a simple linear solve, the non-commutativity of 3D rotations requires a (continuous) non-linear least-squares optimization. Interestingly, despite the non-convex objective function, experiments suggest that the resulting field is independent from the chosen initial configuration. Leveraging a differential frame-field representation for optimizing fields with unconstrained singularities has not been done so far but is an interesting direction for future work. It would require replacing the fixed monodromies by the feasible set of discrete choices from the octahedral group.}

\subsubsection{Frame field optimization.} The optimization problem usually consists in finding the \minrev{\say{best}} frame field, which aligns to the boundary of the domain. Best often means as-smooth-as-possible and is measured with the Dirichlet energy $\int_\Omega ||\nabla \phi||^2 dx$ where $\phi$ is a frame representation. Some earlier methods constrain the complete boundary field to a pre-computed solution~\cite{Li:2012, kowalski2016smoothness}. However, in general, it is preferable to only require boundary alignment and let the rest of the field emerge freely.
Typically, the domain is discretized into a tetrahedral mesh, where frames are located at vertices~\cite{ray2016practical,palmer2019algebraic}, faces~\cite{huang2011boundary}, or cells~\cite{liu2018singularity}. Alternatively, boundary element discretizations have been explored~\cite{solomon2017boundary}, where a triangulation of the boundary is sufficient. Algorithms that are based on a lifted frame representation ($9$- or $15$-dimensional) need to ensure that they do not leave the sub-manifold of frames. This is done either with a projection operator, or with some kind of manifold optimization. 
More specifically, Huang et al.~\shortcite{huang2011boundary} compute an initial field by optimizing a convex relaxation of the actual problem, i.e., minimization of the Dirichlet energy in $\mathbb{R}^9$, followed by a local projection onto closest frames. The boundary alignment of the field is approximated by a single linear constraint per boundary face. The initial field is further improved by a nonlinear optimization restricted to the frame-manifold via Euler angles. Ray et al.~\shortcite{ray2016practical} follow a very similar strategy but discretize the field on vertices, tighten the boundary constraints and improve the performance of the projection. Palmer et al.~\shortcite{palmer2019algebraic} observed that a modified \minrev{Merriman$-$Bence$-$Osher (MBO)} algorithm is beneficial because it is often able to avoid local minima that induce global inconsistencies in the singularity graph. The MBO algorithm alternatively diffuses the $9$- or $15$-dimensional coefficient space and locally projects the coefficients onto frames. The idea of the modified MBO algorithm consists in starting with a large diffusion kernel and iteratively shrinking it in subsequent steps. The rationale behind this strategy is that large diffusion steps sufficiently leave the (non-convex) manifold of frames and thus avoid local minima, while small diffusion steps are required for the accuracy of the solution. The projection of frames can be done approximately with gradient descent~\cite{ray2016practical}, or exactly with a semidefinite relaxation~\cite{palmer2019algebraic}.
More rapid convergence than the MBO algorithm is offered by Riemannian trust-region manifold optimization~\cite{palmer2019algebraic}, which on the downside has a higher risk of getting trapped in local minima.

\subsubsection{\minrev{Generality}}
\minrev{Besides providing directional guidance for hexahedral mesh elements, a key property of a frame field is its network of singularities -- which one commonly aims to adopt for a hexahedral mesh generated based on the frame field. In this context, frame fields are general enough that the singularity network of any hexahedral mesh can be expressed. This means that, in contrast to many other approaches, e.g.~polycube mapping, sweeping, or grid-based approaches, the output is not a priori restricted to a subclass of hexahedral meshes.
Therefore superior mesh quality can be achieved, specifically if complex feature alignment is required. In particular, frame field-based methods are able to express alignment not only to the boundary of a domain but also to arbitrary internal structures, which is, for example, important in multi-material applications or in the simulation of fluid-structure interaction.}

\minrev{The main drawback, on the other hand, is the fact that frame fields are actually over-general for the purpose of mesh generation: Frame fields may exhibit additional types of singularities that cannot occur in hexahedral meshes, cf.~~\cite{viertel2016analysis,liu2018singularity}. Such singularity configurations are said to be ``non-meshable''. A key example are 3-5 singularities~\cite{reberol2019multiple}, which frequently appear in smooth frame fields but are not meshable. Existing approaches are able to automatically repair some locally non-meshable configurations~\cite{Li:2012,jiang2014frame} or involve the user to manually repair the singularity graph~\cite{liu2018singularity} and then generate a frame field with a prescribed singularity network~\cite{liu2018singularity, Corman:2019:SMF}. 
Another option is to optimize a general frame-field in such a way that all singularities are pushed towards the boundary in order to generate a generalized polycube as done in \cite{closedform_pc2016}.
Additional research is required to enable complete repair or to restrict the frame field generation and optimization to the space of meshable configurations in the first place. Little can be learned in this regard from the analogous 2D problem, as the gap, in terms of singularity structure, between 2D frame fields and quad-meshes is significantly smaller.}

\subsubsection{\minrev{Field-guided integer-grid map}}
\minrev{Given a (meshable) frame field, one then aims to conceptually \emph{integrate} it to obtain a parametrization, a map (in particular an integer-grid map) onto part of $\mathbb{R}^3$. As the frame field typically is not integrable, a map whose isocurves are precisely aligned with the frame field's directions does not exist. Approximate alignment, e.g. least-squares alignment, is thus aimed for, as in the Poisson approach described by \cite{Nieser2014}, generalizing the cross-field guided mapping used in the 2D case for quadrilateral meshing \cite{kalberer2007quadcover,Bommes2013}. The frame field's singularities are adopted in this process and define the implied hexahedral mesh's singularity structure.}

\minrev{Unfortunately, this field-guided mapping approach does not guarantee a valid resulting map without flips. While there are heuristics \cite{lyon2016hexex} to recover a valid hexahedral mesh even from some invalid integer-grid maps with flips, no general guarantees are available. In the 2D setting, aiming at quadrilateral mesh generation, a stream of recent work has shown ways to reliably generate flip-free maps with prescribed singularities (for instance implied by frame fields) and boundary alignment \cite{Campen:2017,Campen:2019,Campen:2021,Gillespie:2021}. It is based on phrasing the problem as a constrained metric computation problem; in a specific discrete conformal setting and formulated in per-vertex scale variables, the problem becomes convex and can be solved reliably. Most importantly, these methods employ on-demand mesh refinement (or modification) to ensure feasibility, in the sense that the mesh offers sufficient degrees of freedom to support a valid map, represented in a piecewise-linear manner. Note, however, that field guidance is considered in these works only in the form of adopting the singularities, not in the form of dedicatedly following the field's directions.
Generalization of this general approach to the 3D setting is not straightforward; the space of 3D conformal maps too restricted to be useful. The work by \cite{PRPSL15} may be viewed as a first step: It describes the representation and optimization of discrete metrics on 3D tetrahedral meshes in intrinsic variables (dihedral angles). Using linear constraints in these variables, boundary alignment and singularities can directly be prescribed, for instance those adopted from an optimized 3D frame field. However, the resulting problem in this 3D case is non-convex, and the issue of potentially required mesh refinement is unsolved. While there are known ways to reliably generate flip-free maps in 3D \cite{Campen:2016:Bijective}, they do not support the prescription of arbitrary singularities.}

\minrev{An alternative reliable approach proposed for the 2D setting is based on decomposing the domain into regular pieces based on stream lines of a 2D frame field \cite{Myles:2014}. Also this does not generalize to a 3D stream surface based approach \cite{kowalski2016smoothness} with similar guarantees. }

\subsection{Hex-dominant meshing}
\label{sec:hexdominant}
\todo{\NP{(Nico)}\textcolor{cyan}{(JF)}}
Automatic methods for all-hex meshing are only applicable to a subset of all the possible inputs. In contrast, the grid-based approaches (\cref{sec:grid_based_gen}) can operate on intricate shapes and guarantee all-hex meshes; unfortunately, they produce inferior quality \minrev{results}. High-quality, feature-aligned all-hex meshes are still elusive, so industry still relies on semi-manual block decomposition, a time-consuming process \cite{lu2017evaluation}.

For this reason, other methods focus on the hexahedral-dominant meshing instead of full-hex, aiming to reach the highest possible proportion of hexahedra. Hexahedral-dominant meshing is a relaxation of the problem to significantly improve robustness at the cost of introducing a small number of generic polyhedra. The generation of hex-dominant meshes boosted the use of those datasets in practical contexts such as FEM~\cite{WickeBG07}. Moreover, the recent advancement in the construction of higher-order bases~\cite{SchneiderDGBPZ19} may foster the adoption of hex-dominant meshes in the mechanical analysis.

The first approach to produce hex-dominant meshes agglomerates neighboring tetrahedrons to assemble hexahedral cells. The problem of finding a globally optimal solution is NP-complete; hence the clustering process is usually driven by local heuristics. Meshkat and colleagues~\shortcite{meshkat2000generating} have proposed the first method following this idea. The clustering process relies on an undirected graph representing tetrahedra and their connectivity. The graph is enriched with particular arcs and labels on nodes to calculate the agglomeration heuristic. Given the initial tetrahedral mesh, the algorithm detects and replaces subgraphs with hexahedral nodes.

The method proposed by Yamakawa et al.~\shortcite{yamakawa2002hex} takes as input a general 3D mesh and distributes a set of nodes into the volume by the physical simulation of crystal pattern formation. Then nodes are used to produce a mesh composed of hexes, prisms, and tets, with around 50\% hexahedral cells. This method allows controlling element size and primary orientation. Since it does not require a tetrahedral mesh as input, this method is less sensitive to the input discretization than the approach of Meshkat and Talmor~\shortcite{meshkat2000generating}.

Vyas and Shimada~\shortcite{Vyas09} proposed a more sophisticated method that starts by generating a volumetric tensor field to specify the anisotropy and directionality of the elements. Then, such a field induces an advancing front process where several hexahedral fronts contribute to cover the entire volume.

L\'evy and Liu \shortcite{levy2010p} generalized Centroidal Voronoi Tessellation \cite{DuFabGun99} for hex-dominant meshing, introducing Lp-Centroidal Voronoi Tessellation.  Unlike the standard Voronoi diagram, Lp-CVT favors the formation of cubical cells by using a distance metric that takes into account a predefined background tensor field. The resulting method excels in robustness and controllability.

Despite the considerable advancements in performances, the methods mentioned above cannot obtain a high hexa ratio for the general case. The approach proposed by Sokolov et al. [2016a] produces higher hex ratios by using a guiding frame field. A sampling process generates a point set organized as a regular grid and locally aligned with the frame field. A constrained Delaunay triangulation of the volumetric samples makes a tetrahedral mesh and is finally clustered into hex elements.  They obtain hex-dominant meshes with up to 95\% hexahedral cells (although in the worst case they earned less than 30\%). Despite the result's quality, this method might produce non-conforming meshes containing configurations where a quadrilateral face is adjacent to two separate triangular faces. This issue has been solved by~\cite{ray2018hex}.

The approach by Pellerin et al.~\shortcite{pellerin2017identifying} explores the space of all possible agglomerations of tets. Then a greedy process selects the configurations to agglomerate tets into hexes. Their approach produces hex-dominant meshes with a 60\% ratio of cuboidal elements across all shown examples.

\begin{figure}[tb]
	\includegraphics[width=0.9\linewidth]{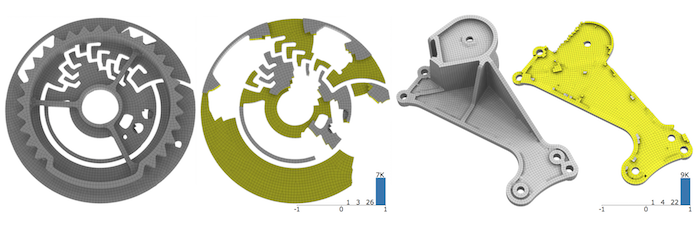}
	\caption{Some challenging examples of hex-dominant re-meshing using~\cite{hybridHexa}.}
	\label{fig:poly_agglo}
\end{figure}

Gao et al.~\shortcite{hybridHexa} directly generate conforming hybrid meshes using polyhedral agglomeration. This method starts from tetrahedral mesh obtained by sampling a guiding frame-field. An iterative process modifies the connectivity utilizing a set of local operators to compose hexahedral cells. The local operators grant the conformity of the final mesh. While this method excels in robustness (as demonstrated by the complex example shown in \cref{fig:poly_agglo} ), it cannot control the class of created polyhedrons (they might have up to 40 facets in some cases, see Table 1 in~\cite{hybridHexa}).

\begin{figure}[tb]
	\includegraphics[width=0.9\linewidth]{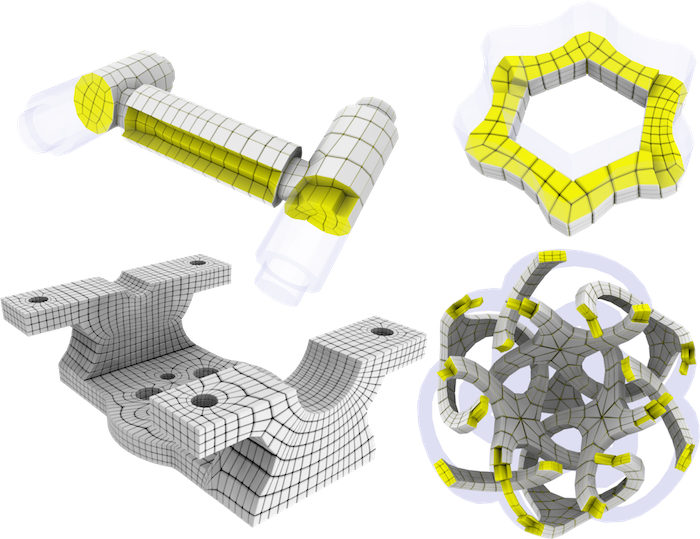}
	\caption{Some CAD models remeshed by~\cite{LoopyCuts2020}. In this case the meshes are hex-only.}
	\label{fig:loopy}
\end{figure}
The recent method proposed by Livesu et al. \shortcite{LoopyCuts2020} produces \emph{strongly hex-dominant meshes}, conforming meshes with less than 2\% non-hexahedral cells. In most cases (76\% of the models tested), this method derives pure hex-meshes. It mimics manual block decomposition. It extracts first a set of well-distributed loops on the surface following a feature-aligned cross-field. Each loop defines a cutting surface that decomposes the volume into simpler polyhedral blocks. The cutting surfaces are added one by one until the quality requirements of the polyhedral blocks are satisfied. These blocks are finally converted into hex-dominant mesh via midpoint subdivision. As shown in \cref{fig:loopy}, this method excels in the preservation of sharp features, which are directly incorporated in the output connectivity.

\minrev{Similarly to LoopyCuts \cite{LoopyCuts2020}, HexDom \cite{yu2021hexdom} produce a block decomposition where each block is either hex, prism, or
tetrahedral cell. This method extends the approach for polycube generation proposed in \cite{cvt_pc_zhang2016} based Voronoi tessellation (CVT) to include non-hex elements. The cells are embedded in 3D using a variation of \cite{yu2020hexgen}. The segmentation and polycube definition process requires manual work.
The methods proposed in \cite{TetraHex2018} and the one used by the commercial package Cubit \cite{Meyers1998TheH} uses advancing front approaches to produce meshes composed of hexahedral and tetrahedral elements. To improve the quality of the final mesh Cubit use some sophisticated cleanup operation based on connectivity editing and geometric measures.\\}

\minrev{The recent approach proposed by Bukenberger and colleagues \cite{bukenberger2021most} generate \emph{At-Most-Hexa Meshes}. At-Most-Hexa Meshes are meshes composed mainly of hexahedral elements, where no cell has more than six faces, and no boundary face has more than four sides. Similarly to tetrahedral and hexahedral meshes, the volume of each cell can be defined by trilinear interpolation from its corners. At-Most-Hexa Meshes meshes are generated by extending to the volume the 2D approaches that use Lloyd relaxation with non-euclidean distance measures \cite{Alejo01}. Using the $L_{\infty}$ norm (instead of the simple euclidean distance), the cells emerging from the volumetric Lloyd relaxation process become more cubical, converging to a hex-dominant mesh. Similarly to most of the meshing methods based on Lloyd relaxation, this method is very permissive on the required input. It works on point clouds, triangular meshes and can be guided by an input orientation field if available.\\}

\minrev{The hex-dominant mesh allows sufficient degrees of freedom to adapt the grid-based methods to conform to sharp features. Trimmed hexahedral meshes are created by intersecting a grid with a closed surface. Non-hexahedral elements emerge along the surface where the surface is not aligned with the edges of the grid. The technique recently proposed by Kim and colleagues \cite{kim2021} extends the trimmed hexahedral methods by creating particular vertices where sharp features intersect with the grid. A feature simplification schema is used when multiple features are concentrated in the same cell.\\}

\minrev{Another class of methods transforms hex-dominant meshes to increase the number of hexahedral elements in the mesh. The approach proposed in \cite{YamakawaS03} increases the number of hexahedral and prism elements by applying sequences of local operations that modify the connectivity. Unfortunately, this method can generate non-conformal meshes.  Instead, HexHoop \cite{Yamakawa2002HEXHOOPMT} converts a mesh composed of hexahedrons, prism and tetrahedrons into a conformal pure-hex mesh. The conversion process is based on the local application of two particular refinement schema, called core and caps. Unfortunately, this method tends to insert a high number of irregular vertices deteriorating the regularity of the tessellation.\\}

\begin{table*}
\footnotesize
\hyphenpenalty=10000
\renewcommand{\arraystretch}{2.0}
\newcolumntype{M}[1]{>{\RaggedRight}m{#1\textwidth}}
\newcolumntype{C}[1]{>{\Centering}m{#1\textwidth}}
\centering
\caption{Summary of the main properties for each class of hex-meshing algorithms reported in \cref{sec:methods}. Some of the columns in this table correspond to items also listed in~\cite{blacker2000meeting}. As a rough indicator for the extent of (ongoing) research activity, we list the number of overall works and recent (published within the last 5 years) works (referenced in this survey) dealing with each class.}
\label{tab:summary}

\begin{tabular}{@{\extracolsep{-.5em}}
M{.065}%
M{.045}%
M{.07}%
M{.05}%
M{.06}%
C{.035}%
M{.085}%
M{.06}%
M{.09}%
C{.05}%
C{.04}%
C{.04}%
M{.135}%
}

\toprule
\textbf{Method} & \textbf{Type} & \textbf{User interaction} & \textbf{Shape class} & \textbf{Feature preserv.} & \textbf{Size contr.} & \textbf{Mesh structure} & \textbf{Element quality} & \textbf{Robustness} & \textbf{Orient. sensitive} & \textbf{Total works} & \textbf{Recent works} & \textbf{Open problems} \\
\midrule

Advancing front & Direct & Automatic & CAD oriented & Surface features only & No & Unstructured & Good at border, poorer inside & Poor & No & 7 & 0 & {\raggedright Improve handling colliding fronts, complex topologies} \\

Dual methods & Both & Automatic, semi-automatic & CAD oriented & Surface features only & No & Unstructured, semi-structured & Good at border, poorer inside & Poor & No & 13 & 2 & Robust handling of self- intersecting sheets \\

Sweeping, decomp. & Both & Semi-automatic &  CAD oriented & Surface features only & No & Semi-structured & Good & Good (manual) & No & 35 & 11 & Automatic definition of sweepable sub-volumes \\ 

Grid based & Indirect & Automatic & Any shape & Yes, (limited valence) & Yes & \minrev{Severely unstructured} & \minrev{Poor at border, optimal inside} & Great (commercial product, demonstrated on many datasets) & Yes & 18 & 3 & Feature preservation, \minrev{mesh size, mapping}\\ 

Polycube maps & Indirect & \minrev{Automatic, semi-automatic} & Any shape & Yes, (limited valence) & Yes & Valence semi-structured & Good (depends on map) & \minrev{Good (demonstrated on medium datasets)} & Yes & 18 & 8 & Polycube topology, mapping, feature preservation \\

Frame fields & Indirect & Automatic, manual fixing & Any shape & Yes & Yes & Valence semi-structured & Good (depends on map) & Poor & No & 17 & 12 & Generation of hexable fields, field aligned mapping \\

Hex-dominant & \minrev{Both} & Automatic & Any shape & Yes & Yes & Valence semi- structured, hybrid & Often good & Good (demonstrated on medium datasets) & No & 20 & 11 & Hybrid elements (topological control, amount, quality) \\

\bottomrule
\end{tabular}
\end{table*}

%% file: 05_topological_operations.tex
\section{Topological operators}
\label{sec:operators}

\minrev{Scientific computing often demands to edit a given hexahedral mesh, e.g. to improve the accuracy of a solution with a posterior refinement~\cite{shen2022hexahedral}, to generate boundary hex layers for CFD applications~\cite{reberol2021robust}, to ensure mesh conformity across surface membranes~\cite{staten2010hexahedral} or to constrain the mesh size \cite{Marechal2009}. Unlike tetrahedral meshes, where changes of the mesh connectivity always have a local footprint, editing the topology of a hexahedral mesh is often a global operation. This makes hexmesh editing significantly more difficult than tetmesh editing. In this section, we revise the most prominent operators for editing the topology of a hexahedral mesh.}

\subsection{Sheet operators}
\todo{\FL{}}
As shown in \cref{sec:dual}, the dual of a hexahedral mesh is a simple arrangement of surfaces. Each surface, i.e., a sheet, corresponds to a layer of hexes in the primal mesh and two surfaces intersect along a chord, which is a column of hexes in the primal mesh. Sheet operators consist in inserting or removing a complete sheet or chord from the mesh. They are used to refine~\cite{tchon02,parrish2007} or coarsen  meshes~\cite{benzley2005, shepherd2010adaptive}, to capture analytic features~\cite{merkley2007}, or to make conforming meshes involved in the assembly of parts. Those parts can result from a volume decomposition during a user-assisted meshing process ~\cite{meshcutting,grafting} or can correspond to two adjacent models sharing contact surfaces~\cite{staten2010hexahedral}.

\begin{figure}[tb]
	\includegraphics[width=0.48\linewidth]{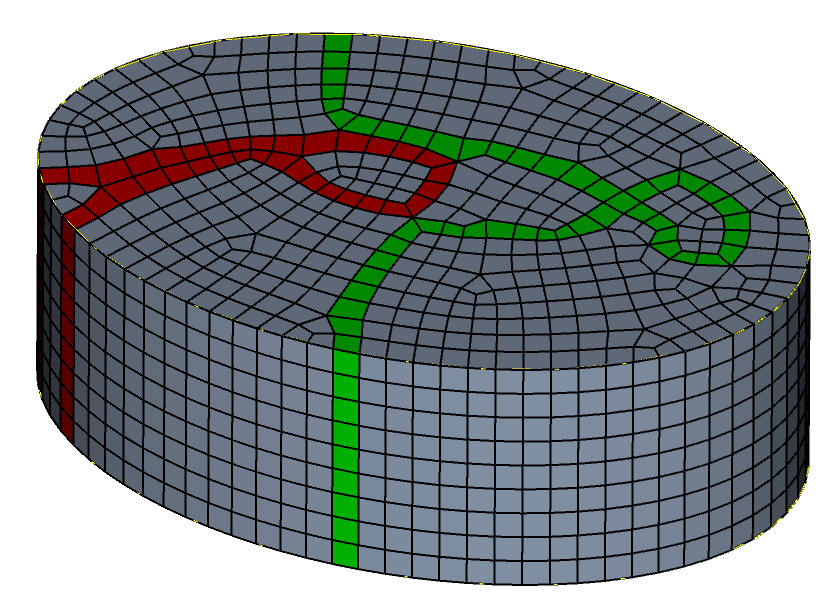}
	\includegraphics[width=0.48\linewidth]{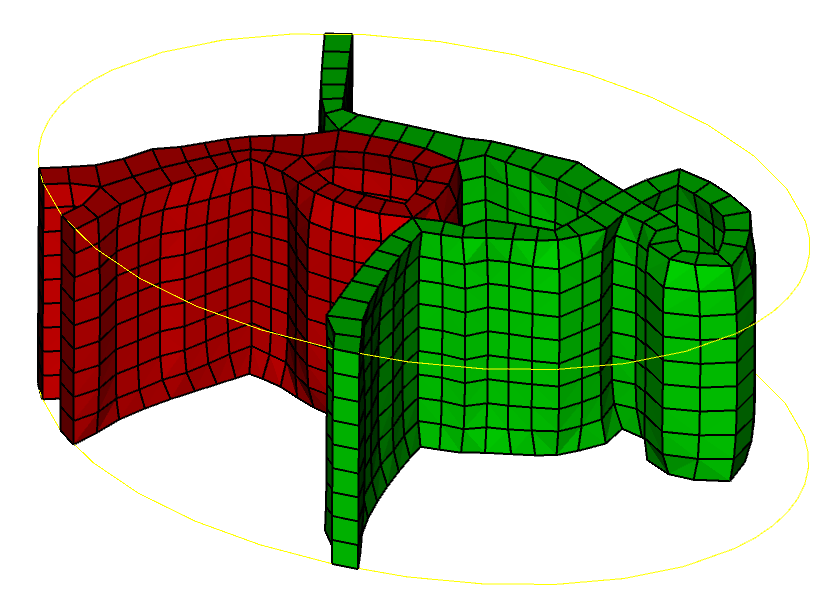}
	\caption{Examples of self-intersecting sheet (green) and self-touching sheet \minrev{(red)}.\label{fig:SelfSheet}}
\end{figure}

Using sheet operations to refine a mesh mainly involves inserting sheets, which is quite easy to control if you avoid self-intersecting and self-touching sheets (see \cref{fig:SelfSheet}). The remaining difficulty is to control the mesh quality, which is connected to the edge valence~\cite{staten2010EdgeValence}. The simplest way of inserting a sheet consists in \minrev{padding} a region of hexahedra by inserting a layer of hexes around it. It is a common post-process to improve meshes obtained with overlay-grid~\cite{Marechal2009,Zhang2010}, where a global \minrev{padding} is performed,  or Polycube-based~\cite{cherchi2019selective} techniques, where the region to \minrev{pad} is selected in such a way that the mesh quality is optimized \minrev{(see \cref{sec:pillowing})}.   

Coarsening is much more tricky since some sheets cannot be collapsed without loosing a part of the geometry - resulting in a non-manifold configuration for instance -  and one might have to deal with self-intersecting and self-touching sheets, which are much more complex to remove. In~\cite{gao2017simplification}, such coarsening is performed to simplify the base complex structure. Generating a  hexahedral block structure can also be seen as coarsening an existing hexahedral mesh. In~\cite{WANG2017}, authors extend the preliminary work of~\cite{kowalski2012fun} where a hexahedral mesh, obtained from converting a tetrahedral mesh by splitting each tetrahedron into four hexahedra, is coarsened by removing all non-funda-mental sheets. They extend the greedy approach proposed in~\cite{kowalski2012fun} by providing much more quality control and sheet selection procedures.

\begin{figure}[tb]
	\includegraphics[width=\linewidth]{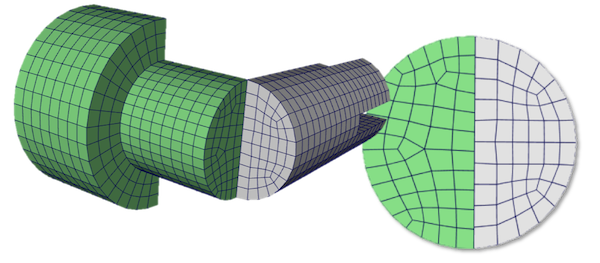}\\
	\includegraphics[width=\linewidth]{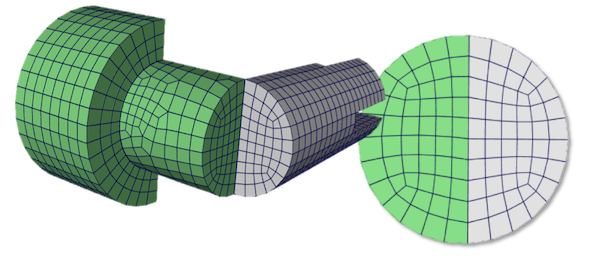}
	\caption{Two meshes generated using sweeping are not conform along a contact surface (top); Performing sheet operations on the green mesh allows to get a conforming interface (bottom). \minrev{Image from \cite{staten2010hexahedral}}.\label{fig:MeshMatching}}
\end{figure}

In order to make two hexahedral meshes of two geometrical parts sharing a surface conforming, the authors of~\cite{staten2010hexahedral} interleave sheet insertions and collapses in both parts. The locality is controlled by performing chord collapses to avoid the propagation of mesh modifications too far from the interface (see \cref{fig:MeshMatching}). Chord collapsing is done by taking care of mesh quality, considering edge valences as an appropriate indicator~\cite{staten2010EdgeValence}.
In~\cite{chen2016approach}, the sheet insertion is enhanced to provide more flexibility, in particular, to handle self-intersecting sheets within a local region while assuring the mesh quality.  It was successfully applied to mesh matching and mesh boundary optimization.

Eventually, in most recent works, like~\cite{shen2019topological}, the chord collapse operation is enhanced to avoid generating poor quality elements, and the chord insertion process is described and used for editing the singularities of a hex-mesh while maintaining its connectivity. In~\cite{wang2018hex}, sheet operations are used in combination with frame fields to improve mesh quality with the ability to handle self-intersecting and self-touching sheets.

\subsection{Flipping operators}
\label{sec:flipping}
\todo{\textcolor{cyan}{(JF)}}
Among the many difficulties of hexahedral meshing, there is one that is unexpected, to say the least. The generation of conforming hexahedral meshes of complex 3D domains is definitively a hard problem. Yet, finding hexahedrizations for small quadrangulations of the sphere is also hard.

\paragraph*{Existence}
Thurston \shortcite{Thurston1993} and Mitchell \shortcite{MitchellCharacterization1996} have shown independently that a ball bounded by a quadrangulated sphere could be meshed with hexahedra \emph{if and only if} the number of quadrangles on the boundary, $n$, is even. 

\begin{figure}[ht]
	\centering
	\includegraphics[width=0.2\linewidth]{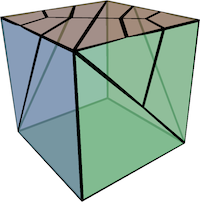}
	\includegraphics[width=0.2\linewidth]{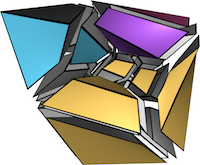}
	\includegraphics[width=0.15\linewidth]{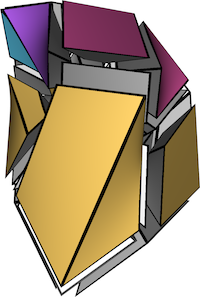}
	\includegraphics[width=0.2\linewidth]{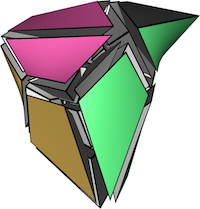}
	\includegraphics[width=0.2\linewidth]{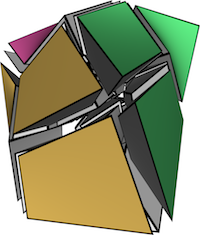}%
	\vspace{.7cm}
	
	\includegraphics[width=0.2\linewidth]{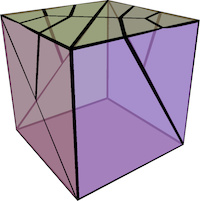}
	\includegraphics[width=0.2\linewidth]{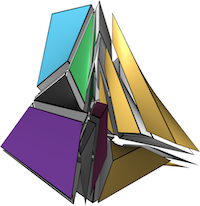}
	\includegraphics[width=0.2\linewidth]{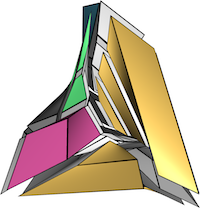}
	\includegraphics[width=0.15\linewidth]{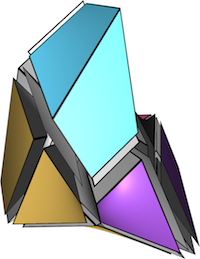}
	\includegraphics[width=0.2\linewidth]{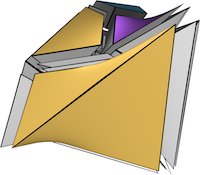}
	\vspace{.2cm}
	
	\caption{The set of the hexahedrizations of the buffer cubes that the~\cite{erickson2014}'s algorithm uses to mesh arbitrary domains. (top) Cells with 20 quadrilaterals are meshed with 37 hexahedra. (bottom) Cells with 22 quadrilaterals are meshed with 40 hexahedra. Elements are colour codeded to show the different sides of the original cubes (top-left and bottom-left). \minrev{Image from~\cite{verhetsel2019finding}.}}
	\label{fig:erickson-buffers-hex}
\end{figure}

\paragraph*{Linear complexity meshing}
Mitchell's construction can necessitate up to ${\mathcal O}(n^2)$ hexahedra. In~\shortcite{eppstein1999linear}, Eppstein proposed a \minrev{\say{semi-constructive}} alternative which guarantees the use of ${\mathcal O}(n)$ hexahedra. 
The algorithm of Eppstein extends the quad-mesh in input into a buffer layer of hexahedra. Then it triangulates the inner of the layer with ${\mathcal O}(n)$ tetrahedra, applies the midpoint subdivision to split each tetrahedron into four hexahedra, and eventually refines the cubes in the buffer into smaller cubes that consistently meet the previously subdivided tetrahedra. \minrev{The inserted buffer layer is mandatory to provide much degrees of freedom to topologically and geometrically modify the inner mesh. It must be remeshed at the end to ensure mesh conformity.} 
The last stage \minrev{\say{only}} requires finding a solution of $20$ or $22$ quadrilaterals buffer cubes. At that point, an explicit solution is required for the buffer cubes. In~\shortcite{shepherd2010adaptive}, Carbonera and Shepherd give the first completely explicit construction of the hexahedrization of the ball. This method, however, requires up to $5396 n$ hexahedra. Using~\cite{shepherd2010adaptive}, a solution for the buffer cubes has been found by Weill and Ledoux~\shortcite{weill2019towards} that involves $76881$ hexes! 
In~\shortcite{verhetsel201944}, Verheltsel introduced an efficient quad flip-based algorithm that allows finding hexahedral meshes for both the types of buffer cells previously described. Furthermore, as depicted in \cref{fig:erickson-buffers-hex}, it provides geometric realizations with a maximum number of 72 hexahedra, thus proving that it is possible to mesh any ball-shaped domain that is bounded by $n$ quadrangles with a maximum number of 78$n$ hexahedra.

\paragraph*{\minrev{Schneiders'} pyramid and octahedral spindle}
The pyramid of Schneiders is square-based, with 8 additional vertices at the edge midpoints, 5 additional vertices at the face midpoints, and its triangular and quadrangular faces divided respectively into 3 and 4 quadrangular faces. To build the Schneiders' pyramid, we can use the octagonal spindle, or tetragonal trapezoid, and add 4 hexahedra to form the pyramid base. 
Meshing this pyramid with all-hexahedral elements is a problem introduced by~\cite{schneiders1995} to show a boundary mesh for which no one hexahedral subdivision was identified.
A good solution to Schneiders’pyramid is considered as the missing piece to transform a hex-dominant mesh into a full unstructured hex-mesh. 
\cite{yamakawa2002hex} introduced, in 2002, the hexhoop template family and built a hexahedral subdivision for the pyramid of Schneiders, composed of 118 hexahedra.  Later, in 2010, they improved their solution by creating a new hexahedral subdivision of 88 elements~\cite{yamakawa201088}. Recently, in 2018, a hexahedral subdivision of 36 elements was built by finding a set of flipping operations allowing to turn the cube into Schneiders' pyramid, by interpreting each operation as the addition of a new hexahedron~\cite{xiang201836}. Verheltsel~\shortcite{verhetsel201944} used quad flips to find another solution with 44 hexahedra.

\paragraph*{Shellings}
In mesh generation, flipping (or swapping) operators convert small cavities of elements into alternative collections of elements having the same boundary. Flips are used extensively in tetrahedral meshing with the aim of \emph{improving the mesh}. 
The \minrev{\say{bistellar flips}}, the most basic operations, operate on a cavity of 5 vertices produced by removing two or three tetrahedra. Instead, the \minrev{\say{edge removal}} operation, a more general transformation $n$-to-$m$ flip, works on a cavity produced by the set of tetrahedra enclosing an edge. Rather than adding more and more operations to the already big set of topological transformations, the \minrev{\say{small polyhedron reconnection (SPR)}}~\cite{liu2007spr} provides an operation that can generalize all the flips. The SPR considers the problem of finding \emph{all the possible triangulations of a cavity} and choosing the best one.

Flipping operators in cubical meshes were introduced by {M. Bern}, {D. Eppstein} and {J. Erickson} in~\shortcite{bern2001flipping} and are analogous to the flipping operators for simplicial meshes. 
Those authors prove that each domain that is simply-connected and has an even number of quadrilateral faces also has a pseudo-shelling. 
A pseudo-shelling is defined as a particular kind of hex-mesh built by adding elements one by one such that the remaining elements always make a ball-shaped domain.

\subsection{Atomic operators}
\todo{\FL{}}
Atomic operators form a set of irreducible local
operations which could be composed to described any
topological modification~\cite{tautgesa2003topology,tau08}. It consists of
three very local atomic operations, which are the atomic pillow,
the face shrink and the face open-collapse. An important feature of those operations is that applying just a single atomic operation does not provide a valid hex-mesh. But it has been demonstrated that flipping operations~\cite{tau08} and sheet operations~\cite{ledoux2010topological} can be obtained as a sequence of atomic operations. Unfortunately, the completeness of this set of operations is not proved, and they are very difficult to be used for writing meshing algorithms in practice.

For instance, those operators do not capture a parity change in the number of hexahedra. 
Therefore, an extra operator was presented in~\cite{jurkova08}, where a Boy’s surface is added in the dual mesh representation. The surface of Boy has the interesting property of having a single vertex. Thus, introducing it, in an appropriate way, into the dual mesh representation, the parity of the hexahedra number changes in the primal mesh. In~\cite{jurkova08}, a sheet diagram is provided, but the primal mesh realization from this insertion is incomplete. Interestingly, utilizing the Carbonera's algorithm~\shortcite{carbonera06} on a single hexahedron, it is possible to perform a parity change. Regardless of the template set of the Carbonera's method, it always replaces a hexahedron with an even number of hex-elements without altering the boundary of the input hex.

%% file: 06_refinement_coarsening.tex
\subsection{Padding} \label{sec:pillowing}
\begin{figure}
	\centering
	\begin{subfigure}[t]{.47\columnwidth}
		\centering
		\includegraphics[width=.7\linewidth]{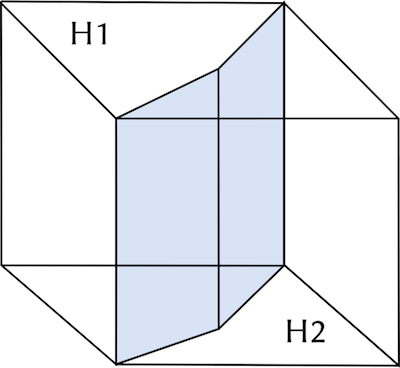}
		\caption{doublets}
		\label{fig:paddingA}
	\end{subfigure}
	\hfill
	\begin{subfigure}[t]{.47\columnwidth}
		\centering
		\includegraphics[width=.7\linewidth]{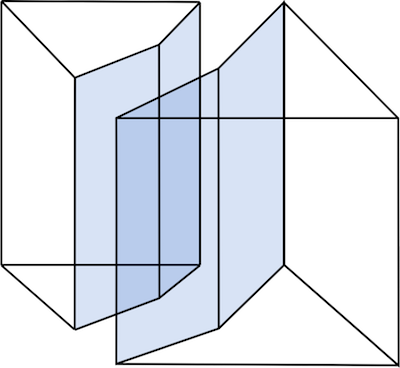}
		\caption{divided shrink set}
		\label{fig:paddingB}
	\end{subfigure}
	
	\medskip
	\medskip
	\medskip
	\medskip
	
	\begin{subfigure}[b]{.47\columnwidth}
		\centering
		\includegraphics[width=.7\linewidth]{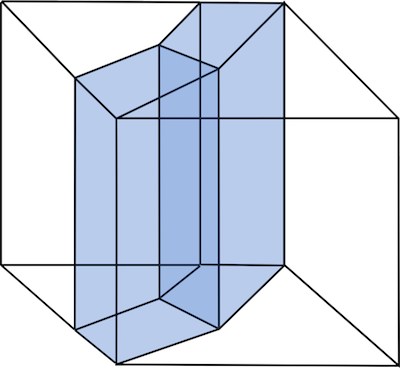}
		\caption{\minrev{padding} layer}
		\label{fig:paddingC}
	\end{subfigure}
	\hfill
	\begin{subfigure}[b]{.47\columnwidth}
		\centering
		\includegraphics[width=.7\linewidth]{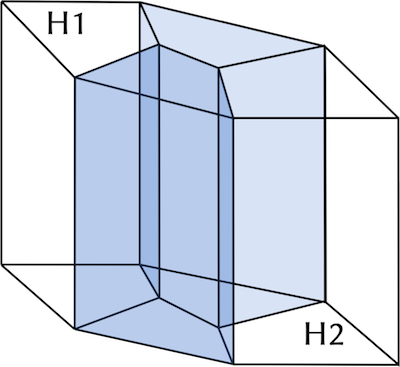}
		\caption{optimized hexahedra}
		\label{fig:paddingD}
	\end{subfigure}
	\caption{A summary of the \minrev{padding} pipeline: (a) The hexahedra H1 and H2 share two faces forming two doublets. (b) The shrink set is disconnected to the other elements in the mesh. (c) The elements of the shrink set are linked to the mesh forming the \minrev{padding} layer. (d) The hexahedra involved in the refinement can now be optimized with a smoothing/untangling step.}
	\label{fig:padding}
\end{figure}

\todo{\textcolor{orange}{(Gianmarco)}}
Sometimes hexahedral meshes (as well as quad-meshes) can contain \emph{doublets}. As described in~\cite{mitchel95_pillowing}, a doublet is defined as two quad faces sharing two edges, and, in the hex-meshes case, this means that two hexahedra share two faces (\cref{fig:paddingA}). If doublets occur in a mesh, any kind of geometric embedding of the faces forming the doublet has a low quality, even if we try to optimize it with some smoothing/untangling step. In fact, one of the involved faces will always have an angle of at least $\pi$. The local connectivity of the mesh requires a refinement step to remove doublets. The \minrev{\emph{padding} refinement operation, also known as \emph{pillowing}, refines the mesh structure in order to provide additional elements, and hence degrees of freedom, for existing approaces of mesh optimization (e.g. untanglers).}
In quad-meshes, this step is trivial. Removing the two shared edges forming a single quadrangular face is sufficient. 
It is not possible to apply the same for hexahedral meshes because we can not ensure that the hexahedra with doublet faces can be matched in a conformal configuration. In this case, it is required to increase the connectivity of the vertex shared by the two edges that form the doublet.

In~\cite{mitchel95_pillowing}, the authors propose a pipeline to face the problem in three steps. First of all, a \emph{shrink set} is defined as a set of hexahedra containing one (and only one) of the doublet faces. Then, the set is separated from its boundary by reducing the size of its elements. In this way, an empty space is created (\cref{fig:paddingB}). Finally, each of the shrink set elements is connected to the boundary through a new layer of hexahedra, filling the previously created empty space (\cref{fig:paddingC}).
After the \minrev{padding}, the original doublet's faces are contained in two different hexahedra, doublets are no longer present in the mesh, and the dihedral angles between faces can now be improved (\cref{fig:paddingD}).

The \minrev{padding} is basically a sheet insertion on a mesh. As described in~\cite{shepherd2007topologic, Shepherd2008}, it is a fundamental step in many applications. Starting from the mesh generation~\cite{ito2009octree, gao2019feature} or the generic refinement of hexahedral meshes~\cite{Tchon2004, benzley2005conformal, Zhang2010, Malone2012TwoRefinementBP, ZHANG201388}, it becomes an essential ingredient in operations like grafting~\cite{grafting}, mesh cutting~\cite{meshcutting}.

In~\cite{gregson_polycube_2011} the \minrev{padding} is identified as a key post-processing step for the hex-meshes obtained from polycubes (see \cref{sec:polycubes}). In this mesh category, the surface edges belonging to the polycubes structure can create configurations similar to doublets. The degrees of freedom of the surface elements are then increased with a \minrev{padding} step performed all around the mesh (all the inner volume becomes the shrink set and is separated from the surface). In this way, a geometric optimization step can enhance the quality of the elements placed in the smooth object parts. Notice that, in almost all polycube-based hex-meshing works, the \minrev{padding} operation is applied as a unique hexahedral layer all over the surface. In the same context, in~\cite{cherchi2019selective}, a smart and localized \minrev{padding} for this hex-mesh category is proposed. The authors demonstrate that selective \minrev{padding} in sporadic surface areas can significantly improve the whole mesh quality compared to the global \minrev{padding} application. 

\subsection{Structure enhancement/simplification}
\label{sec:struct_enahncement}
\todo{\textcolor{cyan}{(Xifeng)}}

\begin{figure}
	\includegraphics[width=\columnwidth]{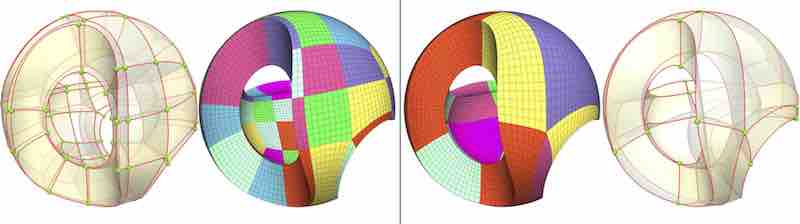}
	\caption{Without controlling of alignment, the same set of singularities can introduce two hex-meshes having base complexes with different complexity. \minrev{Image from~\cite{gao2015hexahedral}.}}
	\label{fig:base_complex_misalignment}
\end{figure}

Hex-meshes with simple structures are preferred for isogeometric analysis~\cite{hughes2005isogeometric} since large components allow the fitting of high-order splines without breaking their smoothness so that accurate PDE solving and fast convergence can be achieved. Note that the base complex of a hex-mesh is not only determined by its singularities, but also the connections between them. Therefore, without careful control, the same set of singularities can lead to dramatically different base complexes (\cref{fig:base_complex_misalignment}). Gao et al.~\shortcite{gao2015hexahedral} propose the first solution to reduce the number of components of the base complex by correcting misalignments of singularities. The misalignment correction is achieved by removing hexahedral sheet defined within the base complex. To maintain singularities, specific conditions are posed for choosing the proper hexahedral sheets for removal. After obtaining the hex-mesh with corrected misalignment issue, they employ an extended version of the parameterization-based optimization from quad-meshes~\cite{tarini2011simple} to hex-meshes to improve the geometric quality of the hex-meshes. To specifically handle misalignment issue for polycube hex-meshes, \cite{cherchi2016polycube} proposes an approach by alternating two steps: (1) computing polycube corner pairs in the integer lattice, and (2) aligning corner pairs through mixed-integer programming. 
\begin{figure}
	\includegraphics[width=\columnwidth]{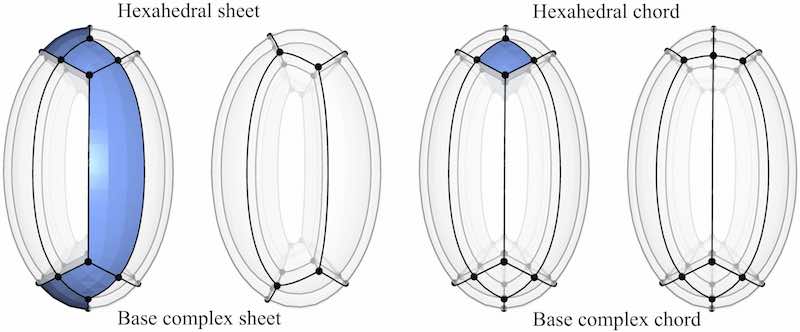}
	\caption{Removing either a base complex sheet (left) or a chord (right) on the global structure of a hex-mesh monotonically reduce the number of components of the base complex. \minrev{Image from~\cite{gao2017simplification}.}}
	\label{fig:base_complex_sheet_chord}
\end{figure}
\begin{figure}
	\includegraphics[width=\columnwidth]{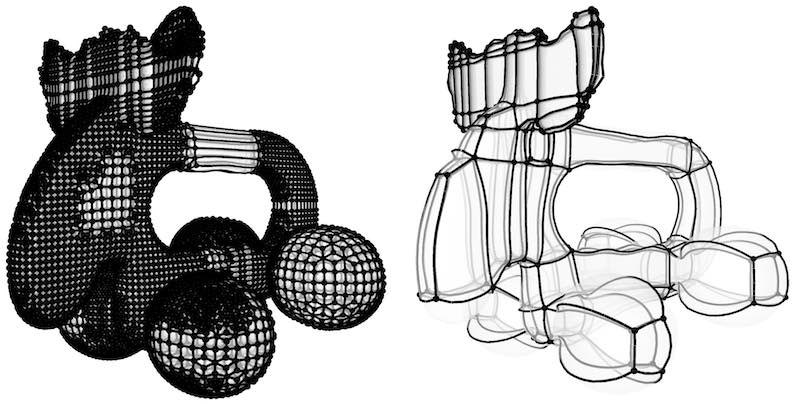}
	\caption{Turn an octree-based hex-mesh with a highly complex structure into a hex-mesh with a coarse structure. \minrev{Image from~\cite{gao2017simplification}.}}
	\label{fig:simp}
\end{figure}

Robustly producing valid hex-meshes with a simple structure remains to be a challenging task. Gao et al.~\shortcite{gao2017simplification} propose a simplification algorithm that can iteratively reduce the structure (i.e., number of components and singularities) complexity of a hex-mesh, while providing several guarantees during the simplification process: (1) topology consistency, (2) inversion-free, (3) the preservation of corner, line, and planar features, and (4) a bounded, user-defined Hausdorff distance from the input surface. The core idea of their approach is to extend the sheet and chord operations on hex element level to the structural level, as illustrated in \cref{fig:base_complex_sheet_chord}. The input to this approach can be an arbitrary hex-mesh. Especially, this approach can be paired with  octree-based methods discussed in~\cref{sec:grid_based_gen} to robustly generate valid (i.e., with no flipped elements) and accurate hexahedral meshes with coarse structures, without user-interactions (\cref{fig:simp}). \minrev{A follow-up work \cite{xu2021singularity} is conducted to improve the ranking scheme of the sheets and chords to be removed. The experiments show that, while being more complex, the introduced ranking leads to better simplification results.}

\minrev{Through proving the equivalence between colorable quad-meshes and Strebel differentials on a manifold closed surface, Lei et al~\shortcite{LEI2017758} propose to first construct a colorable quad-mesh and then partition the surface into sub-volumes where each of them can be swept to generate a hex-mesh. While the theory is elegant, there are several limitations of the work, prohibiting its adoption for practical applications. For example, the required special user inputs are non trivial, only quad vertices with even valences are allowed, and distortions of the hexahedra could be arbitrarily large.}

By adapting the editing operations of singularity pairs for quadrilateral meshes \cite{quad-editing-2011} to hexahedral meshes, Shen et al. \shortcite{shen2019topological} propose to employ chord collapse and insertion to flexibly control the singularities of a hexahedral mesh. The main limitation of this approach is that chord insertions are not always feasible when the structure of the mesh is complex. The authors demonstrate that the proposed editing operations can be used to clear some connectivity inconsistency issues for sweeping based hex-meshing.

%% file: 07_optimization_untangling.tex
\section{Geometric Optimization of Element Quality}
\label{sec:untangling}

The vast majority of hexahedral meshing algorithms employ a two-step process where the first step generates an initial mesh which is expected to be dominated by well-shaped elements, but often also contains some poorly-shaped and even {\em inverted}, or negative Jacobian determinant, elements (cf.~\cref{sec:quality}). This step is typically followed by an optimization step whose goal is to maximize the quality of the mesh elements and specifically to produce an inversion-free mesh, while preserving the meshed domain boundary surface intact. Improvement methods that keep the mesh connectivity fixed while changing only the locations of the mesh vertices, are commonly referred as {\em geometric optimization}, {\em smoothing}, or {\em untangling} methods~\cite{owen1998survey,Shepherd2008}.
The latter terms are commonly used to describe the methods that specifically focus on reducing, ideally to zero, the number of inverted elements. 
As noted by Knupp~\shortcite{Knupp2001}, for hexahedral meshes there can be more than one definition of inverted elements. He identifies four different scenarios: (a) the
integral of the Jacobian determinant over the element is non-positive; (b) the Jacobian determinant is non-positive 
at any of the Gaussian integration points (used, e.g., in FEM) inside the element (c) the Jacobian determinant is non-positive at any of the element's corners, or (d) the Jacobian determinant is non-positive at some
other specific point(s) inside the element. The vast majority of optimization and untangling methods, described below, focus on the scenario (c). 
The general scenario, seeking for positive Jacobian determinant at \emph{every} point, is tackled by a few methods~\cite{johnen2017robust,Marschner2020SOS} based on optimization formulations attempting to maximize lower bounds of the Jacobian determinant, cf.~\cref{sec:detbounds}.
As with mesh generation itself, geometric optimization methods for hex-meshes face some distinctly different challenges from methods for tet-mesh optimization such as 
~\cite{Freitag2006,Sastry14,Erten09,Kelly11,Scherer10}, motivating a distinct line of research dedicated to optimizing hex-mesh geometry.  

One can easily define an objective function whose global minimum (or maximum) constitutes the best quality mesh possible for a given fixed connectivity (and either hard or soft constraints that hold the surface vertices on the surface of the meshed object). However, essentially all such known objective functions are highly non-linear and do not allow for robust global optimum computation. Thus the core challenge in mesh geometry optimization is to obtain a function and a corresponding optimization method such that the optimum obtained has no inverted elements and maximizes as much as possible the mesh Jacobian or other proxy quality metrics (see \cref{sec:quality}).  

Consequently, the main difference between the methods is in the optimization strategy used. A few attempts tried to develop generic global optimization strategies which directly optimize the quality across all mesh vertices (\cref{sec:global}); however existing methods are not widely used and exhibit inferior performance compared to existing alternatives. 
Most existing and widely used methods are based on Gauss-Seidel iterations (\cref{sec:GS}), where vertices are relocated one at a time. The advantage of this strategy is that one can explicitly prevent the quality from dropping locally, e.g., preventing the formation of new inverted elements. 
\minrev{The drawback is in increased likelyhood of converging to a purely local minimum.}
Recent research investigates different local-global approaches for mesh optimization (\cref{sec:LG}).  This line of research shows great promise, with several methods significantly outperforming prior art. Below we review these three families of methods in more detail. 

\subsection{Global optimization}
\label{sec:global}
Several authors aim to optimize mesh quality by simultaneously updating all vertex positions, e.g.~\cite{Yilmaz2009,gaoa2016local}; however they only demonstrate results on simple inputs. It is not clear if these approaches can be extended to a more general setting. 
The use of global non-linear methods for optimizing mesh quality was investigated in depth by 
Sastry et al~\shortcite{sastry2009comparison} and Wilson~\shortcite{WilsonThesis} for tetrahedral and hexahedral meshes respectively.
Both concluded that global methods that directly optimize hex shape metrics as a function of vertex positions are typically less robust than the Gauss-Seidel approach, discussed next, and frequently converge to 
poorer solutions. 

\subsubsection{Gauss-Seidel Iterations}
\label{sec:GS}

Older optimization approaches, reviewed by  ~\cite{Frey2007},  iteratively relocate interior mesh vertices to some weighted average, or center, of their neighbors, one vertex at a time. Earlier methods used simple geometric averages, e.g., positioning vertices so as to minimize the Laplacian 
energy around each vertex
\[\min_i ||v_i - 1/N_i \sum_{j \in N(i)} v_j||^2\]
where $v_i$ are mesh vertex positions, $N_i$ is vertex valence, and $N(i)$ are the vertices adjacent to $i$. This basic framework is often sufficient to produce quality outputs given meshes with simple connectivity and low-detail surface geometry, with quality surface quad-mesh. 
Recent methods, e.g., ~\cite{vartziotis2013improved,Zhang2009,Knupp2003}, use more sophisticated local energy formulations that aim to optimize the size of the solid angles at each vertex. 
For example, Knupp~\shortcite{Knupp2003} encode each hex corner geometry via the condition number of a matrix describing the coordinate system at the corner vertex
\[M=(e_0, e_1,e_2)\]
where $e_i$ are the hex edges emanating from the vertex. The smaller the condition number, the better-shaped the hex is locally (corner solid angle closer to $90^\circ$). His method uses line-search to move the vertices one at a time so as to minimize the worst or average condition number impacted by the position of this vertex. 
Such iterative methods are fairly efficient and can be easily parallelized. Unfortunately, 
applied as-is, such methods do not guarantee an inversion-free output. Specifically, they are often unable to untangle previously inverted element, and when applied as-is are known to frequently introduce new inverted elements near concave features along the boundaries of the meshed domain~\cite{owen1998survey}. 
Several researchers advocate employing these vertex-relocation based methods either pre or post untangling ~\cite{vartziotis2013improved,Knupp2003}. Specifically, they suggest to 
constrain each vertex move so as to avoid new inversions, and rely on the untangling methods to resolve all inverted elements. 
For example, the widely used Mesquite library~\cite{brewer2003mesquite} uses the algorithm of Knupp ~\shortcite{Knupp2001} to first untangle a hex-mesh and then improves its quality iteratively moving one vertex at a time using the method of ~\cite{Knupp2003}. 
Constraining all intermediate solutions to remain in the inversion-free space, can produce sub-optimal, local minimum outputs.

Vartziotis and Himpel ~\shortcite{Vartziotis2014a} have proposed new formulations of vertex-by-vertex optimization designed for mixed element meshes; while these formulations were successfully demonstrated in 2D space, they have yet to demonstrate those on a hexahedral or hex-dominant input. 

Knupp~\shortcite{Knupp2001} proposed an untangling method that focuses solely on correcting inverted hex-elements, while allowing the quality of the non-inverted ones to deteriorate.
The energy function he employs is based on the observation that non-negative numbers are equal to their absolute values. Thus requiring the local volume $\alpha = (v_1-v_0) \cdot (v_2 - v_0) \times (v_3 - v_0)$ at a hex corner $v_0$ (where $v_i,~i=1\ldots 3$ are the corners adjacent to $v_0$) to be non-negative can be cast as minimizing the sum
\[\sum_v (|\alpha| - \alpha)\]
over the eight corners of each hexahedron and over all mesh hexahedra. Notably, the optimum of this function can be minimized while the mesh contains zero volume elements. To prevent this configuration, the author suggests minimizing a modified energy
\[\sum_v (|\alpha - \epsilon \bar V| - (\alpha - \epsilon \bar V))\]
where $\bar V$ is the expected average mesh element size (computed as total mesh volume divided by the number of elements), and $\epsilon$ is a user defined parameter.
This approach demonstrated that this optimized energy is convex as a function of a single vertex position. If a valid solution can be achieved by moving these center vertices, thus using an appropriate convex optimization strategy, this method is guaranteed to untangle all clusters of inverted elements centered around individual vertices. This method fails on many inputs with clusters of connected tangled elements, where only a tandem movement can result in a valid solution.

In some more recent works, e.g.~\cite{WilsonThesis,RuizGirones2014,wilson2012untangling,RuizGirones14a}, iterative local Gauss-Seidel approaches are employed to correct the inverted elements and improve the overall quality of the elements. 
\minrev{In these works, the authors use a specifically designed shape metric to avoid convex elements becoming inverted elements and explictly encouraging untangling of inverted elements. }
Specifically, they start from a metric introduced by~\cite{Knupp2001A} and modify it to avoid division by zero in the presence of zero volume elements. Given the Jacobian matrix $M$, the point-wise distortion is defined via the matrix condition number,
\[ \nu = \frac{|M|_F^2}{3 D(M)^{2/3}}\]
Here $||_F$ is Froebenius norm and $D(\cdot)$ is the determinant. Notably this value goes to infinity as the determinant approaches zero. To avoid instability near zero the authors replace $D(M)$ with $\frac{1}{2}D(M)+\sqrt{D(M)^2+4\delta^2}$ where $\delta$ is a user specified small value.
They propose several different strategies for optimizing the resulting energy. In particular~\cite{RuizGirones2014} indicates that an approach where each Gauss-Seidel update performs only one step of gradient descent toward the local minimum performs best in terms of output quality. Intuitively, this observation is consistent with avoiding premature convergence to a local minimum.
{
Gauss-Seidel methods such as the ones above are widely used in industry.}

\subsection{Local-global optimization}
\label{sec:LG}
Multiple recent methods employ local-global approaches for mesh optimization. They conceptually break the mesh into a collection of local, overlapping sub-meshes, and use those in an iterative optimization process.  
In each iteration, they first optimize each sub-mesh independently, aiming for a solution that is both sufficiently good (inversion free and high quality) and maximally close to the current sub-mesh geometry. They then update the vertex positions globally while striving to maximally retain the geometry of the just computed individual local sub-meshes. The two steps are then repeated until no further improvement is possible.   The main difference between these approaches is in the choice of the local sub-meshes.

\begin{figure}
	\centering
	\includegraphics[width=\columnwidth]{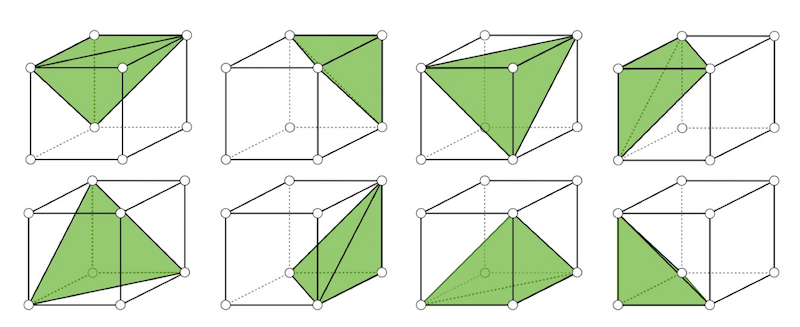}
	\caption{Corner-based approaches optimize hex shape by considering the set of eight tetrahedra formed by each corner and its three incident edges.}
	\label{fig:corner_tets}
\end{figure}
\subsubsection{Corner-Based}
\begin{figure}
	\centering
	\includegraphics[width=\columnwidth]{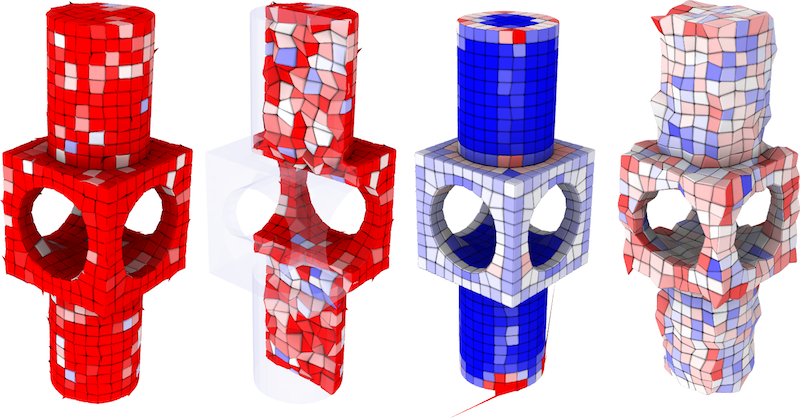}
	\caption{A degraded hex-mesh obtained by randomly displacing interior vertices (left, first two columns), optimized with a state-of-the-art corner based approach~\cite{Aigerman:bd3d:2013}. Hard constraining the surface does not allow to fully untangle the mesh (middle right, see red elements and spikes at the bottom and top). Relaxing it yields a mesh with positive minimum Jacobian, but introduces excessive surface deviation (right). Scaled Jacobian is color coded, from pure red (SJ $\leq 0$) to pure blue (SJ = 1).}
	\label{fig:corner_based_opt}
\end{figure}

Corner-based approaches consider the eight overlapping simplices formed by the corners of each mesh hexahedron (\cref{fig:corner_tets}).
These methods were originally proposed for computation and optimization of maps between simplicial complexes  e.g.~\cite{Aigerman:bd3d:2013,Schueller-13-LIM}, but can be applied as-is for hex-mesh optimization, by treating these meshes as consisting of overlapping corner tetrahedra.  These methods iterate between optimizing each tet's geometry individually, so as to satisfy a lower bound on quality, and solving for vertex positions that best preserve the resulting individual tet shapes. The global optimization step balances tet shape preservation and preservation of the coordinates of the vertices on the outer surface of the input mesh. As discussed by~\cite{Livesu:2015}, on many input this approach fails to adequately control the trade-off between boundary surface preservation and quality optimization: holding the boundaries tightly results in poor quality meshes, while relaxing the boundary constraints so as to obtain adequate quality leads to excessive surface drift (\cref{fig:corner_based_opt}).

\subsubsection{Hex-Based}
Marechal~\shortcite{Marechal2009} proposes a local-global method that is well suited for grid or octree meshes (with or without padding). 
At first, a best matching perfect cube is computed for each individual hexahedron. Since each mesh vertex is shared between multiple hexahedra, each vertex receives as target position the average of all the target positions computed for each of its incident elements. Vertices are then carefully moved towards their target position. Geometric fidelity is balanced with per element quality, and surface vertices are allowed to deviate from the nominal surface to avoid introducing flipped elements.
To avoid excessive deviation from the input boundaries or corruption of surface features, the method frequently terminates with barely convex meshes, with minuscule minimum scaled Jacobian ( $\approx 0.01$). Further quality improvement using this approach leads to significant boundary drift.

\begin{figure}
	\centering
	\includegraphics[width=\columnwidth]{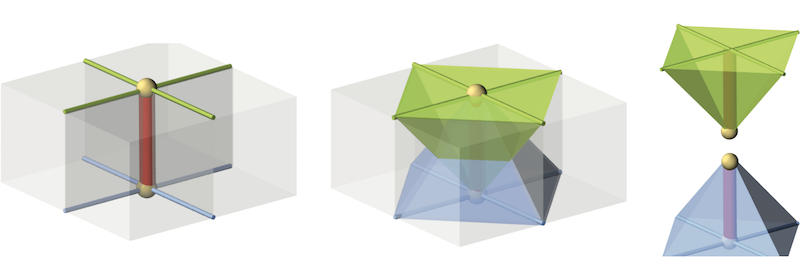}
	\caption{Cone-based methods cast mesh untangling as the problem of finding a valid axis for pairs of oppositely oriented cones associated to mesh edges. If the axis of each cone stays on the positive half space w.r.t. its base, then the Jacobian at the corners of each element incident to such edge is guaranteed to be strictly positive. Image from~\cite{Livesu:2015}.}
	\label{fig:edge_cones}
\end{figure}

\subsubsection{Cone-based}
Livesu et al~\shortcite{Livesu:2015} introduce the notion of cones, sets of hex-mesh corners that surround an {\em oriented} mesh edge (\cref{fig:edge_cones}). They then describe an optimization method that iterates between a local step that optimizes each cone independently to satisfy a minimal quality threshold with minimal changes in cone shape, and a global step that seeks to position all mesh vertices so as to maximally preserve these updated local cone shapes. Both the global and the local steps allow feature vertices to move along the underlying features, and surface vertices to move on the object's surface. The method had been shown to provide  better balance between surface preservation and quality optimization than prior approaches. 
\todo{\alla{cino - add figure?}}
Xu et al.~\shortcite{xu2018untangling} introduce a cone-based two-step optimization strategy whose first step focuses on mesh untangling, potentially at the expense of surface preservation. Their second step then improves surface fidelity and overall mesh quality while preventing the formation of new inverted elements. 

\subsection{Non-linear meshes}
The method of \minrev{Paille et al.~\shortcite{Paille2013}}
aims to compute low-distortion maps between 3D domains and hexahedral meshes with near-perfect element shape. The method progressively increases the order of the hex elements to improve quality and surface fitting. 
\minrev{This approach can be used to optimize the quality of polynomial-basis meshes but is not applicable to standard linear hex-meshes. In particular, while the higher-order elements in the meshes it obtains may be inversion-free and high quality, the underlying linear mesh elements may remain inverted.}
Most recently, Knupp et al. \shortcite{knupp2021adaptive} proposed a high-order hex-mesh optimization method that targets objects with no underlying CAD representation but using on the fly computed implicit surface representations. It specifically targets conforming meshes with interior surfaces and is advertised as well suited for computations with dynamically changing geometry. The authors demonstrate the method on a range of simple to medium complexity inputs. 

\subsection{Simultaneous geometry and topology optimization}
Changing mesh geometry and connectivity in parallel can potentially significantly improve the output quality.
However, robustly changing the topology of meshes with general connectivity can be challenging due to the global impact of such topological changes (cf.~\cref{sec:operators}). Thus such research often focused on meshes with near-regular connectivity. For instance Sun et al.~\shortcite{Sun2012} propose an optimization method specific for grid-based meshes that employs a combination of modified Laplacian smoothing and topological operations on the padding layer of the input mesh. The method is demonstrated to work on simple inputs; thus its applicability in the general case remains to be tested. Guided by a frame field generated from an existing hex-mesh, Wang et al.~\cite{wang2018hex} propose to identify the hexahedral sheets containing hexahedra with the worst scale Jacobian quality, collapsing the identified sheets, and inserting new sheets with possibly higher quality indicated from the frame field. The insertion of new sheets relies on a stream quad surface extraction which may have robustness issues when the mesh structure is complex.

\subsection{Meshes containing hybrid elements}
\begin{figure}
	\includegraphics[width=\columnwidth]{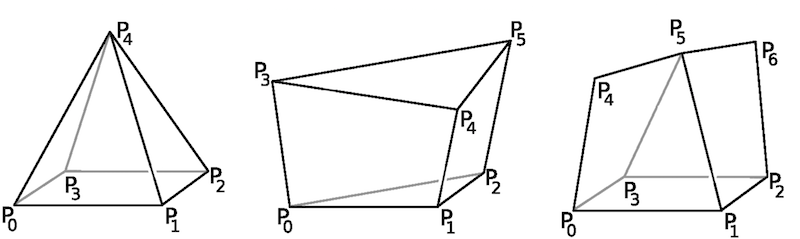}
	\caption{From left to right: pyramid, wedge, and knife -- the only hybrid elements listed in the Verdict Library, a popular reference for the computation of quality metrics of finite elements. Image from~\cite{stimpson2007verdict}.}
	\label{fig:verdict}
\end{figure}

Meshes containing spurious (non-hexahedral) elements demand geometric optimization schemes that are able to improve the quality of arbitrary polyhedral cells. Different from standard finite elements, such as tetrahedra and hexahedra, the literature on general polyhedra is scarce. The manual of the Verdict Library~\cite{stimpson2007verdict} is a prominent reference for quality metrics of finite elements, and briefly reports only about pyramids, wedges, and knives (\cref{fig:verdict}), proposing the signed volume as a unique metric, obtained as the sum of the signed volumes of a tetrahedral decomposition of each element. In a recent work, Lobos and colleagues proposed a novel extension of the scaled Jacobian that applies to pyramids and prisms~\cite{lobos2021measuring}. These elements often occur in hybrid meshes because they are used as \emph{topological bridges} between tetrahedra and hexahedra, or arise when collapsing edges from a regular grid. Nevertheless, some hex-dominant meshing algorithms do not offer any control on the topology of the hybrid elements they create (\cref{sec:hexdominant}), and may even produce cells for which a tetrahedralization does not exist~\cite{goerigk2015indecomposable}. To date, we are not aware of any mesh smoothing or untangling algorithm that can operate on general polyhedral meshes containing elements that do not admit a tetrahedralization.\\

Restricting to elements for which a tet decomposition exists, mesh optimization algorithms are based around the ideas expressed in \cite{VartziotisHimpel}. The authors start from the consideration that per element volume is not a good metric, because it is scale-dependent, and proposed an alternative metric -- called \textit{mean volume} -- which is defined on the tetrahedralization of a general polyhedron.
The mean volume metric exhibits some desirable properties. In fact, it is scale-independent and is maximized by regular tetrahedra, hexahedra, octahedra, pyramids and prisms. Consequently, following the gradient of the mean volume allows optimizing hybrid meshes made of these elements~\cite{vartziotis2013improved}. Also, more general elements can be deformed following the gradient, but it remains unclear whether this improves the mesh or not, because of the lack of a canonical reference element. 
Alternatively, one could tetrahedralize each element and smooth the resulting simplicial mesh, optimizing the shape of each tet. Schemes to convert pyramids, hexahedra and prisms in a globally consistent manner are reported in~\cite{dompierre1999subdivide}, and are implemented in open-source tools like CinoLib~\shortcite{cinolib}. 
Also in this case, it is not clear to what extent optimizing the tetrahedralization of a hybrid mesh improves the original elements. The topic is indeed under-investigated, and with the proliferation of hex-dominant meshing techniques we expect more and more contributions to be released in the near future.\\

A parallel line of works is devoted to the study of shape regularity criteria for general polyhedral elements. Shape regularity plays a central role in FEM analysis, as it allows to define precise error estimates on the solution of a PDE, which depends solely on geometric properties of mesh elements. These criteria are well known for triangles, quads, tetrahedra and hexahedra, but the problem hasn't been taken into consideration for general polygons and polyhedra until recently. As of today, shape regularity for arbitrary polygonal and polyhedral elements is relatively strict: concavities are admitted, but elements must have a bounded number of faces and edges, and be star shaped~\cite{lipnikov2013shape,mu2015shape}. Some numerical schemes for hybrid meshes (e.g., the Virtual Element Method  \cite{beirao2014hitchhiker}) have empirically shown to be resilient to meshes containing even large amounts of elements that spectacularly violate these criteria, suggesting that more permissive shape regularity criteria could be devised. The problem is still open, and various research groups are working on it. Note that shape regularity criteria are not quality metrics, and involve the assessment of geometric properties that are computationally expensive to evaluate (e.g., being star-shaped) hence they can hardly plugged into mesh optimization schemes. Recent studies are trying to discover new connections between basic geometric properties of mesh elements and the performances of PDE solvers (e.g., approximation error, the condition number of the stiffness matrix), with the ultimate goal to isolate geometric quantities that can be used to drive mesh generation and optimization algorithms in a \emph{PDE-aware} manner~\cite{attene2019benchmark}. 

%% file: 08_visualization.tex
\section{Visualization}
\label{sec:visualization}
\todo{\NP{(Nico)}}
Researchers involved in mesh generation made extensive use of tools to explore the structure of a volumetric mesh interactively. There are several reasons for the volumetric investigation of hex-meshes:
\begin{description}
	\item[Evaluation]
	Finite Element analysis often requires the visualization of the result of an experiment, such as the stress distribution or heat propagation.
	
	\item[Visualization]
	An interactive visualizer of volumetric meshes is a powerful tool for the visual inspection of a model. Hence, researchers might use it to assess meshing algorithms' performance in terms of element quality or global element arrangement. Interactivity becomes more challenging for high-resolution datasets or meshes with intricate 3D structures.
	
	\item[Assessment]
	Secondarily, automated techniques might perform numerical measurements and plotting histograms to assess the quality of a hex model or some 3D field embedded in the elements.
	
	\item[Presentation]
	Finally, if the produced images are of high quality, they constitute a valuable resource for dissemination, e.g., scientific publications. 
\end{description}

Visualizing a volumetric dataset in an effective and user-friendly way is a challenging task. The main challenge is to render the internal elements efficiently, even if the external shell occludes them. Volumetric rendering~\cite{RodriguezGGMMPS14} overcomes this limitation by integrating the inner field along a particular view direction. However, besides their practical use in the exploration of biomedical data or FEM, they cannot be effectively used to visualize the mesh's connectivity and the quality of its elements.

An alternative trend enables an interactive user-guided visual exploration of hex-meshes directly via cell filtering or using transparency to reveal mesh internal structure. This approach allows for a detailed analysis of the mesh structure, isolating weak points, or degenerate elements. Some basic library offers a set of essential tools to filter and visualize the elements selectively~\cite{cinolib}. Other advanced geometry processing~\cite{graphite} or mesh generation~\cite{geuzaine2009gmsh,zheng1995feview} tools offer necessary instruments to examine the internal cells, such as sweeping planes. Similarly, Paraview~\cite{ayachit2015paraview} and~\cite{ansys} provide some methods for the visualization and exploration of volumetric datasets, including additional tools to plot and elaborate statistics on volumetric fields embedded in the elements. Besides their use in most application contexts, none of the tools mentioned above is tailored to hexahedral meshes. A different generation of software like Hexalab~\cite{bracci2019hexalab} or the method presented in~\cite{xu2018hexahedral} have been designed explicitly for hexahedral mesh visualization.

\begin{figure}[t]
	\centering
	\begin{tabular}{@{}c@{}c@{}c@{}c@{}}
		\includegraphics[height=0.3\linewidth]{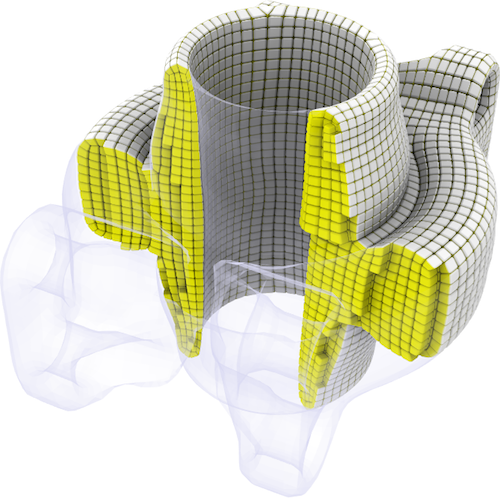}&
		\includegraphics[height=0.3\linewidth]{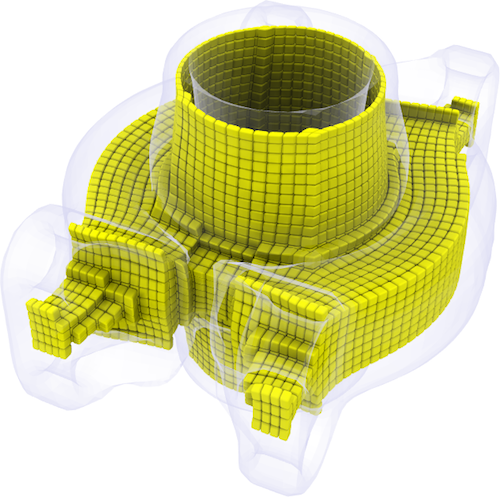}&
		\includegraphics[height=0.3\linewidth]{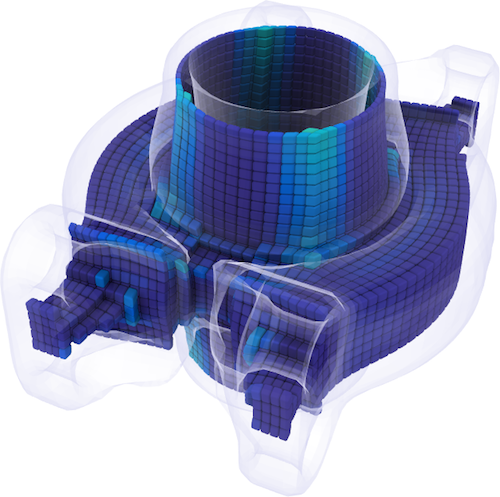}&
		\includegraphics[height=0.3\linewidth]{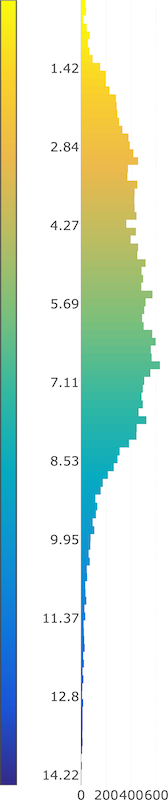}\\
		(a)&(b)&(c)&\\
	\end{tabular}
	\caption{The interactive visualization tools offered by Hexalab: Internal exploration using a sweeping plane (a) or the peeling tool (b);  Coloring elements by their quality and the resulting histogram (c).}
	\label{fig:hexalab}
\end{figure}

Hexalab offers a set of interactively controlled tools to reveal the internal structure of the mesh. The user can either use an interactively controlled sweeping plane (see \cref{fig:hexalab}~a) or peel the object surface layer-by-layer from the outside (see \cref{fig:hexalab}~b). Even if removed from the visualization, the outer surface can be visualized with some transparency effect. Hexalab also provides a high-quality rendering, including non-photorealistic effects on the GPU, like ambient occlusion, to enhance the internal structure and the arrangement of the elements and better communicate the shape of the cells. It also implements all the quality measures in~\cite{Gao2017Evaluation}, offering automated techniques to numerical assess the quality of a mesh and plot histograms for the inspected model (see \cref{fig:hexalab}~c). Hexalab is also an easily accessible portal online repository of hex-meshes, including a variety of results from various state-of-the-art techniques. This characteristic makes this tool an excellent platform to compare the performance of the different meshing techniques.

\begin{figure}[b]
	\centering
	\begin{tabular}{ccc}
		\includegraphics[height=0.38\linewidth]{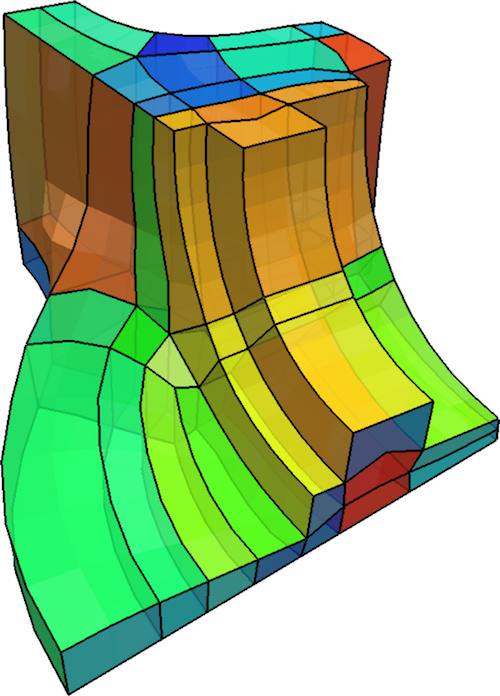}&
		\includegraphics[height=0.38\linewidth]{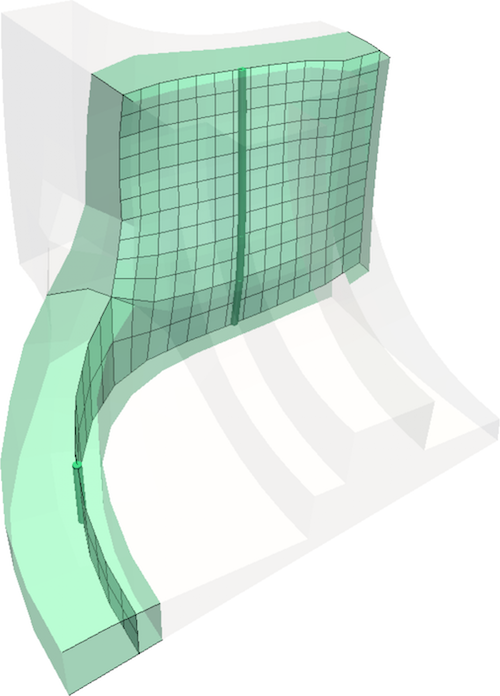}&
		\includegraphics[height=0.38\linewidth]{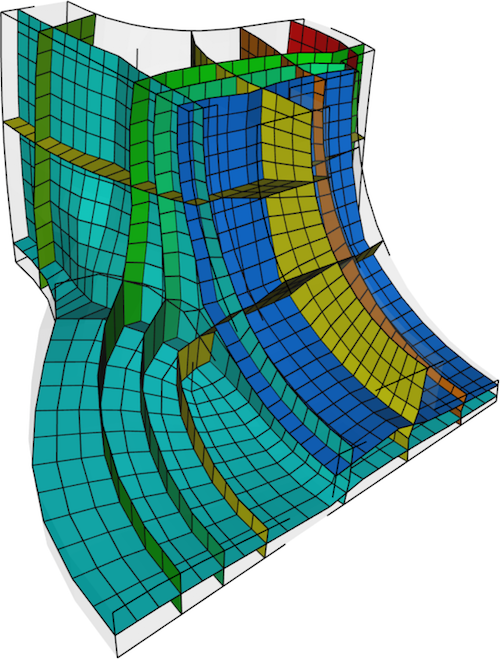}\\
		(a)&(b)&(c)
	\end{tabular}
	\caption{The pre-processing pipeline of~\cite{xu2018hexahedral}: (a) Extracting the base complex; (b) One Dual sheet layer; (c) The final visualization.}
	\label{fig:base_complex}
\end{figure}

While Hexalab provides essential tools to visualize the global connectivity, such as the location and valence of singularities, the approach of~\cite{xu2018hexahedral} provides a sophisticated method for the exploration and the visualization of the volumetric structure. This method exploits the connectivity between base complexes composing the hex-mesh. The structure and connectivity of base complexes provide an excellent overview of the hexahedral elements' underlying flow. The base complexes are groups of cells delimited by the sheets emanating from singularities (\cref{fig:base_complex}~a).  Adjacent base complexes following the same frame field directions compose dual-sheets (\cref{fig:base_complex}~b). A global optimization process uses the connectivity between the dual-sheets to select the most significant ones and finally provide an efficient instrument to navigate the high-level overview of the underlying structure (\cref{fig:base_complex}~c). The recent approach proposed by Neuhauser and colleagues \cite{Neuser2021} makes use of GPU shader functionality to generate an advanced volumetric rendering that focuses on a subset of elements. This strategy allows, for example, to visualize poorly shaped elements or the elements surrounding a singularity and, at the same time, gradually blend the visualization with the surrounding structure.\\

\minrev{Most of the current visualization tools specialize in exploring the mesh connectivity and assessing elements' quality and arrangement. However, most of the tools presented in this section show their limits when employed in an actual industrial application.} 

\minrev{The volumetric visualization systems discussed in this section do not scale directly to massive datasets composed of millions of elements. As the opposite, the industrial systems (such as Paraview ~\cite{ayachit2015paraview} and Ansys ~\cite{ansys} ) allow for the visualization of meshes composed of millions of elements. However, the rendering quality provided by such commercial packages is usually not as informative as the one provided by the recent tools proposed in academia. Because meshes composed of millions of elements are the standard in several FEM contexts, open-source tools like Hexalab must bridge this gap to have a chance of significant impact in industry. 
}

\minrev{When a dataset becomes massive, current exploration tools based on transparency, ambient occlusion, or slicing planes might become inadequate to ensure full visual access to the volume. We believe that future visualization tools could overcome these limitations by exploiting current VR developments and possibly interactive gesture tracking.}

\minrev{Finally, most application scenarios require the visualization of complex fields defined over each volume element (such as stress or tensor fields), exploring their variation, and doing some statistics. While Paraview ~\cite{ayachit2015paraview}  offers already some advanced tools to this scope, renderings are still not adequate to the state-of-the-art techniques. The modern GPU architecture that supports real-time ray-tracing can trace a new path for the advanced volumetric rendering of hexahedral meshes with complex embedded fields.}

%% file: 09_outro.tex
\section{Current trends and future perspectives}
\label{sec:future}
\minrev{In this section, we report on the status of hex-meshing as a whole and its future perspectives. Specifically, in \cref{sec:theoret_challenge} we discuss open theoretical problems that are relevant for hex-mesh generation algorithms. In \cref{sec:algo_challenge} we report on algorithmic issues of existing approaches. Unlike the previous section, in this case, the theoretical formulation of the problem is clear, but current solutions are unsatisfactory, mainly for the intrinsic complexity of the problem tackled. In \cref{sec:pratical_challenge} we indicate various activities that, even though they do not directly advance the state of the art, may foster new research and facilitate the development of better techniques.}

\subsection{Theoretical challenges} 
\label{sec:theoret_challenge}

\paragraph{Characterization and synthesis of hex-meshable  frame fields} \minrev{The fully automatic applicability of frame field based techniques (\cref{sec:frame_fields}) is currently limited to rather simple shapes. User intervention to cure imperfections in the guiding field are required to handle general shapes. As reported in the dedicated section, the space of frame fields is topologically larger than the space of hexahedral meshes. This raises two fundamental theoretical questions: 
	\begin{enumerate}
		\item Given a frame field, is it possible to algorithmically verify whether a valid hexahedral mesh of identical topology exists?
		\item Is it possible to restrict hexmeshing algorithms to operate in the space of fields that admit a hexahedralization?
	\end{enumerate}
	Unfortunately, both questions are still unanswered and demand further research to understand these geometrical entities at a deeper level. For the characterization of hex-meshable fields, in~\cite{liu2018singularity} the authors enumerate all local conditions for hex-meshes having singular edges with valence 3 and 5. Also they report a global condition which is a discrete version of the Poincar\'{e}-Hopf index formula. While their characterization can be easily extended to a broader set of singular edges, as reported by the authors, the global condition is necessary but not sufficient, meaning that there may still exist fields that obey all these criteria but do not admit a valid mesh (e.g., due to limit cycles~\cite{viertel2016analysis}, or other global inconsistencies~\cite{sokolov2015fixing}).  The second question cannot be answered either due to our inability to exhaustively characterize hex-meshable fields. To sidestep this issue, many authors start by computing an unconstrained frame field, then cure it with manual fixing~\cite{liu2018singularity} or adopting local heuristics~\cite{Li:2012,jiang2014frame,reberol2019multiple,viertel2016analysis}, and then use the corrected singular graph to bootstrap methods such as~\cite{liu2018singularity,Corman:2019:SMF} which -- being able to produce a field that conforms to a given singular structure -- are guaranteed to produce a hex-meshable result. Not only this is a cumbersome pipeline, but considering the inability to precisely state whether a field is hex-meshable or not, failure is always possible, even for semi-automatic methods that put the user in the loop. }  

\begin{figure}
	\centering
	\includegraphics[width=\columnwidth]{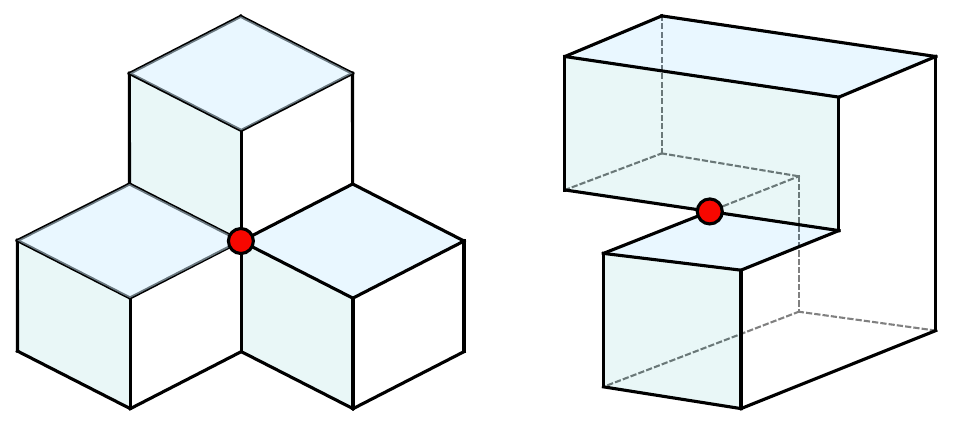}
	\caption{\minrev{Two polycubes having non 3-connected graphs. The red nodes at the left is 6-connected; the one at the right is 4-connected. These orthogonal polyhedra are not included in the graph characterizations provided in~\cite{eppstein2010steinitz} and~\cite{pc_shapespace}.}}
	\label{fig:non_3_connected}
\end{figure}

\paragraph{Graph characterization of polycubes} \minrev{Many existing polycube methods (\cref{sec:polycubes}) exploit the graph characterization described in~\cite{eppstein2010steinitz} to ensure that the topological structures they define correspond to orthogonal polyhedra, hence to polycubes. While this process could potentially permit to find the most suitable shape without ever leaving the feasible space of valid polycubes, current implementations do not fully exploit this potential and are limited to heuristically restricting the number of incoming polycube edges at a given vertex, or the number of edges at a given polycube face. More stringent (and robust) checks cannot be currently implemented because the graph characterization proposed by Eppstein and Mumford is limiting in many ways. Firstly, they restrict to 3-connected graphs, meaning that each vertex has exactly three incident axis-aligned edges according to their definition of an orthogonal polyhedron. As a result, objects such as the ones shown in \cref{fig:non_3_connected} are not considered valid polycubes. Secondly, the authors only characterized the graph of three subclasses of orthogonal polyhedra, thus further (and severely) restricting the space of legal polycubes. Among the various restrictions, the most limiting one regards the genus of the polycube, which must be zero. In a recent paper~\cite{pc_shapespace} the authors were able to extend the graph characterization in~\cite{eppstein2010steinitz}, also comprising polycubes with higher genus, but the limitation to 3-connected graphs still holds. Extending the polycube characterization to 4- and 6-connected graphs has high practical importance, especially for meshing CAD-like shapes, where more than three sharp features often meet at the same point and the inability to catch this layout in the polycube would inevitably result in a lack of feature preservation and -- most likely -- in severe and unnecessary geometric distortion. To date, the generation of a comprehensive graph characterization of polycubes has been elusive, and this remains an open problem not only for practitioners in mesh generation but also for the graph community.}\\	

\subsection{Algorithmic challenges} 
\label{sec:algo_challenge}

\paragraph{Volume mappings} \minrev{Many indirect methods that operate on a supporting tetrahedral mesh rely on a mapping between the input object and a parametric space embedding the mesh connectivity. A common desideratum to promote element validity is that this mapping is (globally or locally) injective. The generation of injective volumetric maps is a broad topic that finds application in many fields. While there are reliable approaches for the computation of such maps~\cite{Campen:2016:Bijective}, they are not versatile enough for hex-mesh generation. In recent years some more flexible methods with improved success rates have been introduced~\cite{du2020lifting,garanzha2021foldover} but none of them can actually guarantee an injective result and failure cases are easily encountered in the hex-meshing context.
	Besides injectivity, it is also important that the map has low geometric distortion. Elements should preserve their good shape through the map so that a regular grid in parametric space translates into a well-shaped uniform hex-mesh of the target object. To this end, current relatively robust methods such as~\cite{du2020lifting} fall short, because they are focused more on the validity of the solution than on the distortion of the map, and may therefore produce valid meshes that are unusable in practice. 
	Recent literature has shown that adaptively sampling the parametric space can be used to counterbalance map distortion, even in extreme cases (\cref{fig:polycube_hand}). Nevertheless, devising new algorithms that provide guarantees of robustness and minimize geometric distortion (possibly editing mesh connectivity)
	will be highly beneficial for many hex-meshing algorithms.\\
	Volume mappings become even more complex when integer constraints are added to the formulation, leading to a mixed-integer problem. These problems typically arise in frame-field based methods (\cref{sec:frame_fields}) due to the presence of integer transition functions and integer alignment conditions, but may also arise in certain polycube methods (\cref{sec:polycubes}), to ensure that input features and polycube edges map to integer isolines in parametric space. As the resulting mixed-integer problems are very hard to solve to the optimum~\cite{bommes2010practical}, heuristics are commonplace, e.g. based on rounding \cite{Nieser2014,jiang2014frame,Li:2012} or reduction to linearly constrained integer programs \cite{Brueckler:2021,cherchi2016polycube}. Yielding a map that is not only of low distortion on average, but strictly locally injective can be even more challenging in this integer-constrained setting.}

\paragraph{Feature transfer} \minrev{By their very definition, direct methods are guaranteed to conform to the input geometry and all its features. Conversely, indirect methods can only produce an approximation of the target shape. Many indirect methods insert all (or a part) of the input features after the mesh generation stage, resolving a \emph{feature transfer} problem. This happens for all grid-based methods, but may also happen for methods based on domain decomposition or polycube methods (e.g., to restore features that do not map to polycube edges). Feature transfer is primarily a topological problem because feature lines must be assigned to chains of edges in the hex-mesh, ensuring that no spurious overlaps or intersections are introduced. Known methods operate heuristically, inserting one feature curve at a time along some pre-computed sequence (e.g., sorting features by their length). While the first insertion is free to occupy any mesh edge, the subsequent ones are constrained by previous insertions, clearly designing a combinatorial problem with exponential complexity. Recent literature has shown that a greedy processing sequence may lead to catastrophic results (see, e.g. Figs. 1,2,11 in~\cite{born2021layout}). Besides the intrinsic complexity of the general problem, the sparse hex-mesh connectivity and impossibility to apply local refinement to increase the valence of a vertex (e.g. to accommodate more incoming arcs) makes this problem much more challenging on structured meshes than on unstructured ones. Current methods such as~\cite{gao2019feature} are limited to simply discarding a feature line if a non-intersecting chain of edges can be computed, and insert it otherwise, regardless of its geometric deviation from the target curve. The use of more sophisticated heuristics such as~\cite{born2021layout} may significantly increase the number of features successfully transferred and also help reduce the geometric distortion due to a wiser choice of the feature edges. Alternative methods tailored to operate on the connectivity of hexahedral meshes may also be developed, and coupled with (as local as possible) hex-mesh refinement techniques to resolve intricate configurations. Considering the limitations of the hex connectivity application-aware de-featuring may also play an important role~\cite{buffa2020analysis}.}

\paragraph{Volumetric modeling}\minrev{The prominent idea at the basis of Isogeometric Analysis (IGA) is to establish a unified geometric representation for both modeling and simulation, thus avoiding the need to iteratively convert from one representation to the other throughout the product development cycle~\cite{hughes2005isogeometric}. Differently from Computer-Aided Design (CAD), which is historically concerned with boundary representations (B-Rep), numerical simulation often necessitates an explicit volumetric representation of the product (V-Rep), which typically comes in the form of a finite tetrahedral or hexahedral mesh. Therefore, the fulfillment of the IGA principles passes through the adoption -- also for the design part -- of V-Reps. Not only IGA, but also modern manufacturing techniques call for this transition: composite materials and internal microstructures do not scale well with B-Reps, and would benefit from an explicit volumetric representation.
	Volumetric modeling that operates with tensor products~\cite{antolin2019isogeometric,antolin2019optimizing,massarwi2016b} has close analogies with the definition of structured hexahedral meshes that endow a coarse block decomposition. To this end, advances in this field will go hand in hand with advances in user interactive tools for the generation of semi-structured hexahedral meshes. A few proposals already exist in the literature, but the topic is quite new and under-investigated.}

\paragraph{User interaction} \minrev{Since no existing hex-meshing method combines robustness, quality, and generality in a fully satisfactory way, manual or semi-manual hexahedral mesh generation is still a prominent approach in industry~\cite{lu2017evaluation}. Professional software such as~\cite{cubit,ansys} and many others are based on interactive pipelines where the user provides a high-level understanding of the object, and is required to instruct the program on how the shape can be split into simpler parts. If and when parts become sufficiently simple, direct methods such as sweeping and advancing front are launched to complete the discretization. Parts that are not sufficiently simple will remain empty, and the user is required to modify the current partitioning or split the non-meshed elements into simpler sub-components. This process is extremely tedious, and requires the user to \say{understand} (and overcome) the limitations of direct meshing approaches, in order to provide a decomposition that nicely combines the necessity to keep the number of parts low and at the same time simple enough to be processed separately in an automatic fashion~\cite{coreformcourse}. Since these tools follow a divide-and-conquer approach, direct hexmeshing techniques are preferred to indirect ones, because they ensure that the meshes of all sub components will be conforming. In recent years, academic literature has started to explore the possibility to couple user guidance with indirect approaches that operate on a supporting tetrahedral mesh~\cite{interactive_polycubes,Kenshi:2019,yu2020hexgen,yu2021hexdom}. These methods are not based on the typical divide-and-conquer paradigm, and their ability to scale on complex shapes it yet to be demonstrated.\\
	The usefulness of interactive approaches is twofold: from the perspective of mesh users, they allow to hex-mesh objects that would not be possible to produce automatically. From the perspective of practitioners mesh generation, these interactive pipelines often permit to spot the weak parts of the pipeline, isolate corner cases, and interactively explore alternative solutions. To this end, these tools may be highly important for the development of better (i.e., more robust) fully automatic methods.}

\paragraph{Hex-mesh booleans}\minrev{In medicine, the simulation of human organs often relies on templated hexahedral or hex-dominant meshes that well capture biological structures such as separation tissues or the alignment of muscular fibers, effectively reproducing their activation~\cite{rohan2017finite,schonning2009hexahedral,buchaillard2009biomechanical,gerard20063d,takhounts2008investigation}. 
	Considering the important information encoded in the connectivity of these meshes, when simulating complex body dynamics that involve multiple organs it becomes important to create composite simulation domains that preserve as much as possible the connectivity of each original mesh. Blending multiple meshes into a single one is a widely studied problem in the literature, especially for the case of unstructured meshes composed of triangles or tetrahedra~\cite{cherchi2020fast,zhou2016mesh,diazzi2021convex,hu2018tetrahedral}. For structured meshes made of quads or hexahedra, the problem is more complex because the necessary changes of the local connectivity have a global footprint. Recently, in~\cite{NHESCP19} the authors introduced a method to blend quadrilateral meshes with minimal topological impact. Extending this idea to volumetric hex-meshes remains an appealing avenue for future research with a significative potential impact for bio-medical applications.}

\paragraph{Scalability} With the recent advancement of additive manufacturing and topology optimization strategies, mechanical shapes are rapidly growing in complexity. Consequently, hex-meshing methods need to comply with this trend by providing the ability to process large datasets at a reasonable computational overhead. Scalability has not been a central point for most of the proposed methods, but it will increase in importance in the years to come.\\

\subsection{Practical challenges} 
\label{sec:pratical_challenge}

\paragraph{PDE-aware volume meshing}\minrev{As discussed in \cref{sec:outputreq}, a deeper fundamental understanding of the connection between a hex-mesh and the final application is required. Most of the hex-meshing methods strive to ensure every element to have a positive Jacobian determinant. While this is already hardly achievable reliably with most of the proposed methods, even a positive Jacobian determinant throughout the entire mesh only avoids the presence of degenerate elements. Still, it does not ensure the mesh fits with the target application. The precise connection between a hex-mesh and its final application is usually elusive. In \cref{sec:quality}, we presented several quality measures for individual hexes. Still, even for Finite Element Analysis, it is not clear how those metrics impact the accuracy of the simulation in detail. Other applications might prefer the alignment of the elements to a particular vector field over their individual quality. Recent literature has started to investigate the correlation between geometric quality and the accuracy of numerical solvers at a deeper level. A whole line of research is devoted to the evaluation of the Virtual Element Methods for polygonal and polyhedral meshes~\cite{attene2019benchmark,sorgente2021vem,sorgente2021role,cabiddu2021pemesh}. More related to the topic of this survey is the study published in~\cite{Gao2017Evaluation}, who conducted a statistical correlation analysis between hexahedral meshes obtained with various techniques, and the resolution of a few representative PDEs. While it remains difficult to design mesh generation algorithms that can address geometric quality criteria at the mesh generation stage, a few exceptions exist. For example, the VoroCrust algorithm~\cite{abdelkader2020vorocrust} is designed to intrinsically satisfy the orthogonality criterion required by CFD solvers, obtained with a wise positioning of the Voronoi seeds that fully avoids the necessity to cut (or \emph{clip}) Voronoi cells. It would be interesting to investigate similar ideas to obtain a tighter coupling of mesh generation and its downstream applications.}

\paragraph{Tets vs Hexa}\minrev{Meshes made of hexahedral elements were traditionally considered superior to tetrahedral meshes, both in terms of performance and accuracy. Tuchinsky and Clark observed that since a hexahedral mesh can cover the same volume of a tetrahedral mesh with roughly one-quarter of the elements, there is a 75\% saving in terms of computational cost~\cite{tuchinsky1997hextet}. This estimate is based on the assumption that \say{analysis setup and pre-processing requires the same time for hex- and tet-based work}, which does not reflect the current state of mesh generation because creating and processing tetrahedral meshes is significantly easier and more robust than creating hexahedral ones~\cite{hu2018tetrahedral,hu2020fast,diazzi2021convex}. Regarding accuracy, it seems to be well understood and established that linear tetrahedra are to be avoided because they introduce artificial stiffness in the problem (i.e., they \say{lock})~\cite{wang2004back}, whereas linear hexahedral elements do not introduce such artifact. Typically locking depends on the number of degrees of freedom~\cite{francu2021locking} and disappears when higher order bases are used~\cite{wang2004back}. This makes hexahedral meshes particularly suited for problems where linear elements are used, such as in the interactive simulation of hyper-elastic and plastic phenomena (e.g. in surgical simulation~\cite{gao2021improving}) and in fast transient dynamic phenomena that employ explicit time integration (e.g. crash and impact simulation)~\cite{gravouil2009explicit} because higher-order basis functions would necessarily demand a reduction of the time step to achieve numerical stability, according to the Courant$-$Friedrichs$-$Lewy condition~\cite{weber2021stability,courant1967partial}. In recent years some scientists have questioned the superiority of hexahedral elements over tetrahedra and advocate the use of tetmeshes with quadratic basis functions as general purpose simulation domains~\cite{schneider2019large}. The topic is somewhat orthogonal to this survey, which focuses only on hex-mesh generation aspects. Whether it is for their (uncertain) superiority or because of the presence of highly trusted legacy code that runs only on hexahedral grids, the interest for hexahedral meshes is still high both in industry and in academia. This is witnessed by the growing number of scientific articles published in recent years \cite{beaufort2021hex}, by the central role that hexahedral grids occupy in industrial and commercial software, and ultimately by the interest that the industry manifests for each advancement in hex-mesh generation.}

\paragraph{File formats}\minrev{Many algorithms for hex- and hex-dominant mesh generation necessitate the ability to process general polyhedral meshes, either at the intermediate steps of the pipeline~\cite{Marechal2009,livesu2021optimal,gao2019feature} or directly in the output mesh~\cite{LoopyCuts2020,hybridHexa}. In general, these methods put no constraints on the topology of each cell, which can therefore contain any amount of vertices, edges and faces. While data structures capable of handling these entities exist (\cref{sec:resources}), we are not aware of any widely accepted file format that allows to save and load output hexahedral dominant meshes or intermediate steps of hex-meshing pipelines. To our knowledge, popular tools such as VTK~\shortcite{VTKformats} only support file IO of \emph{canonical} finite elements, such as tetrahedra, hexahedra, pyramids and wedges, while methods that produce meshes with more complex elements all rely on ad-hoc formats that were released alongside the algorithms themselves~\cite{LoopyCuts2020,hybridHexa}, thus limiting the possibility to exchange material and ultimately triggering a proliferation of alternative proposals. Considering the growing importance of hex-dominant meshes, it would be important to define a file format for general polyhedral meshes, so that groups working in the field can store and release their data in a way that is intelligible by the other groups, and that can be easily supported by third party software such as~\cite{bracci2019hexalab} (e.g., for visualization, comparison, and analysis).}

\paragraph{Datasets, Benchmarks}\minrev{In recent years the computer graphics community has released multiple databases that have been extremely useful for practitioners in the field, raising the bar for new algorithms in terms of scalability and ability to handle a variety of inputs with different complexity, from easy ones to highly challenging. To make a practical example, the Thingi10K~\cite{zhou2016thingi10k} dataset has quickly become a popular means to empirically validate the robustness of surface mesh generation and processing algorithms~\cite{PNACT21,hu2018tetrahedral} and some of its models are so pathological that being able to process them is an achievement by itself, with authors reporting both running times and memory consumption (see e.g. Fig. 17 in~\cite{hu2020fast} and Fig.1 in~\cite{cherchi2020fast}). To this end, new methods for hexahedral and hex-dominant meshing can greatly benefit from the release of similar databases. The Hexalab project~\cite{bracci2019hexalab} collects output data from the most prominent mesh generation methods in the literature, but it is not meant to be a validation database for novel methods.
A few contributions in this direction have been proposed very recently: \cite{reberol2019multiple,MAMBO} propose input CAD models, while~\cite{beaufort2021hex} offers input tetrahedral meshes with tagged feature entities. Specifically, hard constraints on the preservation of feature curves and (boundary and interior) surfaces can be very challenging for meshing algorithms.}

\paragraph{Beyond PDEs: novel applications}\minrev{The numerical resolution of Partial Differential Equations (PDEs) is by far the most prominent application of volumetric meshes in general, and of hexahedral meshing in particular. Nevertheless, in recent years both meshes of this kind and techniques that were originally developed in the field have been used in alternative applications, such as topology optimization and advanced manufacturing~\cite{Arora:michell:scf:2019,Wu_2019,cyril2021synthesis}. To this end, current themes in automatic hex-mesh generation are beneficial not only for the numerical resolution of PDEs, but may also reach a broader audience.}

\minrev{\section{Available resources} 
\label{sec:resources}
Besides various professional and semi-professional tools such as VTK~\cite{schroeder1998visualization} and its front end ParaView~\cite{ayachit2015paraview}, Cubit~\cite{cubit}, MeshGems \cite{meshgems}, Gmsh \cite{geuzaine2009gmsh}, CoreForm \cite{coreform}, CGAL \cite{fabri2009cgal} and many others, over the years academics have released both data and a variety of open-source tools to aid not only their research, but also the activities of other practitioners in the field. This section summarizes the most prominent available resources for hexahedral and hex-dominant meshing. Note that the list of authors releasing their data, code and toolkits is in constant evolution.
\paragraph{Datasets} In~\cref{tab:datasets} we list datasets released by authors of the methods surveyed in \cref{sec:methods}. This includes in particular sets of example hexahedral meshes generated by these various methods, but also hex-dominant meshes  (e.g.,~\cite{hybridHexa}) as well as challenging input models (e.g.,~\cite{beaufort2021hex,MAMBO,reberol2019multiple}).
 The HexaLab project~\cite{bracci2019hexalab} is a unified portal to visualize hexahedral meshes directly in a web browser as well as to download them. It collects meshes produced with the most recent techniques in the field, with a focus on pure hexahedral meshes, and includes most of the output data listed in \cref{tab:datasets}.
\paragraph{Algorithms} In recent years, more and more authors are releasing their source code, either through activities for the reproducibility of scientific experiments, such as~\cite{bonneel2020code,repstamp}, or simply by publishing their code on Github or similar portals. In \cref{tab:algorithms} we report on all the implementation of algorithms surveyed in this article, both in the form of source code or pre-compiled binaries.
\paragraph{Toolkits} While there exist countless open source libraries for the processing of surface (e.g. triangular) meshes, the number of tools that offer data structures for volume meshes is scarce. Besides, most of these tools are dedicated to tetrahedral meshes only. They do not support alternative cells, such as hexahedra or general polyhedral elements that may arise at the intermediate steps of the meshing pipeline~\cite{Marechal2009,PLCGS21,LoopyCuts2020}, or in hex-dominant methods. In \cref{tab:toolkits} we report on the most prominent existing software tools for volume mesh processing, summarizing their main features w.r.t. the scope of this survey.}

\begin{table}[]
	\minrev{
	\caption{\minrev{List of input/output datasets for the hex and hex-dominant methods surveyed in this article available at the time of writing. Meshes that are included also in the HexaLab database~\cite{bracci2019hexalab} are flagged accordingly.}}
	\label{tab:datasets}
	\resizebox{\columnwidth}{!}{%
		\begin{tabular}{lcccl}
		\hline
			\multicolumn{1}{c}{\textbf{Method}} &
			\textbf{\begin{tabular}[c]{@{}c@{}}Data\\Available\end{tabular}} &
			\textbf{\begin{tabular}[c]{@{}c@{}}File\\Formats\end{tabular}} &
			\textbf{\begin{tabular}[c]{@{}c@{}}On\\Hexalab\end{tabular}} &
			\textbf{URL} \\ \hline
			\cite{gregson_polycube_2011} &
			output hex-meshes &
			.MESH &
			yes &
			\href{https://www.cs.ubc.ca/labs/imager/tr/2011/HexMeshingPolycubeDeformation/HexMeshing_files/HexMeshSGP2011_sample_data.zip}{zip} \\ \hline
			\cite{Li:2012} &
			output hex-meshes &
			.VTK &
			yes &
			\href{https://app.box.com/s/1yng7m1eky0tgo3q9b39msrrxmiwoa7a}{link} \\ \hline
			\cite{polycut_livesu} &
			output hex-meshes &
			\begin{tabular}[c]{@{}c@{}}.VTU \\ .MESH\end{tabular} &
			yes &
			\href{http://www.cs.ubc.ca/labs/imager/tr/2013/polycut/}{webpage} \\ \hline
			\cite{l1pc2014} &
			\begin{tabular}[c]{@{}c@{}}input tet-meshes\\ polycube maps\\ output hex-meshes\end{tabular} &
			.VTK &
			yes &
			\href{http://www.cad.zju.edu.cn/home/hj/14/l1-poly/l1-poly-dat.zip}{zip} \\ \hline
			\cite{Livesu:2015} &
			\begin{tabular}[c]{@{}c@{}}input hex-mesesh\\ output hex-meshes\end{tabular} &
			.MESH &
			yes &
			\href{http://www.cs.ubc.ca/labs/imager/tr/2015/untangler/downloads/untangler_res.zip}{zip} \\ \hline
			\cite{closedform_pc2016} &
			\begin{tabular}[c]{@{}c@{}}input tet-meshes\\ output hex-mesh\end{tabular} &
			.VTK &
			yes &
			\href{http://www.cad.zju.edu.cn/home/hj/16/closed-form-polycube/closed-form-polycube.7z}{zip} \\ \hline
			\cite{fu2016efficient} &
			\begin{tabular}[c]{@{}c@{}}input tet-meshes\\ polycube maps\\ output hex-meshes\end{tabular} &
			.VTK &
			yes &
			\begin{tabular}[c]{@{}l@{}}\href{http://pan.baidu.com/s/1o7YUFYq}{link}\\ \href{http://pan.baidu.com/s/1bp5fiRP}{link}\end{tabular} \\ \hline
			\cite{cherchi2016polycube} &
			\begin{tabular}[c]{@{}c@{}}input polycubes (hex)\\ output polycubes (hex)\\ intput hex-meshes\\ output hex-meshes\end{tabular} &
			.MESH &
			yes &
			\href{https://www.gianmarcocherchi.com/dataset/pc_simpl_dataset.zip}{zip} \\ \hline
			\cite{LMPS16} &
			\begin{tabular}[c]{@{}c@{}}input surface meshes\\ input curve-skeletons\\ output hex-meshes\end{tabular} &
			\begin{tabular}[c]{@{}c@{}}.OBJ\\ .SKEL\\ .MESH\end{tabular} &
			yes &
			\href{http://pers.ge.imati.cnr.it/livesu/papers/LMPS16/LMPS16.zip}{zip} \\ \hline
			\cite{Gao16} &
			\begin{tabular}[c]{@{}c@{}}input surface meshes\\ output hex-meshes\end{tabular} &
			\begin{tabular}[c]{@{}c@{}}.OFF\\ .MESH\end{tabular} &
			yes &
			\href{https://gaoxifeng.github.io/papers/2015/TVCG_data.zip}{zip} \\ \hline
			\cite{wu2017global} &
			output hex-mesh &
			.MESH &
			yes &
			-- \\ \hline
			\cite{LAPS17} &
			output hex-meshes &
			.MESH &
			yes &
			\href{http://pers.ge.imati.cnr.it/livesu/papers/LAPS17/LAPS17.zip}{zip} \\ \hline
			\cite{Shang2017} &
			output hex-meshes &
			.VTK &
			yes &
			\href{https://ndownloader.figstatic.com/articles/5020001/versions/1}{link} \\ \hline
			\cite{gao2017simplification} &
			\begin{tabular}[c]{@{}c@{}}input hex-meshes\\ output hex-meshes\end{tabular} &
			.VTK &
			no &
			\href{https://cims.nyu.edu/gcl/papers/Robust-Hex-2017.zip}{zip}\\ \hline
			\cite{hybridHexa} &
			output hex-dominant meshes &
			.HYBRID &
			no &
			\href{https://cims.nyu.edu/gcl/papers/Robust-Meshes-2017-data.zip}{zip}\\ \hline
			\cite{wu2018fuzzy} &
			output hex-meshes &
			.MESH &
			yes &
			-- \\ \hline
			\cite{cherchi2019selective} &
			output hex-meshes &
			.MESH &
			yes &
			\href{https://www.gianmarcocherchi.com/dataset/sel_padding_dataset.zip}{zip} \\ \hline
			\cite{Corman:2019:SMF} &
			output hex-meshes &
			.VTK &
			yes &
			\href{http://www.cs.cmu.edu/~kmcrane/Projects/SymmetricMovingFrames/data.zip}{zip} \\ \hline
			\cite{Kenshi:2019} &
			output hex-meshes &
			.VTK &
			yes &
			\href{http://research.nii.ac.jp/~takayama/dual-sheet-meshing/dual-sheet-meshing.zip}{zip} \\ \hline
			\cite{gao2019feature} &
			\begin{tabular}[c]{@{}c@{}}input surface mesh\\ input features\\ output hex-meshes\end{tabular} &
			\begin{tabular}[c]{@{}c@{}}.OBJ\\ .FGRAPH\\ .MESH\\ .VTK\end{tabular} &
			yes &
			\href{https://cims.nyu.edu/gcl/papers/2019-OctreeMeshing.zip}{zip} \\ \hline
			\cite{reberol2019multiple} &
			input CAD models &
			.GEO &
			no &
			\href{https://mxncr.github.io/data/ff_correction_models.zip}{zip} \\ \hline
			\cite{yang2019computing} &
			output hex-meshes &
			.VTK &
			no & 
			\href{http://rec.ustc.edu.cn/share/e6a45a90-d2f2-11e9-a4e0-a95fc4721cf6}{link}\\ \hline
			\cite{palmer2019algebraic} &
			input tet-mesh &
			.OVM &
			no &
			\href{http://cgg.unibe.ch/media/papers/7/InputFiles.zip}{zip} \\ \hline
			\cite{LoopyCuts2020} &
			\begin{tabular}[c]{@{}c@{}}input surface meshes\\ input rosy fields\\ input features\\ cutting loops\\ refined surface meshes\\ output hex-meshes\end{tabular} &
			\begin{tabular}[c]{@{}c@{}}.OBJ\\ .ROSY\\ .SHARP\\ .TXT\\ .MESH\end{tabular} &
			yes &
			\href{https://github.com/mlivesu/LoopyCuts/tree/master/test_data}{github} \\ \hline
			\cite{guo2020cut} &
			\begin{tabular}[c]{@{}c@{}}input surface meshes\\ input features\\ polycubes (surface)\\ polycubes CE (surface)\\ output hex-meshes\end{tabular} &
			\begin{tabular}[c]{@{}c@{}}.OBJ\\ .FEA\\ .VTK\end{tabular} &
			yes &
			\href{https://drive.google.com/file/d/1g1RwWSkPRhl4HpcstE5hJALF3Zpq61pQ/view}{link} \\ \hline
			\cite{xu2021singularity} &
			output hex-meshes &
			.MESH &
			yes &
			\href{https://github.com/ohehe/HexMeshSimplification}{github} \\ \hline
			\cite{bukenberger2021most} &
			output hex-meshes &
			.MESH &
			yes &
			-- \\ \hline
			\cite{PLCGS21} &
			output conforming grids &
			.MESH &
			yes &
			-- \\ \hline
			\cite{MAMBO} &
			input CAD models &
			.STEP &
			no &
			\href{https://gitlab.com/franck.ledoux/mambo}{gitlab}\\ \hline
			\cite{beaufort2021hex} &
			\begin{tabular}[c]{@{}c@{}}input tet-meshes \\ input features\end{tabular} &
			.VTK &
			no &
			\href{https://cgg.unibe.ch/hexme/}{webpage}\\ \hline
		\end{tabular}%
	}
}
\vspace{-3mm}
\end{table}

\begin{table}[]
	\minrev{
	\caption{\minrev{List of available implementation of hex-mesh processing algorithms.}}
	\label{tab:algorithms}		
	\resizebox{\columnwidth}{!}{%
		\begin{tabular}{lcccl}
			\hline			
			\textbf{Method} &
			\textbf{Type} &
			\textbf{\begin{tabular}[c]{@{}c@{}}Sample\\Input\\Available\end{tabular}} &
			\textbf{License} &
			\textbf{URL} \\ \hline
			\cite{levy2010p} &
			C++ &
			yes &
			-- &
			\href{https://app.box.com/s/vh9mz9eody9xuxj7xtp2f19d8d7gya97}{link} \\ \hline
			\cite{huang2011boundary} &
			executable &
			yes &
			-- &
			\href{http://www.cad.zju.edu.cn/home/hj/11/SH-cross-frame-1607-JiongCHEN.7z}{zip} \\ \hline
			\cite{baudouin2014frontal} &
			\begin{tabular}[c]{@{}c@{}}C++\\ (Gmsh branch)\end{tabular} &
			no &
			GPL 2 &
			\href{https://gitlab.onelab.info/gmsh/gmsh/-/tree/quadMeshingTools}{gitlab} \\ \hline
			\begin{tabular}[c]{@{}l@{}}\cite{gregson_polycube_2011}\\ \cite{polycut_livesu}\\ \cite{Livesu:2015}\end{tabular} &
			executable &
			yes &
			\begin{tabular}[c]{@{}c@{}}patented,\\ one month trial\end{tabular} &
			\href{http://www.cs.ubc.ca/labs/imager/tr/2018/HexDemo/}{webpage} \\ \hline
			\cite{closedform_pc2016} &
			C++ &
			yes &
			-- &
			\href{http://www.cad.zju.edu.cn/home/hj/16/closed-form-polycube/closed-form-polycube-release-V1.0.7z}{zip} \\ \hline
			\cite{lyon2016hexex} &
			C++ &
			yes &
			GPL 3 &
			\href{https://www.graphics.rwth-aachen.de/software/libHexEx/}{webpage} \\ \hline
			\cite{fu2016efficient} &
			\begin{tabular}[c]{@{}c@{}}C++\\ (incomplete)\end{tabular} &
			no &
			-- &
			\href{http://staff.ustc.edu.cn/~fuxm/projects/EfficientPolyCube/PolyCube.zip}{zip} \\ \hline
			\cite{hybridHexa} &
			C++ &
			yes &
			-- &
			\href{https://github.com/gaoxifeng/robust_hex_dominant_meshing}{github} \\ \hline
			\cite{Gao2017Evaluation} &
			C++ &
			no &
			-- &
			\href{https://github.com/gaoxifeng/Evaluation_code_SGP2017}{github} \\ \hline
			\cite{gao2017simplification} &
			C++ &
			no &
			MPL 2 &
			\href{https://github.com/gaoxifeng/Robust-Hexahedral-Re-Meshing}{github} \\ \hline
			\cite{xu2018hexahedral} &
			C++ &
			no &
			GPL 3 &
			\href{https://github.com/Cotrik/CotrikMesh}{github} \\ \hline
			\cite{xu2018untangling} &
			C++ &
			yes &
			-- &
			\href{https://cotrik.github.io/research/projects/2017_hexmesh_optimization/files/AngleBased_HexOpt_Materials.zip}{github} \\ \hline
			\cite{liu2018singularity} &
			C++&
			yes &
			GPL 3 &
			\href{https://gitlab.vci.rwth-aachen.de:9000/SCOF/SingularityConstrainedOctahedralFields}{gitlab}\\ \hline
			\cite{yang2019computing} &
			C++ &
			yes &
			-- &
			\href{http://rec.ustc.edu.cn/share/dcd64240-c881-11e9-af13-3b5e4718f6cd}{link}\\ \hline
			\cite{bracci2019hexalab} &
			\begin{tabular}[c]{@{}c@{}}C++\\ Javascript\end{tabular} &
			yes &
			MIT &
			\href{https://github.com/cnr-isti-vclab/HexaLab}{github} \\ \hline
			\cite{Kenshi:2019} &
			C++ &
			no &
			BSD 3 &
			\href{https://bitbucket.org/kenshi84/dual-sheet-meshing/src/master/}{bitbucket} \\ \hline
			\cite{gao2019feature} &
			C++ &
			yes &
			-- &
			\href{https://github.com/gaoxifeng/Feature-Preserving-Octree-Hex-Meshing}{github} \\ \hline
			\cite{reberol2019multiple} &
			\begin{tabular}[c]{@{}c@{}}C++\\ (Gmsh branch)\end{tabular} &
			no &
			GPL 2 &
			\href{https://gitlab.onelab.info/gmsh/gmsh/-/tree/hexbl}{gitlab} \\ \hline
			\cite{palmer2019algebraic} &
			Matlab &
			yes &
			MIT &
			\href{https://github.com/dpa1mer/arff}{github} \\ \hline
			\cite{verhetsel2019finding} &
			C &
			no &
			GPL &
			\href{https://www.hextreme.eu/Download/topological-hex-0.3.0.tar.gz}{zip} \\ \hline
			\cite{LoopyCuts2020} &
			C++ &
			yes &
			GPL 3 &
			\href{https://github.com/mlivesu/LoopyCuts}{github} \\ \hline
			\cite{guo2020cut} &
			C++ &
			yes &
			MIT &
			\href{https://github.com/msraig/CE-PolyCube}{github} \\ \hline
			\cite{Marschner2020SOS} &
			Matlab &
			yes &
			MIT &
			\href{https://github.com/zoemarschner/SOS-hex}{github} \\ \hline
			\cite{yu2020hexgen} &
			C++ &
			yes &
			-- &
			\href{https://github.com/CMU-CBML/HexGen_Hex2Spline}{github} \\ \hline
			\cite{PLCGS21} &
			C++ &
			no &
			MIT &
			\href{https://github.com/cg3hci/Gen-Adapt-Ref-for-Hexmeshing}{github} \\ \hline
			\cite{livesu2021optimal} &
			\begin{tabular}[c]{@{}c@{}}C++\\ (inside Cinolib)\end{tabular} &
			no &
			MIT &
			\href{https://github.com/mlivesu/cinolib/blob/master/include/cinolib/hex_transition_schemes.h}{github} \\ \hline
			\cite{Neuser2021} &
			C++ &
			yes &
			BSD 2 &
			\href{https://github.com/chrismile/HexVolumeRenderer}{github} \\ \hline
			\cite{xu2021singularity} &
			executable &
			yes &
			-- &
			\href{https://github.com/ohehe/HexMeshSimplification}{github} \\ \hline
			\cite{interactive_polycubes} &
			C++ &
			no &
			MIT &
			\href{https://github.com/lingxiaoli94/interactive-hex-meshing}{github}\\ \hline
			\cite{yu2021hexdom} &
			C++ &
			yes &
			-- &
			\href{https://github.com/CMU-CBML/HexDom}{github}\\ \hline
		\end{tabular}%
	}
}
\vspace{-3mm}
\end{table}

\begin{table*}[]
	\caption{\minrev{List of existing open source toolkits for visualization and processing of hex and hex-dominant meshes. Note that while some of these tools endow a broader set of facilities (e.g. for surface mesh processing), the table summarizes only the aspects that are relevant for the scope of this article.}}
	\label{tab:toolkits}
	\minrev{
	\resizebox{\linewidth}{!}{%
		\begin{tabular}{llllllllll}
			\textbf{Name} &
			\multicolumn{1}{c}{\textbf{Type}} &
			\multicolumn{1}{c}{\textbf{\begin{tabular}[c]{@{}c@{}}Supported\\geometries\end{tabular}}} &
			\multicolumn{1}{c}{\textbf{\begin{tabular}[c]{@{}c@{}}File\\formats\end{tabular}}} &
			\multicolumn{1}{c}{\textbf{\begin{tabular}[c]{@{}c@{}}Rendering\\facilities\end{tabular}}} &
			\multicolumn{1}{c}{\textbf{\begin{tabular}[c]{@{}c@{}}Visual\\inspection\end{tabular}}} &
			\textbf{\begin{tabular}[c]{@{}l@{}}Mesh\\attributes\end{tabular}} &
			\multicolumn{1}{c}{\textbf{\begin{tabular}[c]{@{}c@{}}Tools for\\volume\\processing\end{tabular}}} &
			\multicolumn{1}{c}{\textbf{License}} &
			\multicolumn{1}{c}{\textbf{URL}} \\ \hline
			\begin{tabular}[c]{@{}c@{}}CinoLib\\\cite{cinolib}\end{tabular} &
			\begin{tabular}[c]{@{}c@{}}C++ library\\(header only)\end{tabular} &
			\multicolumn{1}{c}{\begin{tabular}[c]{@{}c@{}}tetrahedra,\\hexahedra,\\general\\polyhedra\end{tabular}} &
			\multicolumn{1}{c}{\begin{tabular}[c]{@{}c@{}}.MESH,\\.VTU,\\.VTK,\\.HEDRA$^1$,\\.HEXEX$^2$,\\.HYBRID$^3$,\\Tetgen$^4$\end{tabular}} &
			\multicolumn{1}{c}{yes} &
			\begin{tabular}[c]{@{}c@{}}plane slicing\\(axis aligned),\\thresholding,\\manual selection,\\ambient occlusion\end{tabular} &
			\begin{tabular}[c]{@{}c@{}}generic\\attributes\\ for all\\mesh\\elements\end{tabular} &
			\begin{tabular}[c]{@{}c@{}}grid hex-meshing\\facilities\\
			(schemes~\cite{livesu2021optimal},\\surface mapping,\\faeture mapping~\cite{gao2019feature}),\\
			hex-to-tet\\ conversion\\\cite{dompierre1999subdivide},\\ 
			extraction of\\coarse block layouts,\\ volume smoothing,\\subdivision schemes,\\padding,\\all quality metrics in\\\cite{stimpson2007verdict}\end{tabular} &
			MIT &
			\href{https://github.com/mlivesu/cinolib}{github} \\ \hline
			\begin{tabular}[c]{@{}c@{}}HexaLab\\\cite{bracci2019hexalab}\end{tabular} &
			web app &
			hexahedra &
			\begin{tabular}[c]{@{}l@{}}.MESH,\\ .VTK\end{tabular} &
			yes &
			\begin{tabular}[c]{@{}l@{}}plane slicing,\\thresholding,\\peeling,\\manual selection,\\ambient occlusion\end{tabular} &
			\begin{tabular}[c]{@{}l@{}}scalar attributes\\(per cell)\end{tabular} &
			\begin{tabular}[c]{@{}l@{}}all quality metrics in\\\cite{stimpson2007verdict}\end{tabular} &
			GPL 3 &
			\begin{tabular}[c]{@{}l@{}}\href{https://www.hexalab.net}{webpage}\\ \href{https://github.com/cnr-isti-vclab/HexaLab}{github}\end{tabular} \\ \hline
			\begin{tabular}[c]{@{}l@{}}GEOGRAM\\Graphite\\\cite{geogram}\end{tabular} &
			C++ library &
			\begin{tabular}[c]{@{}l@{}}tetrahedra,\\hexahedra,\\pyramids,\\wedges,\\connectors\\(between solid elements)\end{tabular} &
			\begin{tabular}[c]{@{}l@{}}.GEOGRAM$^5$\\.MESH\end{tabular} &
			\begin{tabular}[c]{@{}l@{}}yes\end{tabular} &
			plane slicing &
			\begin{tabular}[c]{@{}l@{}}generic\\ attributes\\for all\\mesh elements\end{tabular} &
			\begin{tabular}[c]{@{}l@{}}quality metrics,\\histograms\end{tabular} &
			BSD 3 &
			\href{https://members.loria.fr/BLevy/PACKAGES/geogram_1.7.7.zip}{zip} \\ \hline
			\begin{tabular}[c]{@{}c@{}}OpenVolumeMesh\\\cite{openvolume}\end{tabular} &
			C++ library &
			general polyhedra &
			.OVM$^6$ &
			no &
			-- &
			\begin{tabular}[c]{@{}l@{}}generic\\ attributes\\for all\\mesh elements\end{tabular} &
			-- &
			GPL 3 &
			\href{https://www.graphics.rwth-aachen.de/software/openvolumemesh/}{webpage} \\ \hline
			\begin{tabular}[c]{@{}c@{}}PolyScope\\\cite{polyscope}\end{tabular} &
			C++ library &
			\begin{tabular}[c]{@{}l@{}}tetrahedra,\\hexahedra,\\hybrid\\(soup of\\non conforming\\tets and hexa)\end{tabular} &
			.MESH &
			yes &
			plane slicing &
			\begin{tabular}[c]{@{}l@{}}scalars and vectors\\for all\\mesh elements\end{tabular} &
			-- &
			MIT &
			\href{http://polyscope.run}{webpage} \\ \hline
			\begin{tabular}[c]{@{}c@{}}PyMesh\\\cite{pymesh}\end{tabular} &
			Python library &
			\begin{tabular}[c]{@{}l@{}}tetrahedra,\\hexahedra\end{tabular} &
			\begin{tabular}[c]{@{}l@{}}.MESH,\\.MSH,\\Tetgen$^4$\end{tabular} &
			yes &
			plane slicing &
			\begin{tabular}[c]{@{}l@{}}scalars and vectors\\for all\\mesh elements\end{tabular} &
			-- &
			-- &
			\href{https://pymesh.readthedocs.io/en/latest/}{webpage} \\ \hline
			\begin{tabular}[c]{@{}c@{}}Py3DViewer\\\cite{py3dviewer2019}\end{tabular} &
			Python library &
			\begin{tabular}[c]{@{}l@{}}tetrahedra,\\hexahedra\end{tabular} &
			.MESH &
			yes &
			plane slicing &
			\begin{tabular}[c]{@{}l@{}}integer attributes\\(per cell)\end{tabular} &
			-- &
			MIT &
			\href{https://github.com/cg3hci/py3DViewer}{github} \\ \hline
		\end{tabular}%
	}
\begin{tiny}
$^1:$ \url{https://github.com/mlivesu/cinolib/blob/master/include/cinolib/io/write_HEDRA.cpp}\\
$^2:$ \url{https://www.graphics.rwth-aachen.de/media/resource_files/hexex_input_examples.zip}\\
$^3:$ \url{https://github.com/gaoxifeng/robust_hex_dominant_meshing/blob/master/src/meshio.cpp}\\
$^4:$ \url{https://wias-berlin.de/software/tetgen/fformats.html}\\
$^5:$ \url{http://alice.loria.fr/software/geogram/doc/html/geofile_8h_source.html}\\
$^6:$ \url{https://www.graphics.rwth-aachen.de/media/openvolumemesh_static/Documentation/OpenVolumeMesh-Doc-Latest/file_format.html}\\
\end{tiny}
}
\end{table*}